\documentclass[twocolumn]{aastex63}

\usepackage[caption=false]{subfig}
\usepackage{xcolor}
\usepackage{graphicx}

\received{July 1, 2021}
\revised{September 29, 2021}
\accepted{October 11, 2021}

\submitjournal{AJ}

\shorttitle{2nd-Order Resonances}
\shortauthors{Bailey, Gilbert, \& Fabrycky}

\graphicspath{{./}{figures/}}

\begin{document}

\title{Period Ratio Sculpting Near Second-Order Mean-Motion Resonances}

\correspondingauthor{Nora Bailey}
\email{norabailey@uchicago.edu}

\author[0000-0001-7509-0563]{Nora Bailey}
\affiliation{Department of Astronomy \& Astrophysics, University of Chicago, Chicago, IL 60637}

\author[0000-0003-0742-1660]{Gregory Gilbert}
\affiliation{Department of Astronomy \& Astrophysics, University of Chicago, Chicago, IL 60637}

\author[0000-0003-3750-0183]{Daniel Fabrycky}
\affiliation{Department of Astronomy \& Astrophysics, University of Chicago, Chicago, IL 60637}

\begin{abstract}

Second-order mean-motion resonances lead to an interesting phenomenon in the sculpting of the period ratio distribution due to their shape and width in period-ratio/eccentricity space. As the osculating periods librate in resonance, the time-averaged period ratio approaches the exact commensurability. The width of second-order resonances increases with increasing eccentricity, and thus more eccentric systems have a stronger peak at commensurability when averaged over sufficient time. The libration period is short enough that this time-averaging behavior is expected to appear on the timescale of the Kepler mission. Using N-body integrations of simulated planet pairs near the 5:3 and 3:1 mean-motion resonances, we investigate the eccentricity distribution consistent with the planet pairs observed by Kepler. This analysis, an approach independent from previous studies, shows no statistically significant peak at the 3:1 resonance and a small peak at the 5:3 resonance, placing an upper limit on the Rayleigh scale parameter, $\sigma$, of the eccentricity of the observed Kepler planets at $\sigma=0.245$ (3:1) and $\sigma=0.095$ (5:3) at 95\% confidence, consistent with previous results from other methods.

\end{abstract}


\section{Introduction} \label{sec:intro}

The interactions between two planets orbiting a star become especially strong when the planets' periods are integer multiples of one another. These period commensurabilities are called mean-motion resonances (MMRs), and their order is determined by the difference in the integer multiples; e.g. a first-order resonance has an $m:(m-1)$ period ratio. Different order MMRs have different characteristics, and in this paper we will examine how the symmetrical shape of second-order MMRs affects the aggregate period ratio distribution for a population of planets.

A general overview of MMRs is given in \cite{1999Murray} and a recent detailed analytical treatment can be found in \cite{2019Hadden}. An analysis of the period ratio distribution of Kepler multi-transiting systems, including features related to MMRs, was in \cite{2014Fabrycky}.

When planets are in resonance, their orbital periods vary. The typical rate at which this variation occurs is faster than the orbital precession by a factor of $ ( \frac{M_\star}{m_p}  )^{1/2}$ \citep{1999Murray}. Therefore, to keep the resonant angles in a librating state, the orbital period ratio averages to very close to the ratio of integers that defines the resonance. 

Orbital eccentricities play a role in the maintenance of a resonance, because it is the eccentricity that supplies torque between two planets \citep{1976Peale}. Therefore, the more eccentric the planets of a system are, the more likely they will be in resonance, and the more likely their period ratios will average out to the nominal value of the resonance. The high-order resonances are more sensitive to eccentricity; the strength of the coupling is proportional to the eccentricity to the power of the order of the resonance \citep{1999Murray}. 

Eccentricities are challenging to measure for transiting exoplanets. The transit timing variation (TTV) method is the most precise for measuring eccentricities, but it favors planets near first-order MMRs and presents a degeneracy between mass and eccentricity \citep{2012Lithwick}. 

Examining the effect of eccentricity on transit durations is a more general but less precise technique. \cite{2014Fabrycky} (building on \cite{2011Moorhead}) compared transit durations between planets in the same system to derive a typical Rayleigh scale. \cite{2008Ford} proposed a similar technique with dependence on the stellar density; later studies were able to apply this technique as improved stellar parameters became available \citep{2015VanEylen,2016Xie,2019VanEylen,2019Mills}.

\cite{2016Shabram} used transit and occultation detections to measure projected eccentricity via transit duration ratios and phase offsets, independent of the stellar density, but this probes generally the Hot Jupiter population.

Alternatively, forward modeling can use observed populations to constrain intrinsic population distributions, e.g. \cite{2019He,2020He}.

The general picture that has emerged from these studies are that multi-planet systems of small (i.e. non-giant) exoplanets tend to have low eccentricities. Both multi-planet giant systems, such as those probed by radial velocity measurements, and single-planet systems tend to have a distinctly excited eccentricity distribution compared to multi-planet sub-giant systems.

We develop a new method to measure the eccentricity distribution of multi-planet systems observed by Kepler using second-order MMRs. This method has different associated systematics and does not depend on precisely measured stellar properties or specific transit observations. Although this method relies on MMRs, given that migration likely doesn't easily trap planets in second-order MMRs \citep{2014Delisle,2015Migaszewski,2017Xu}, we expect it traces post-scattering systems.

The structure of this paper is as follows: Section~\ref{sec:methods} discusses the methodology used to generate and analyze our simulated systems; the results are covered in Section~\ref{sec:results}; an analytical approximation is discussed in Section~\ref{sec:analytics}; Section~\ref{sec:disc} discusses the importance of the results and further considerations, including consideration of the TTVs of the observed planets in K02261; and Section~\ref{sec:conc} summarizes our key findings.

\section{Methods}\label{sec:methods}

For our investigation, our goal was to create a synthetic population of planet pairs that was similar to those found in the Kepler sample. We intentionally emulate the properties of the sample as observed; i.e., we seek to probe the observed distribution and not the intrinsic distribution. To isolate the effects of eccentricity, we use a wide variety of planet pairs like those observed by Kepler to marginalize over other factors, particularly the mass which has a strong degeneracy with eccentricity in MMRs (see Section~\ref{sec:masses}). We then seek to quantitatively compare the resulting second-order MMR peaks at 3:1 and 5:3 for various eccentricity distributions, particularly in comparison to that of the Kepler sample.

\subsection{System Generation}\label{sec:datagen}

In order to numerically investigate planetary behavior near the second-order MMRs, we needed a population of planet pairs. Our intent was to create a population that was similar to that of Kepler, although without being rigorously Kepler-like.

To develop a population of planet pairs near the 5:3 and 3:1 MMRs, we first randomly generated a large number of planet pairs evenly in period ratio near the commensurability location. The mutual inclination of the two planets was chosen randomly from a Rayleigh distribution with scale 0.032 radians \citep{2014Fabrycky}. The inner planet period, each planet mass, and the stellar mass were randomly assigned using distributions based on the population of multi-planet systems observed by Kepler. Arguments of pericenter were chosen uniformly between $-\pi$ and $\pi$. 

This set of generated planet pairs is intended to be used to probe pairs at various eccentricities. Therefore, to ensure a wide coverage of eccentricity, several methods of assigning eccentricities were used, including uniformly between 0 and 1, uniformly between 0 and 0.5, and linearly decreasing from 0 to 0.5. For each planet pair, multiple systems were created varying only in the mean longitude of the planets.

This initial population of planet pairs was then integrated for $10^6$ orbits of the outer planet to remove any ``unstable'' pairs. In this context, pairs were considered unstable when they departed from the period ratio range of interest and not as a strict determination of instability. The bounds at which pairs were removed were $<$2.5 or $>$3.5 for the 3:1 resonance and $<$1.55 or $>$1.88 for the 5:3 resonance. The integrations were computed using the IAS15 integrator in the \texttt{REBOUND} package \citep{2012Rein,2015Rein}.

At the end of the first stability run, 50 surviving systems were randomly chosen for each resonance to be integrated for ten times longer. Of these 50 systems, an additional 4 went ``unstable'' (departed the specified period ratio bounds) for the 3:1 systems and an additional 8 for the 5:3 systems, with no apparent preferential location in period ratio. These low percentages indicated that the initial stability run was sufficient to ensure a robust sample that would not be dominated by planet pairs on the verge of instability.

From the planet pairs that remained in the sample after the stability cut, one pair was randomly chosen from each initial set of mean-longitude systems to be used in the analysis, and thus each pair used in the analysis has unique parameters. Finally, of these pairs, those that had an average period ratio in close proximity to exact commensurability were selected for the analysis. This last cut was done between 2.94 and 3.06 for the 3:1 resonance ($3.00 \pm 0.06$) and between 1.62 and $1.71\bar{3}$ for the 5:3 resonance ($\frac{5}{3} \pm \frac{7}{150}$). The limits are chosen to be wide enough to capture the expected resonant behavior but narrow enough not to include additional higher-order resonances. Specifically, the limits around the 5:3 were chosen to exclude the 8:5 and 7:4 mean-motion resonances.

Detailed distributions of the properties of the final generated sets of 913 (near 3:1) and 670 (near 5:3) planet pairs are shown in Appendix~\ref{app:pop}.

\subsection{Analysis}\label{sec:analyis}

The final results of simulations described in Section~\ref{sec:datagen} provide an instantaneous snapshot of the orbital properties of the planet pairs. These properties change over time, however, as the planets perturb each other, particularly planets that are caught in the mean-motion resonance and librate around a fixed point configuration. Given the time baseline of the Kepler mission, we can actually observe the average orbital properties. Thus the first step of the analysis was to integrate each planet pair for 3.5 years and to time-average the osculating orbital properties over this period. This is done by saving the orbital properties at 2000 steps during the 3.5 year integration and then taking the mean.

The approximate TTV period of planets near a $m:n$ mean-motion resonance is given by \cite{2016Deck}:

\begin{equation}\label{eqn:P_TTV}
    P_{TTV} = \left | \frac{m}{P_2}-\frac{n}{P_1} \right | ^{-1},
\end{equation}

and pairs actually in the resonance will experience libration around the center of the resonance on a timescale shorter than that \citep{2016Nesvorny}. Calculating the TTV period for the final planet pairs in our set of generated systems and comparing to the time-averaging period of 3.5 years, we find that the majority of these planet pairs, 54.2\% for the 3:1 and 81.8\% for the 5:3, will have completed at least one libration cycle. We can also calculate an analytical approximation of the libration periods with the Andoyer module of the \texttt{celmech} package. This predicts a libration period of less than 3.5 years for 31.2\% for the 3:1 pairs and 55.7\% for the 5:3 pairs. Note however that this approximation generally assumes low eccentricities, which is not the case for many of these pairs. Thus we expect that for both generated populations, the average period ratio would fall much closer to the nominal resonance period ratio than the instantaneous period ratio.

\begin{figure*}[h]
	\centering
	\subfloat{
	    \label{fig:PR_distro_pdf-a}
	    \includegraphics[width=.5\linewidth]{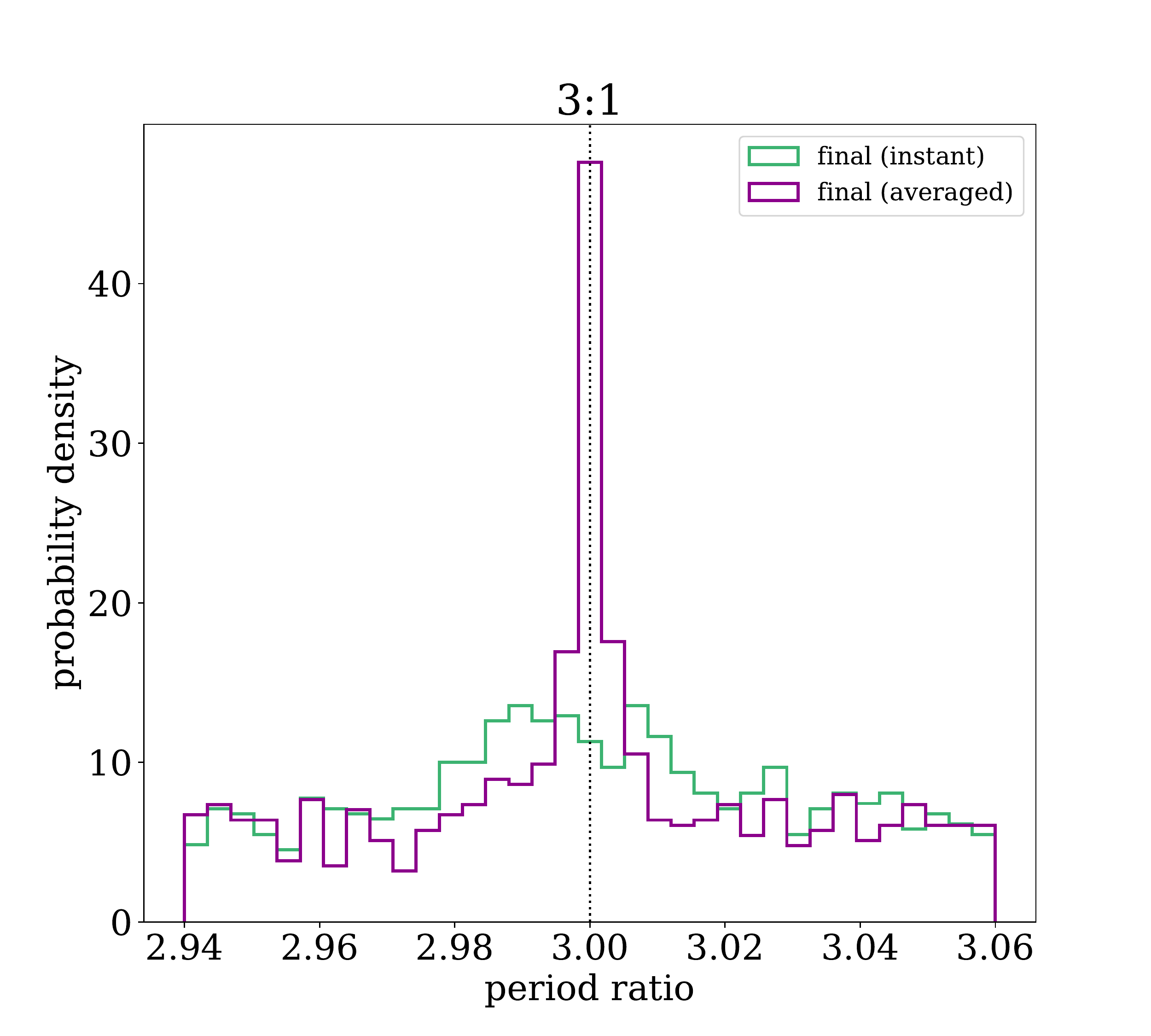}
	}
	\subfloat{
	    \label{fig:PR_distro_pdf-b}
	    \includegraphics[width=.5\linewidth]{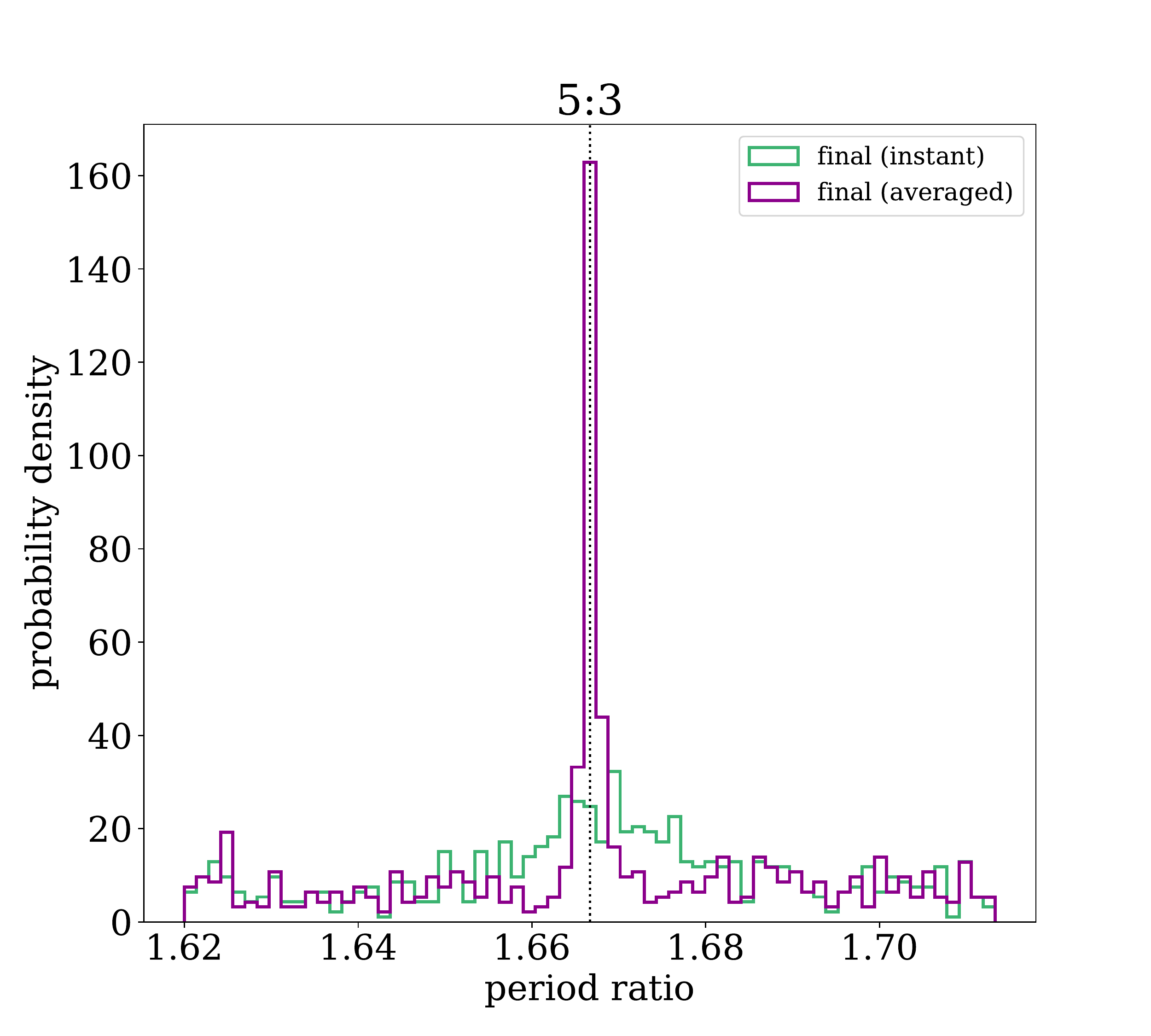}
	}
	\caption{The probability density of period ratios near the \subref{fig:PR_distro_pdf-a} 3:1 and \subref{fig:PR_distro_pdf-b} 5:3 mean-motion resonances. The nominal resonance value is marked with a dotted line. The instantaneous final and averaged final distributions are shown for the specified set of generated systems as described in Section~\ref{sec:datagen}. There is an excess of planet pairs due to the resonance seen in the instantaneous distribution (a result of stability sculpting), but the signal of interest is the narrow peak that is clearly revealed by time-averaging. }
	\label{fig:PR_distro_pdf}
\end{figure*}

This peaking can be seen in Figure~\ref{fig:PR_distro_pdf}. There is also some stability sculpting from the initial uniform period ratio distribution (see Figure~\ref{fig:PRif}). This sculpting is expected, as being near resonance can increase instability, but being in resonance can provide protection against instability, thus leading to an excess of planet pairs near each resonance. However, at any random point, the peak of the excess planet pairs is much more spread out and indistinct than the time-averaged peak. This narrow time-averaged peak is a distinct signature, and is even expected when there is no excess planet pairs (see Figure~\ref{fig:analytical_population}). In our generated systems, we find that no more than approximately one-third of the planet pairs located in the time-averaged peak can be attributed to a stability-induced excess. Thus the effect of sculpting from instability is both expected in real populations and not necessary for the appearance of the time-averaged resonant peak.

\subsection{Model}\label{sec:analysis_model}

To quantify the size and shape of this peak, we use a model comprised of a linear background and a Gaussian peak centered on the nominal resonance value. The normalized equation for this model is

\begin{equation}\label{eqn:model_pdf}
    P(r) = \frac{c_1(r-r_0)+ c_2\mathcal{N}(r_0,c_3)+1}{\int_{r_{min}}^{r_{max}} (c_1(r-r_0)+ c_2\mathcal{N}(r_0,c_3)+1) \, dr},
\end{equation}

and it has three parameters to be fitted: $c_1$, $c_2$, and $c_3$, where $c_1$ is the slope of the linear background distribution, $c_2$ is the amplitude of the Gaussian peak, and $c_3$ is the width of the Gaussian peak (standard deviation). The $\mathcal{N}$ indicates a normal distribution for the given center and width. The period ratio is denoted by $r$, and $r_0$ is the nominal period ratio of the mean-motion resonance. The limits on the period ratio range, $r_{min}$ and $r_{max}$, are taken to be 2.94 and 3.06 for the 3:1 resonance ($3.00 \pm 0.06$) and 1.62 and $1.71\bar{3}$ for the 5:3 resonance ($\frac{5}{3} \pm \frac{7}{150}$).

\subsection{Fitting}\label{sec:analysis_fitting}

To fit for these parameters to our simulated period ratio distribution, we implemented a Markov chain Monte Carlo (MCMC) sampler using the \texttt{emcee} package \citep{2013Foreman-Mackey}. We treat each period ratio in our sample as a Poisson process, calculating their probabilities as individual independent events, which requires no binning of the data. Given that we expect to find a well-defined peak for this set of systems, the priors are uninformative, using a flat log prior for the peak height and width and a prior for the slope that is flat in the angle of the line with the axis, to avoid biasing the slope fit towards higher slopes.

We initialized the MCMC analysis with 32 walkers based on an initial least squares fit to the data. The chains were considered converged when the number of steps was more than 100 times the maximum autocorrelation time calculated using \texttt{emcee} and the autocorrelation time changed less than one percent over 100 steps ($<$10,000 steps in total). We used a burn-in of twice the maximum autocorrelation time and thinned the sample by half the minimum autocorrelation time. 

After fitting for the full set of systems, we then use the resulting posterior of the peak width as a prior for the remaining fittings, including fits to the observations. As some of these fits do not have well-defined peaks, keeping this strict prior allows us to compare whether similar peaks are present in other sets of systems. Without this prior, for a set of systems with no peak, the peak height and width are highly correlated and the peak width is unconstrained. Additionally, this narrow width is a signature of the peak that we are interested in, and using this strict prior ensures we are investigating the peak arising from the time-averaging effect and not, for example, from stability-enhanced MMR protection. 

We fit for the Kepler data set around each commensurability to find values for the three parameters. The KOIs included in this data set are listed in Table~\ref{tab:KOIs}. We take the KOIs from the NASA Exoplanet Archive\footnote{https://exoplanetarchive.ipac.caltech.edu} for all candidate and confirmed KOIs. Including candidate KOIs may mean some pairs are not genuine planets. Looking at planet pairs is an intrinsically multi-planet phenomena, which leads to some detection biases from masking of transits in the light curve \citep{2019Zink}. For planets with periods $<$200 days, this is only a 5.5\% efficiency loss, so we neglect this potential bias. Additionally, eccentricity affects the probability of transit \citep{2008Burke}. However, these effects are offsetting, and any net effect would be to make it more likely for eccentric planets to be detected.

To quantify how many additional planets are present due to the time-averaged resonance peak, we find the fractional area of the Gaussian curve that is in excess from the background.

\subsection{Sampling and Fitting Eccentricity Scales}\label{sec:analysis_eccscales}

\begin{figure*}[h]
	\centering
	\subfloat{
	    \label{fig:eif-a}
	    \includegraphics[width=.5\linewidth]{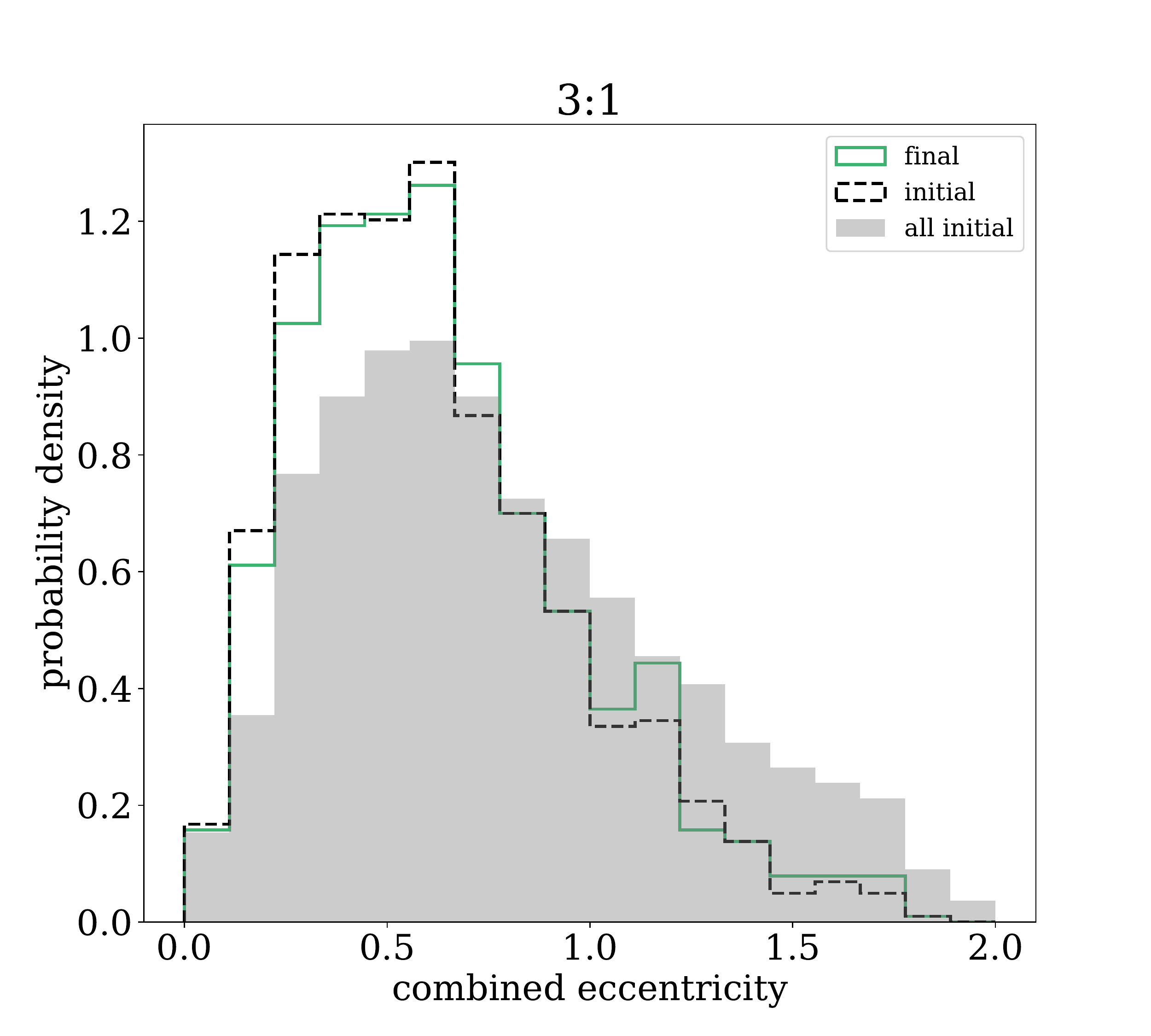}
	}
	\subfloat{
	    \label{fig:eif-b}
	    \includegraphics[width=.5\linewidth]{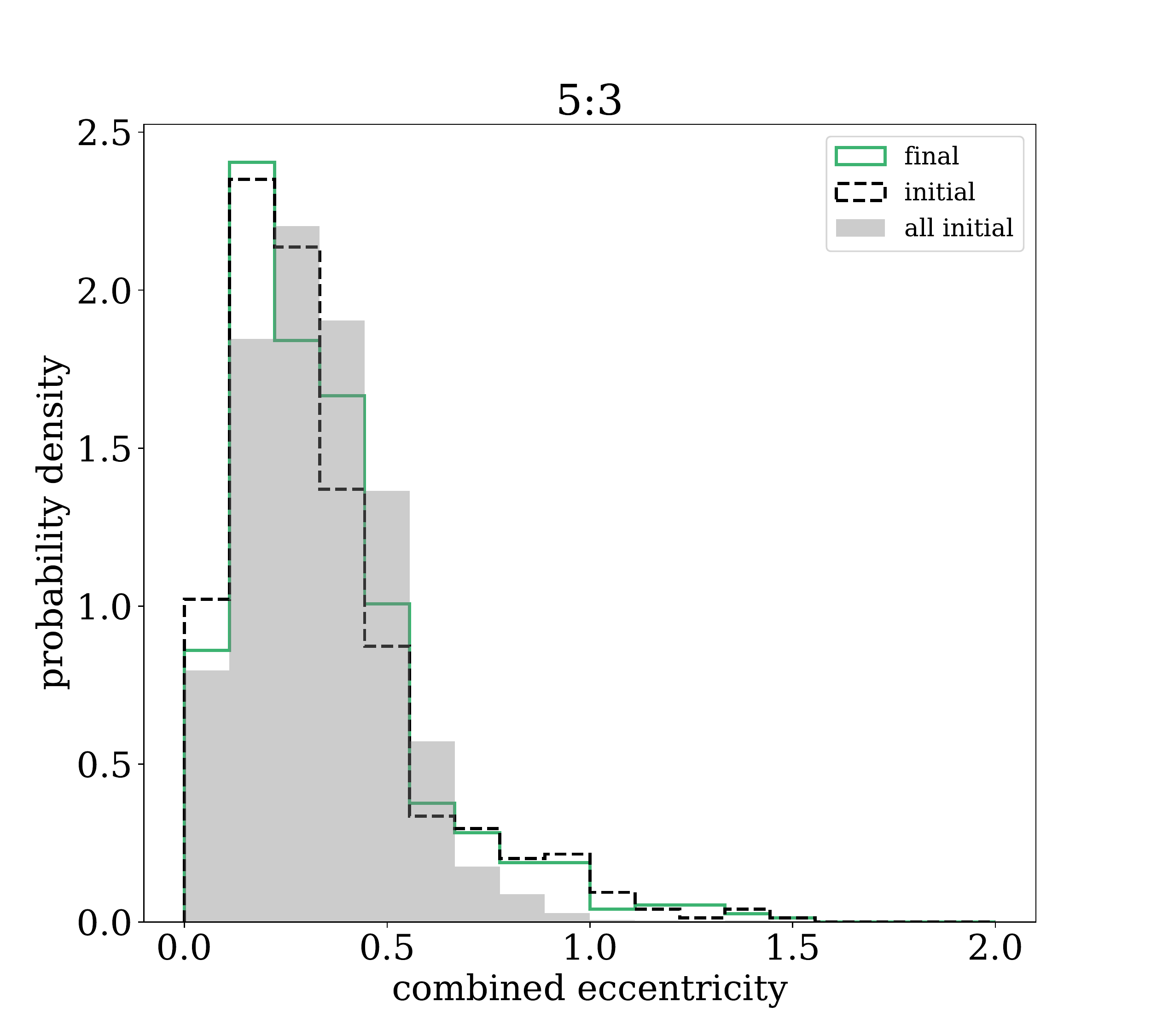}
	}
	\caption{Distribution of the instantaneous eccentricities of the final selected planet pairs at the beginning (dashed black) and end (solid green) of the stability run. Their similarity shows they are in statistical steady-state. The gray background shows the initial combined eccentricity for all planet pairs, including those that went unstable and were excluded from the final selection. There is some stability sculpting, particularly at higher eccentricities, but many highly eccentric planet pairs remain. Distributions are shown for the \subref{fig:eif-a} 3:1 and \subref{fig:eif-b} 5:3 resonances.}
	\label{fig:eif}
\end{figure*}

Figure~\ref{fig:eif} shows the eccentricity distributions of the generated system sets. While there is a clear amount of stability sculpting in the eccentricity distributions, there are many highly-eccentric planet pairs that remain in our stable samples. In order to compare the Kepler distributions to various eccentricity distributions, we take sub-samples of our generated system sets that are consistent with a given eccentricity distribution. This sampling is done by calculating a probability for each eccentricity using the desired eccentricity probability distribution function and then using those probabilities as weights for drawing planet pairs without replacement. The sub-sample size was fixed at the same number of planet pairs as in the Kepler samples, 30 for the 3:1 resonance and 73 for the 5:3 resonance to allow for the same amount of systematic error from the sample size. The sub-samples were checked against a sample of the same size generated from the desired distribution using a 2-sample Kolmogorov–Smirnov test and a 2-sample Anderson-Darling test and accepted only if both p-values were $\geq0.1$.

\begin{figure}[h]
	\centering
    \includegraphics[width=\linewidth]{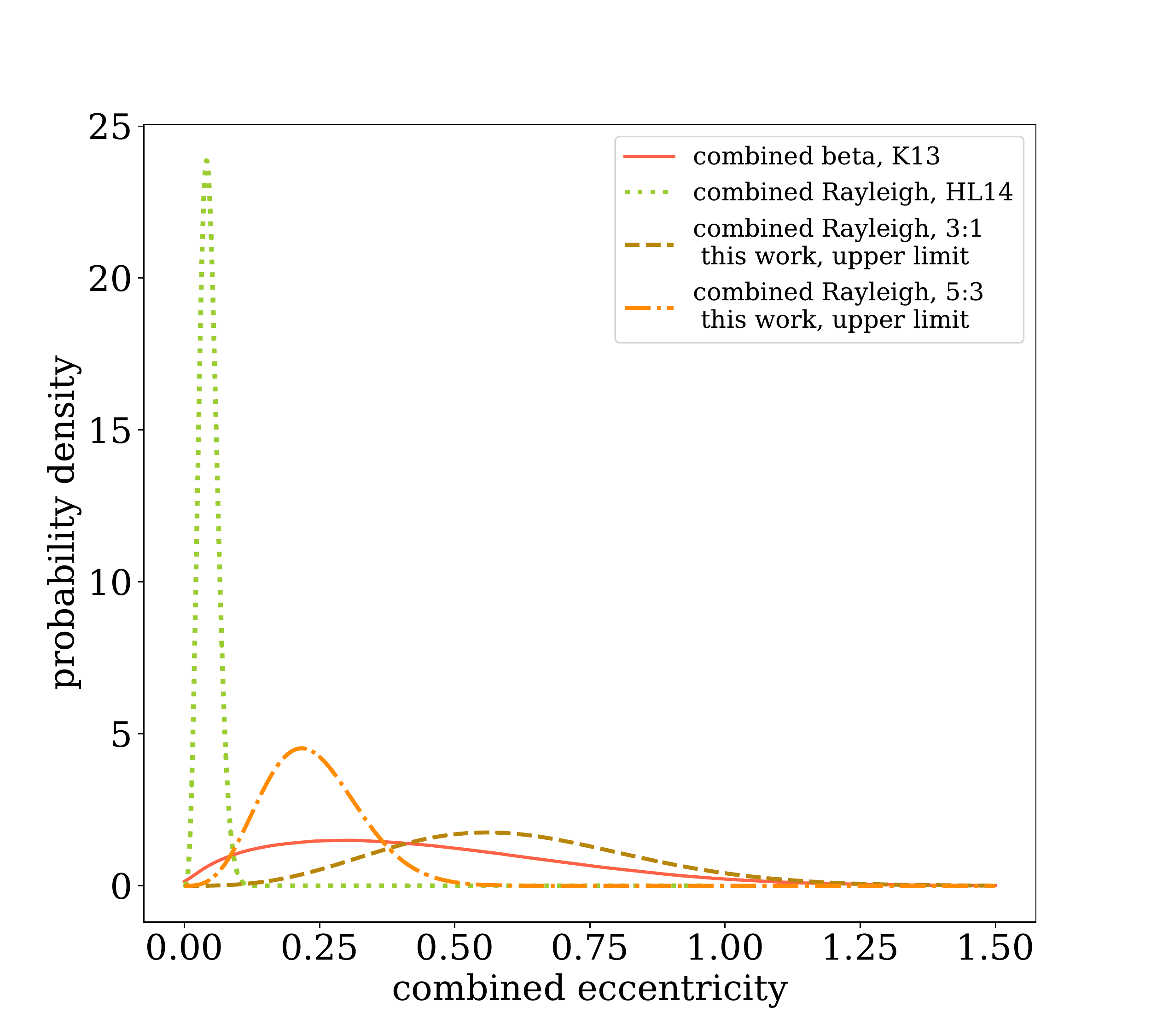}
	\caption{Several probability density functions with example parameters for the combined eccentricity of the planet pairs ($e_1+e_2$). The combined beta distribution was derived numerically from samples taken in accordance with \cite{2013Kipping} parameters and plotted using a Gaussian kernel-density estimate. Combined Rayleigh distributions for various scales were calculated analytically from Equation~\ref{eqn:doubleRayleigh} and plotted for several samples scales: the best-fit value from \cite{2014Hadden}, $\sigma=0.018$; the upper limit for 3:1 from this work, $\sigma=0.245$; and the upper limit for 5:3 from this work, $\sigma=0.095$.}
	\label{fig:eccs_pdfs}
\end{figure}

The eccentricity distributions used are visualized in Figure~\ref{fig:eccs_pdfs}. Given that we are considering planet pairs, the eccentricity distributions are those for combined eccentricities, $e_1+e_2$. We considered the individual eccentricities to be independent.

The combined Rayleigh distributions were calculated analytically. The distribution of the sum of two independent variables is expressed by the convolution of the two distributions, $P(x_1+x_2)=P(x_1) * P(x_2)$. Thus, the sum of two independent Rayleigh random variables with the same scale parameter $\sigma$ has a probability density function of the form

\begin{equation}\label{eqn:doubleRayleigh}
    P(x,\sigma) = \frac{e^{-x^2/2\sigma^2}}{4\sigma^3}\left (2\sigma x - e^{x^2/4\sigma^2}\sqrt{\pi}(2\sigma^2-x^2) \textup{erf}(\frac{x}{2\sigma}) \right ).
\end{equation}

The combined beta distribution was calculated numerically, incorporating the uncertainties in the distribution parameters from \cite{2013Kipping}. A set of $10^6$ eccentricities were drawn, each time with parameters drawn from a normal distribution with the mean and sigma given by the error bars from \cite{2013Kipping}, $\alpha = 0.867 \pm 0.044$ and $\beta = 3.03 \pm 0.17$ (this is slightly modified from $\beta = 3.03^{+0.17}_{-0.16}$ as listed by \cite{2013Kipping} for ease of computation, with negligible effects). The probability density function was then calculated using a Gaussian kernel-density estimate.

We repeated this ten times, using a new random sub-sample, for each eccentricity distribution and combined the resulting chains to get the final posterior for a given eccentricity distribution. For the beta distribution, we increased the number of calculations to fifty, but found negligible improvement in the fit and so only ten were used for the remaining sub-samples. Finally, we used the MCMC analysis on the sub-samples of the generated systems and calculated the fitted fractional peak area at each resonance. The peak area was found by calculating the area due solely to the continuum distribution and subtracting that from one (the normalized area). The results of this fitting and comparison with the Kepler data are discussed next.

\section{Results}\label{sec:results}

\begin{figure*}[h]
	\centering
	\subfloat{
	    \label{fig:alldata_cdf-a}
	    \includegraphics[width=.5\linewidth]{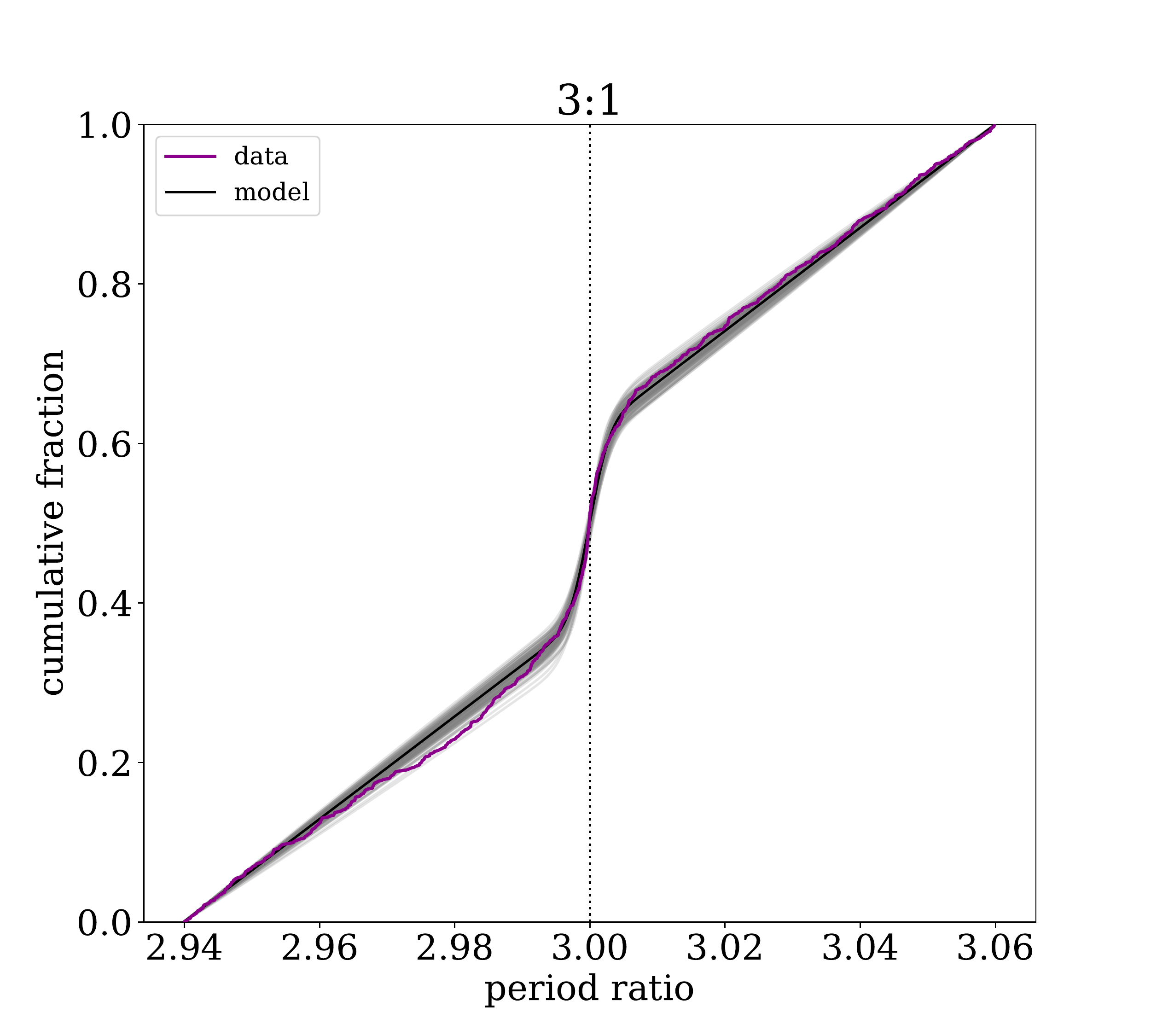}
	}
	\subfloat{
	    \label{fig:alldata_cdf-b}
	    \includegraphics[width=.5\linewidth]{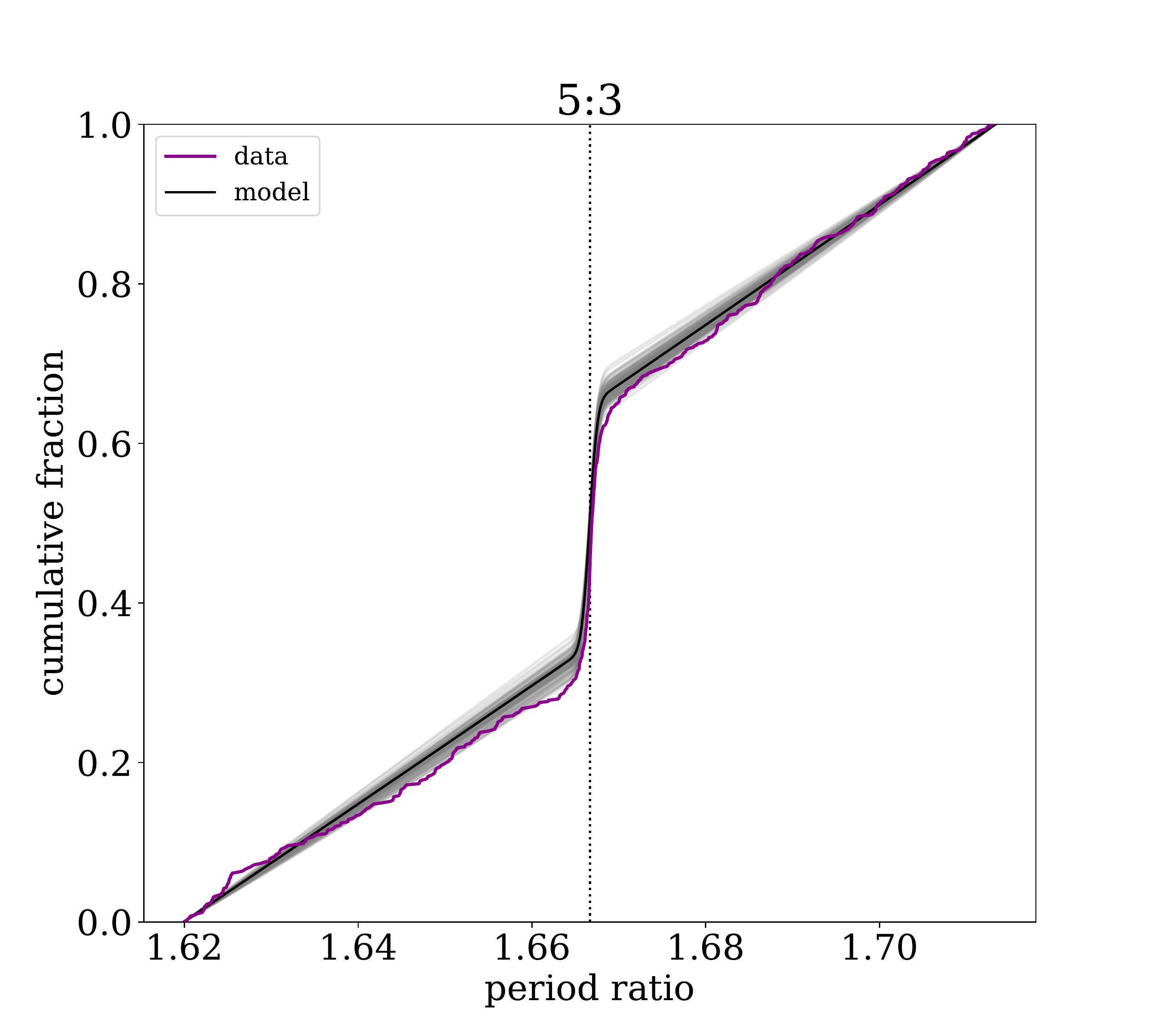}
	}
	\caption{The cumulative distribution functions of the generated system set (magenta) along with the median model of the posteriors (black) and 100 sample draws from the MCMC posteriors (gray) for the \subref{fig:alldata_cdf-a} 3:1 and \subref{fig:alldata_cdf-b} 5:3 mean-motion resonances. The model is tightly fit to the data and shows a clear narrow peak at the resonance.}
	\label{fig:alldata_cdf}
\end{figure*}

Our first result is the detection and quantification of the presence of an excess in planet pairs near the 3:1 and 5:3 resonance in our simulated systems. The fitted parameters are shown in Table~\ref{tab:parameter_results}, with errors listed from the 68.2\% interval, along with the derived parameter of peak area. The data and median models, along with a sample of posterior draws, are shown in Figure~\ref{fig:alldata_cdf}.

\begin{table*}[]
\centering
\begin{tabular}{|l|c|c|c|c|}
\hline
\textbf{} & \textbf{$c_1$} & \textbf{$c_2$} & \textbf{$c_3$} & \textbf{$A_{pk}$} \\ \hline
\multicolumn{5}{|c|}{\textbf{3:1}} \\ \hline
\textbf{Kepler} & $0.1^{+1.0}_{-0.8}$ & $-0.003^{+0.005}_{-0.002}$ & $0.0024 \pm 0.0003$ & $-0.03^{+0.04}_{-0.02}$ \\ \hline
\textbf{Generated Sys (All)} & $0.1 \pm 0.6$ & $0.035 \pm 0.004$ & $0.0023 \pm 0.0003$* & $0.23 \pm 0.03$ \\ \hline
\textbf{Generated Sys (beta)} & $0.0^{+1.0}_{-0.9}$ & $0.02^{+0.02}_{-0.01}$ & $0.0024 \pm 0.0003$ & $0.16^{+0.11}_{-0.09}$ \\ \hline
\textbf{Generated Sys (Rayleigh, $\sigma$=0.07)} & $-0.1^{+0.9}_{-1.0}$ & $0.02^{+0.02}_{-0.01}$ & $0.0024 \pm 0.0003$ & $0.11^{+0.12}_{-0.08}$ \\ \hline
\textbf{Generated Sys (Rayleigh, $\sigma$=0.245)} & $0.0 \pm 1.0$ & $0.03 \pm 0.02$ & $0.0024 \pm 0.0003$ & $0.2 \pm 0.1$ \\ \hline
\textbf{Generated Sys (Rayleigh, $\sigma$=0.34)} & $0.0 \pm 0.9$ & $0.06^{+0.03}_{-0.02}$ & $0.0023 \pm 0.0004$ & $0.3 \pm 0.1$ \\ \hline
\textbf{Generated Sys (Rayleigh, $\sigma$=0.57)} & $-0.1^{+0.9}_{1.0}$ & $0.13^{+0.09}_{-0.05}$ & $0.0024 \pm 0.0003$ & $0.5 \pm 0.1$ \\ \hline
\textbf{Generated Sys (Rayleigh, $\sigma$=0.70)} & $0.0 \pm 0.9$ & $0.17^{+0.12}_{-0.07}$ & $0.0023 \pm 0.0003$ & $0.6 \pm 0.1$ \\ \hline
\multicolumn{5}{|c|}{\textbf{5:3}} \\ \hline
\textbf{Kepler} & $0.2^{+1.1}_{-0.8}$ & $0.004^{+0.004}_{-0.003}$ & $(72 \pm 6) \times 10^{-5}$ & $0.04 \pm 0.03$ \\ \hline
\textbf{Generated Sys (All)} & $0.4^{+1.0}_{-0.7}$ & $0.041 \pm 0.004$ & $(72 \pm 6) \times 10^{-5}$* & $0.31 \pm 0.03$ \\ \hline
\textbf{Generated Sys (beta)} & $0.1^{+1.0}_{-0.9}$ & $0.04^{+0.02}_{-0.01}$ & $(73 \pm 6) \times 10^{-5}$ & $0.32 \pm 0.08$ \\ \hline
\textbf{Generated Sys (Rayleigh, $\sigma$=0.06)} & $-0.1^{+0.8}_{-1.0}$ & $0.012^{+0.007}_{-0.005}$ & $(72 \pm 6) \times 10^{-5}$ & $0.11^{+0.05}_{-0.04}$ \\ \hline
\textbf{Generated Sys (Rayleigh, $\sigma$=0.095)} & $0.1 \pm 0.9$ & $0.024^{+0.012}_{-0.008}$ & $(73 \pm 6) \times 10^{-5}$ & $0.21^{+0.07}_{-0.06}$ \\ \hline
\textbf{Generated Sys (Rayleigh, $\sigma$=0.125)} & $0.3^{+1.5}_{-0.8}$ & $0.033^{+0.01}_{-0.009}$ & $(73 \pm 6) \times 10^{-5}$ & $0.26 \pm 0.06$ \\ \hline
\textbf{Generated Sys (Rayleigh, $\sigma$=0.16)} & $0.2^{+1.3}_{-0.9}$ & $0.06^{+0.02}_{-0.01}$ & $(73 \pm 6) \times 10^{-5}$ & $0.38^{+0.07}_{-0.06}$ \\ \hline
\textbf{Generated Sys (Rayleigh, $\sigma$=0.205)} & $0.2^{+1.1}_{-0.9}$ & $0.09^{+0.03}_{-0.02}$ & $(75 \pm 6) \times 10^{-5}$ & $0.48 \pm 0.08$ \\ \hline
\end{tabular}
\caption{Summary of the fitted parameters from our MCMC posteriors. $c_1$ is the slope of the linear background distribution, $c_2$ is the amplitude of the Gaussian peak, and $c_3$ is the width of the Gaussian peak (standard deviation). The prior on $c_3$ used on the all generated systems set (*) was different from the other fittings; see text for details. The fractional peak area, $A_{pk}$, is derived from the fitted parameters to quantify the additional excess or lack at the resonance location. We include here only the Rayleigh distribution results corresponding with  1-, 2-, 3-, 4-, and 5-$\sigma$ discrepancy with Kepler.}
\label{tab:parameter_results}
\end{table*}

\begin{figure*}[h]
	\centering
	\subfloat{
	    \label{fig:Kepler_cdf-a}
	    \includegraphics[width=.5\linewidth]{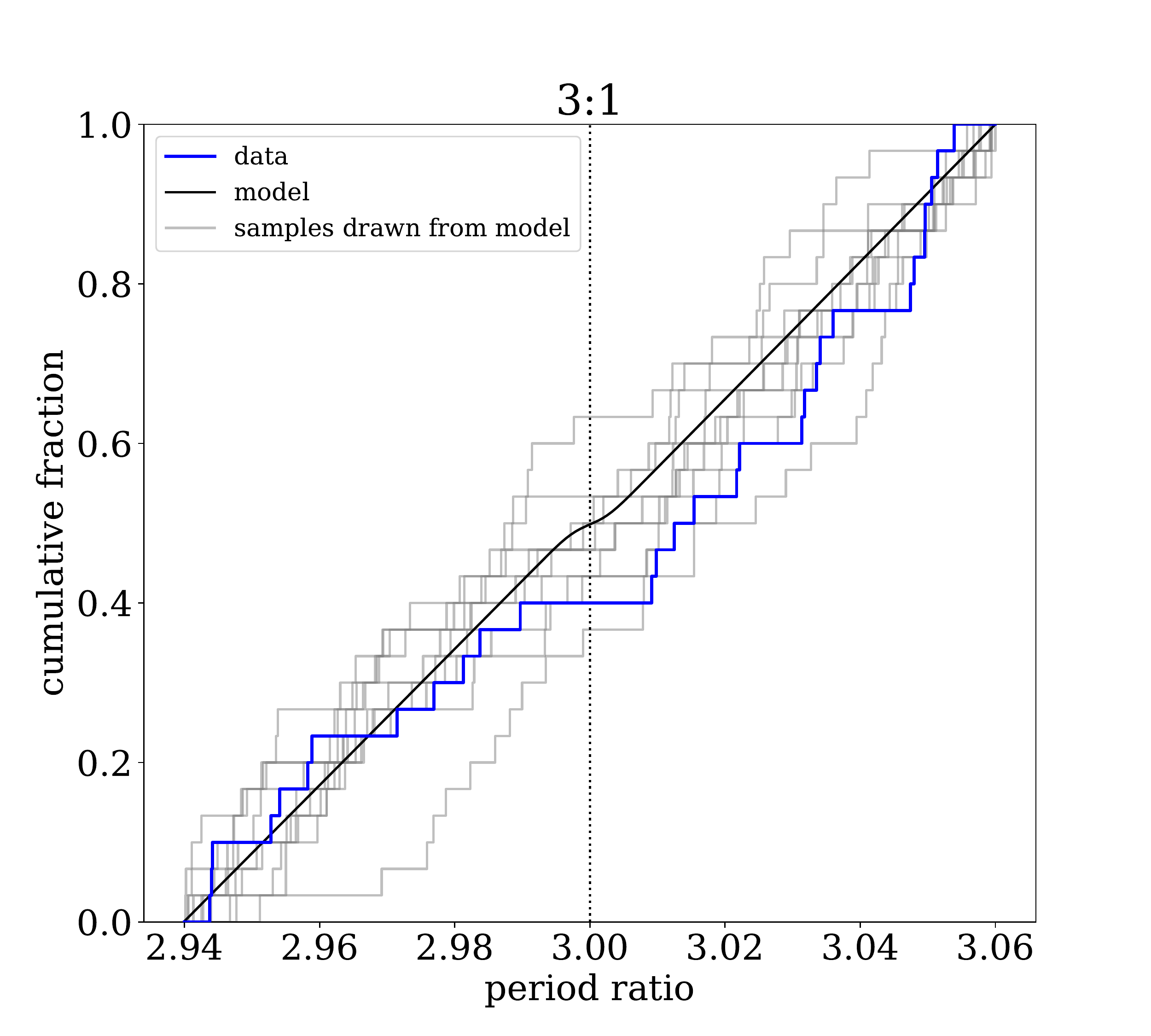}
	}
	\subfloat{
	    \label{fig:Kepler_cdf-b}
	    \includegraphics[width=.5\linewidth]{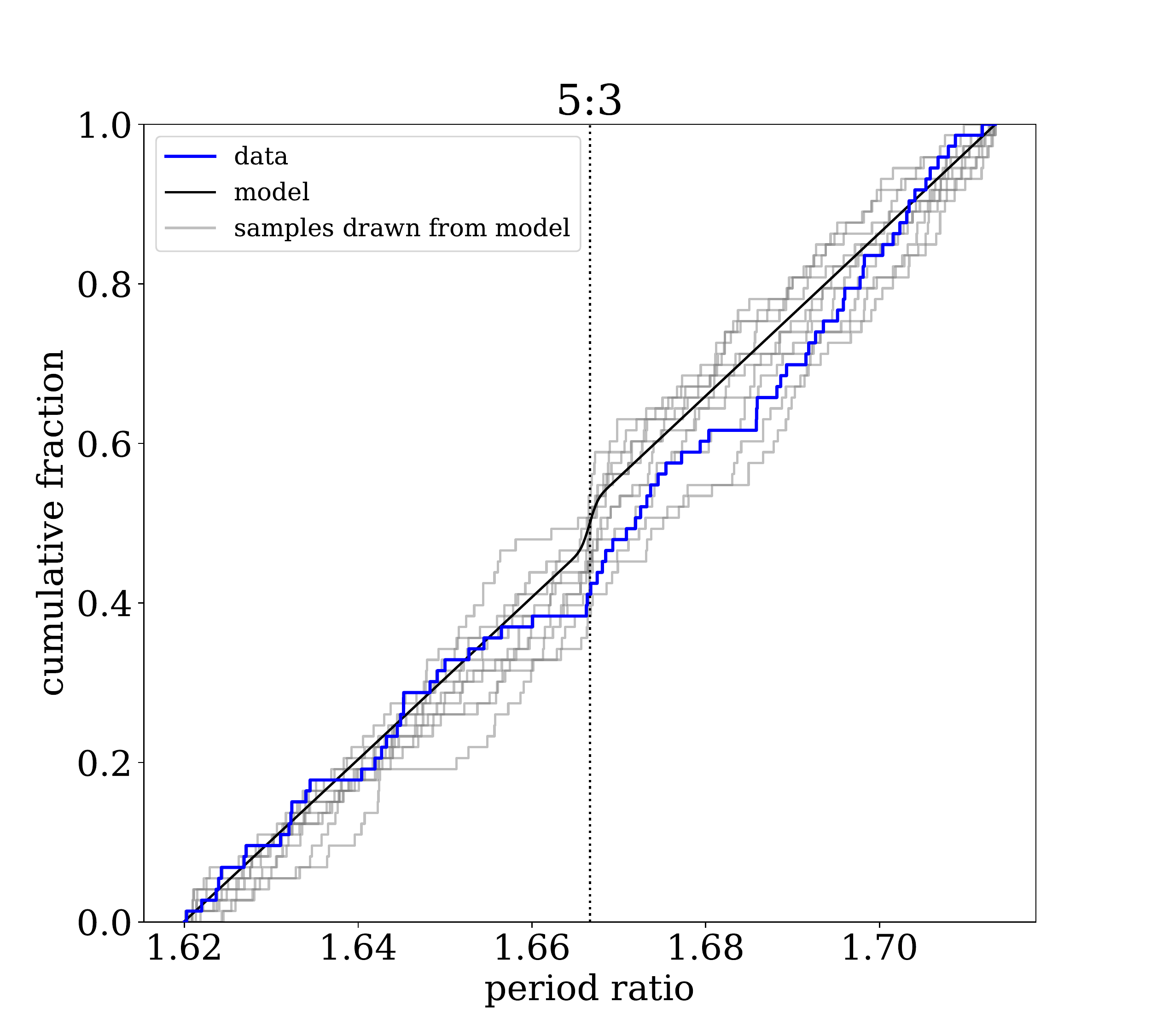}
	}
	\caption{The cumulative distribution functions of the Kepler data (blue) along with the median model of the posteriors (black) and ten samples with size equal to that of the Kepler data drawn from the median model (gray) for the \subref{fig:Kepler_cdf-a} 3:1 and \subref{fig:Kepler_cdf-b} 5:3 mean-motion resonances. The difference in the data and model can be explained by the small sample size of the Kepler data.}
	\label{fig:Kepler_cdf}
\end{figure*}

Next, we classify the presence of a peak near the 3:1 and 5:3 resonances in the Kepler data. The 3:1 for Kepler has no indication of a peak, and in fact the median model is a slight dip at the resonance location; however, the model fit is consistent with a flat background and zero peak. The 5:3 for Kepler has a small peak at the resonance location. The fitted parameters are shown in Table~\ref{tab:parameter_results}, with errors listed from the 68.2\% interval, along with the derived parameter of peak area. The data and median models are shown in Figure~\ref{fig:Kepler_cdf}. While the models are not as closely fit to the data as with the generated systems, this can be attributed to the much smaller sample size. For illustration, we drew ten samples with size equal to that of the Kepler data from the median model for each resonance, also shown in Figure~\ref{fig:Kepler_cdf}. For both resonances, the Kepler distributions fall within the range of these samples. 

\begin{figure*}[h]
	\centering
	\subfloat{
	    \label{fig:Kepbeta_pkarea-a}
	    \includegraphics[width=.5\linewidth]{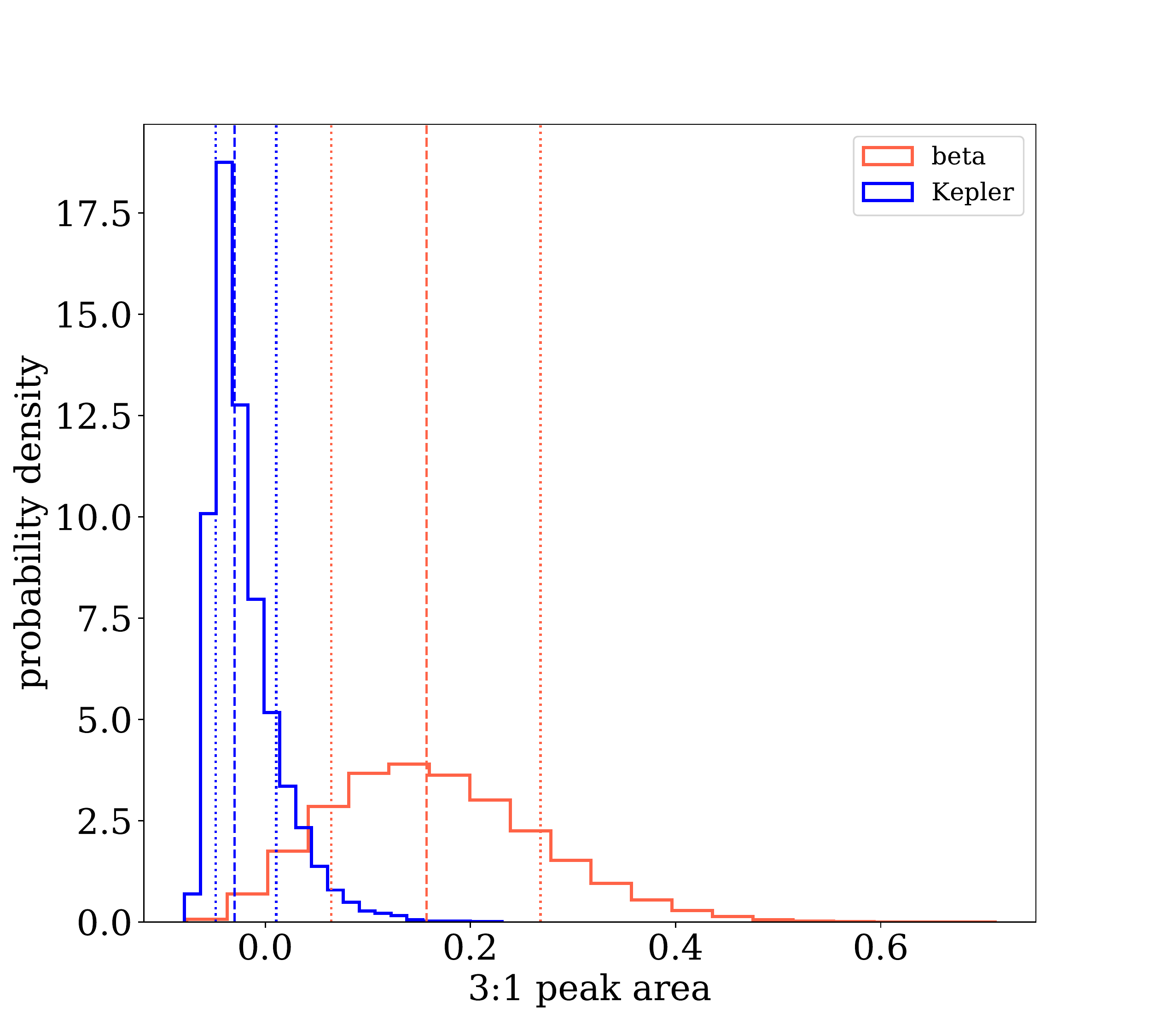}
	}
	\subfloat{
	    \label{fig:Kepbeta_pkarea-b}
	    \includegraphics[width=.5\linewidth]{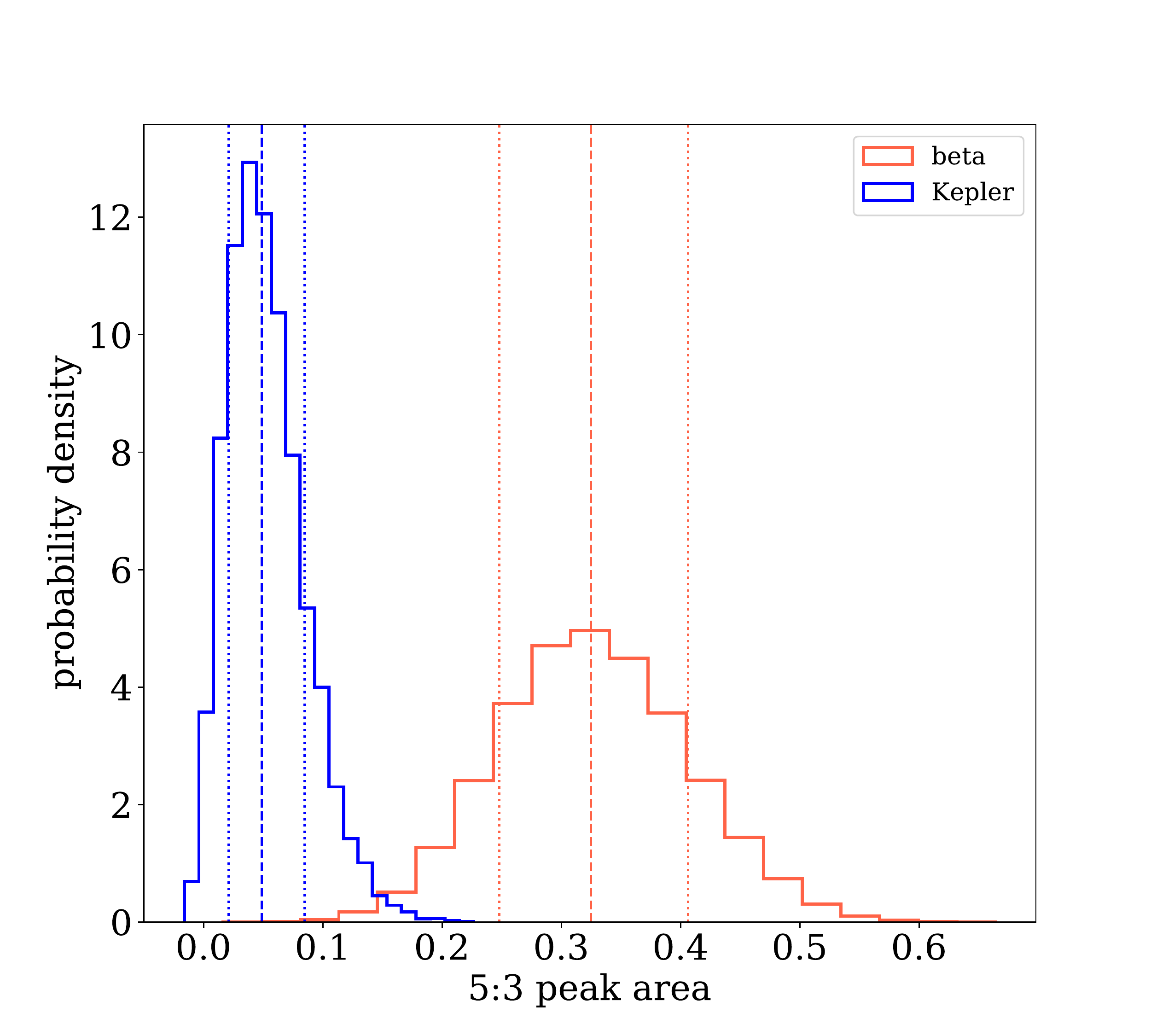}
	}
	\caption{Comparison of the posteriors of the fractional peak area of the \subref{fig:Kepbeta_pkarea-a} 3:1 and \subref{fig:Kepbeta_pkarea-b} 5:3 mean-motion resonances. The Kepler sample is in blue and the simulated combined beta distribution from \cite{2013Kipping} sample in red. The median values are shown with a dashed line and the 68.2\% percentile intervals with dotted lines. The Kepler sample is not consistent with the beta distribution, which was derived from radial velocity planets.}
	\label{fig:Kepbeta_pkarea}
\end{figure*}

We use our sub-sampling as described in Section~\ref{sec:methods} to generate a posterior for the peak area of a set of planet pairs with eccentricity consistent with a beta distribution as described in \cite{2013Kipping}. These posteriors, along with the posteriors of the peak area for the Kepler samples, are shown in Figure~\ref{fig:Kepbeta_pkarea}. The Kepler samples are not consistent with the beta distribution to a level of 1.8$\sigma$ (92.0\% confidence) for the 3:1 and 3.2$\sigma$ (99.9\% confidence) for the 5:3. The fitted parameters for the beta distribution sub-samples are shown in Table~\ref{tab:parameter_results}, with errors listed from the 68.2\% interval, along with the derived parameter of peak area. 

\begin{figure*}[h]
	\centering
	\subfloat{
	    \label{fig:Rayleigh_models-a}
	    \includegraphics[width=.5\linewidth]{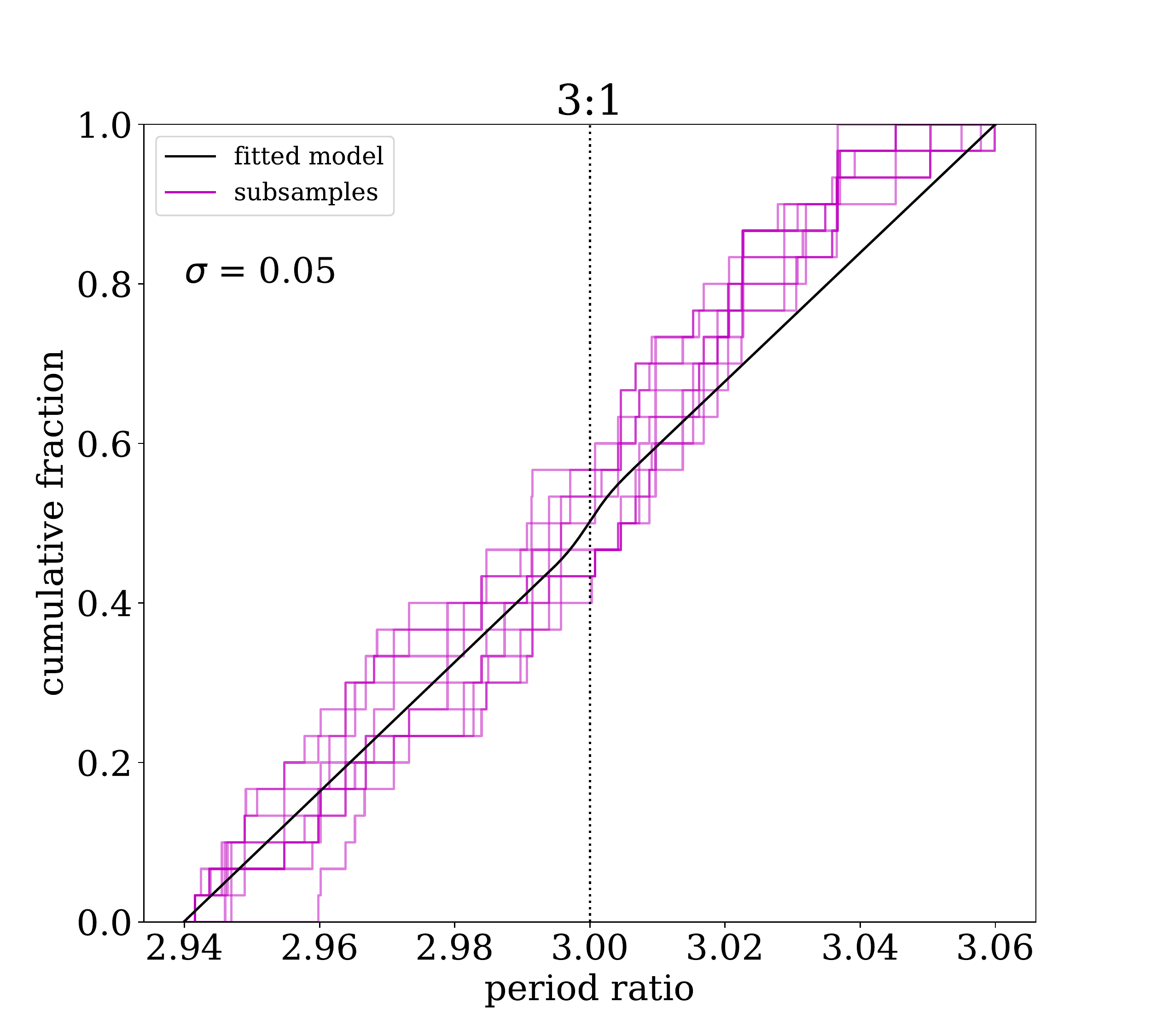}
	}
	\subfloat{
	    \label{fig:Rayleigh_models-b}
	    \includegraphics[width=.5\linewidth]{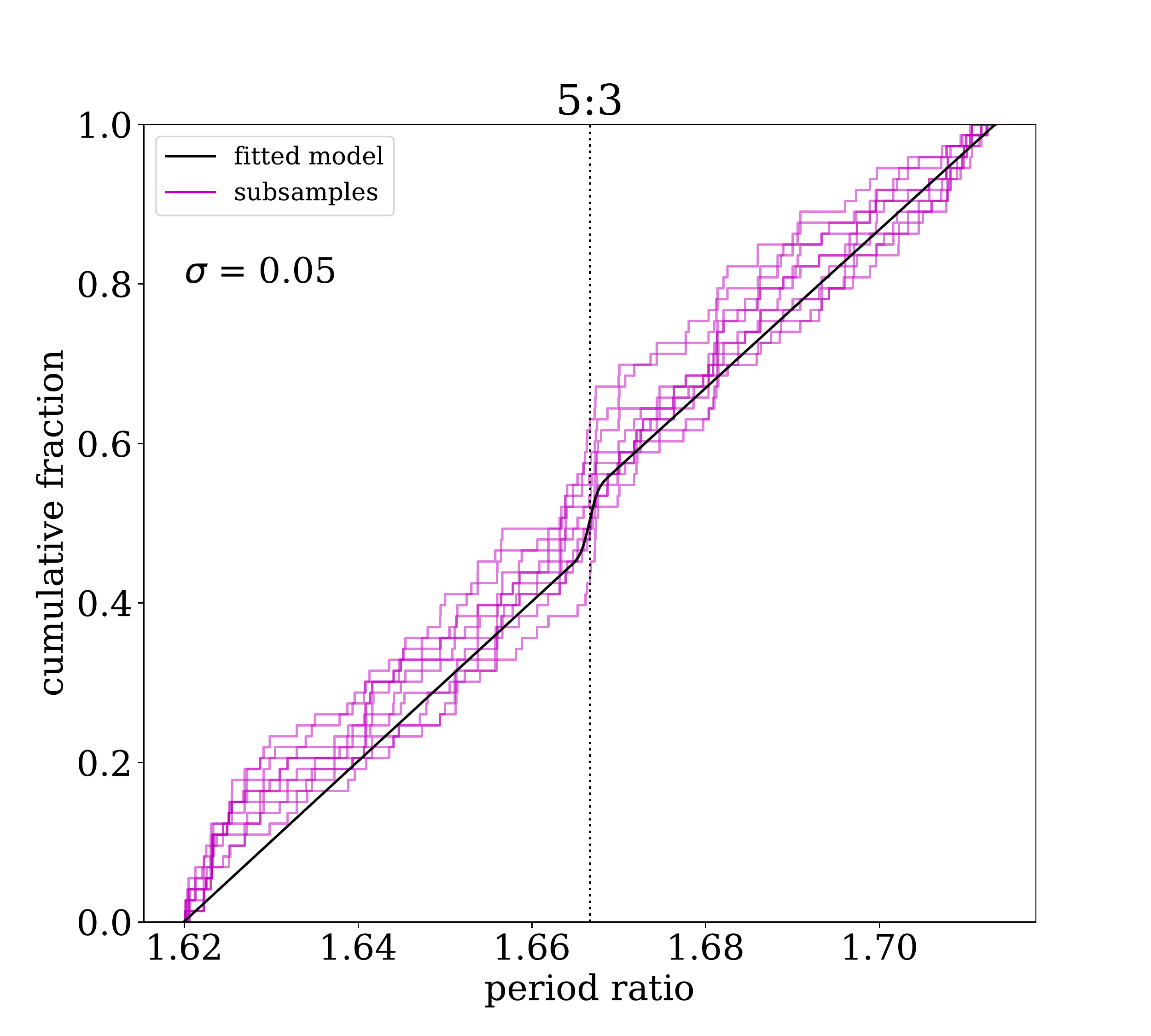}
	}\\
	\subfloat{
	    \label{fig:Rayleigh_models-c}
	    \includegraphics[width=.5\linewidth]{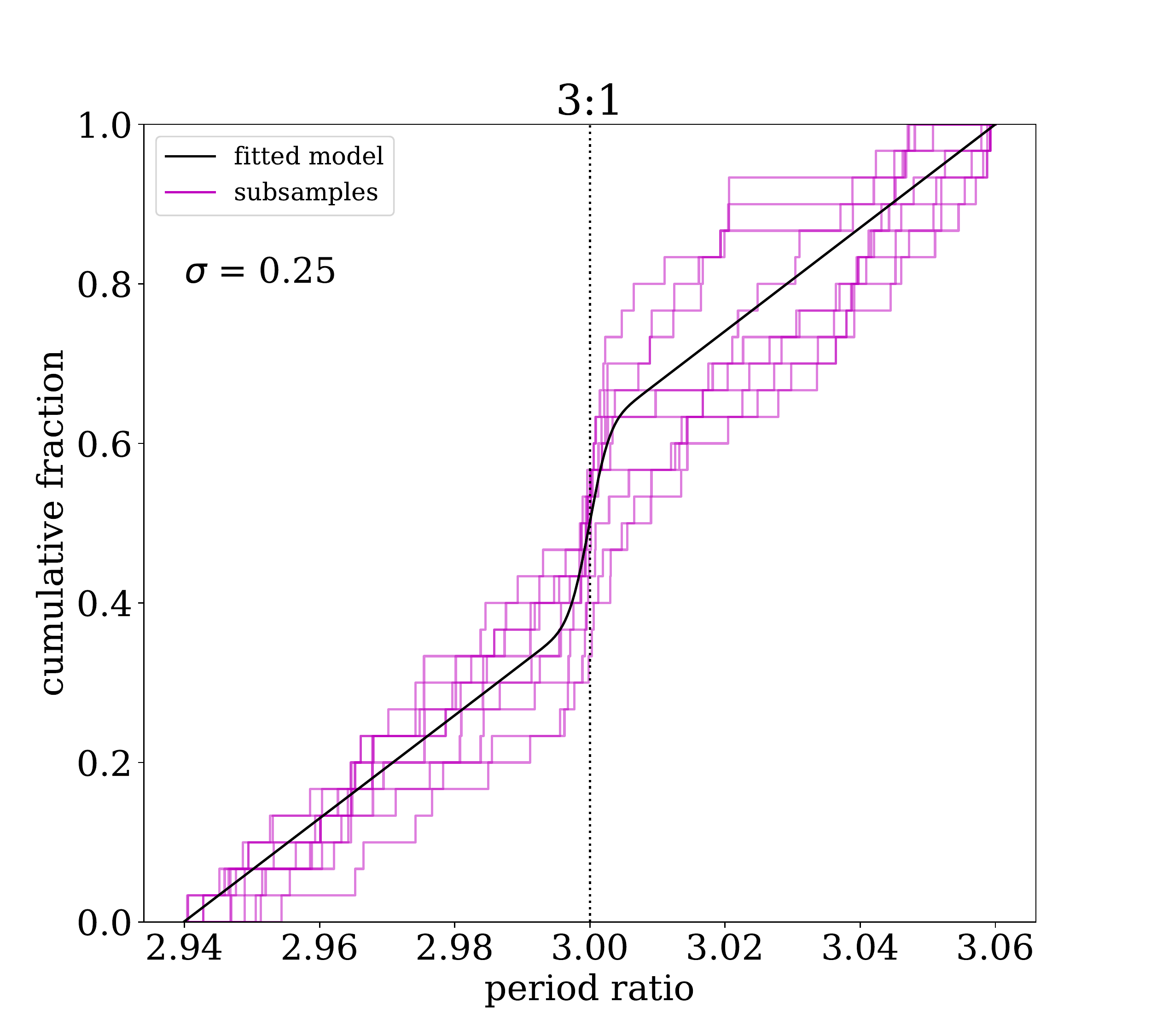}
	}
	\subfloat{
	    \label{fig:Rayleigh_models-d}
	    \includegraphics[width=.5\linewidth]{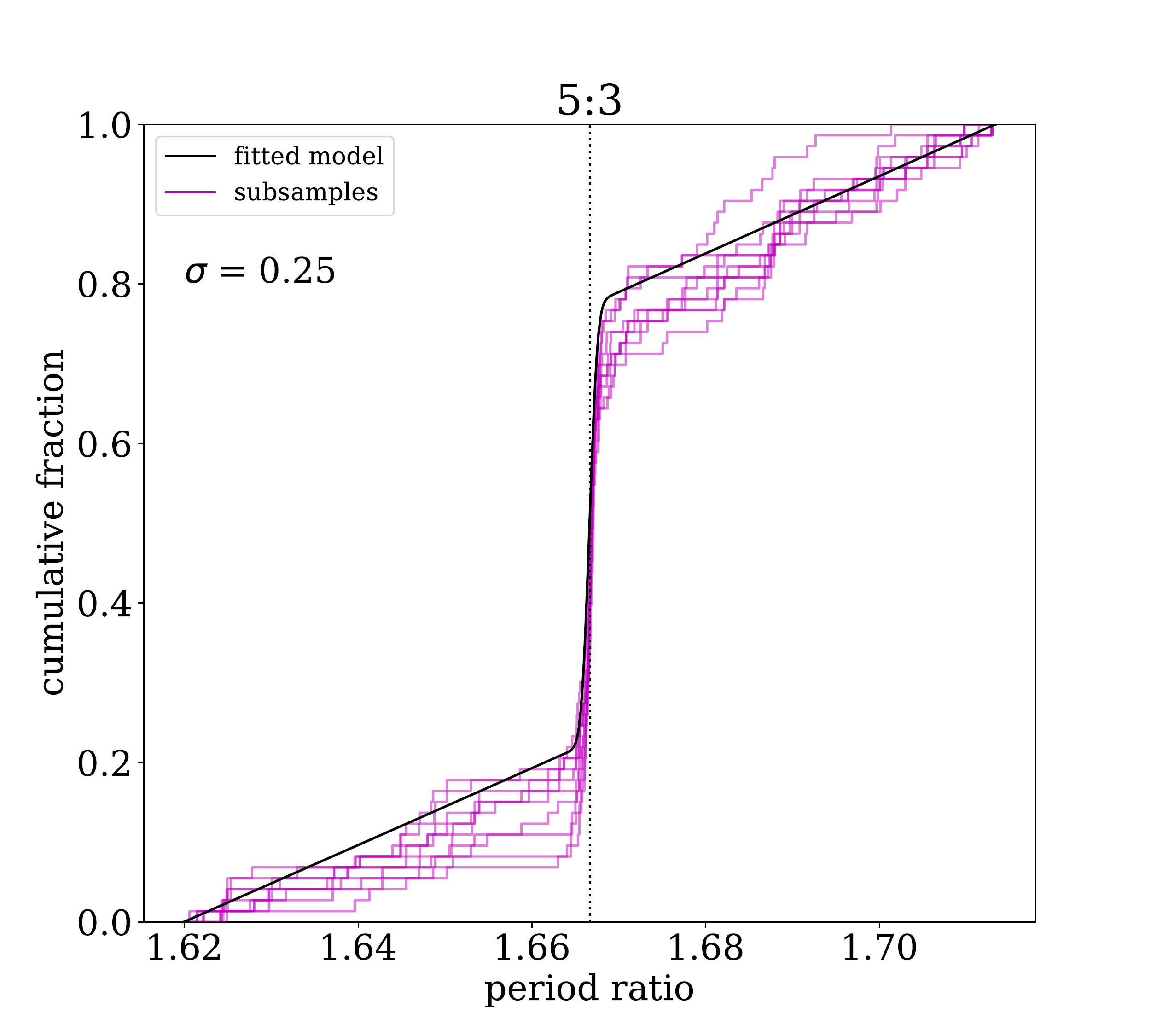}
	}
	\caption{The sub-samples (magenta) and corresponding median model (black) for two Rayleigh example eccentricity scales. The low-eccentricity $\sigma=0.05$ for the \subref{fig:Rayleigh_models-a} 3:1 and \subref{fig:Rayleigh_models-b} 5:3 resonances and the moderate-eccentricity $\sigma=0.25$ for the \subref{fig:Rayleigh_models-c} 3:1 and \subref{fig:Rayleigh_models-c} 5:3 resonances. At low eccentricities, the distribution shows very little peak, while the peak appears more clearly as the eccentricity increases. Our model captures the behavior of the peak at various eccentricity scales.}
	\label{fig:Rayleigh_models}
\end{figure*}

We do a similar comparison for sub-samples of sets of planet pairs with eccentricity consistent with a Rayleigh distribution of various scales, from $\sigma=0.005$ to 1.000 in steps of 0.005. Some scales were poorly sampled by our generated systems and were not analyzed or included in the results. Figure~\ref{fig:Rayleigh_models} shows the results for two of the eccentricity scales for each resonance, including the cumulative distribution for each of the ten random sub-samples as well as the model resulting from the median of the combined MCMC posteriors. At low eccentricities, the distribution has very little peak, as expected. This shows up as almost entirely continuum in our model, which describes the distribution well. At the higher eccentricity, the peak starts to become clear. Our model captures this changing behavior of the distribution, allowing us to quantify the fractional peak area for comparison with the model fitted from the Kepler data.

\begin{figure*}[h]
	\centering
	\subfloat{
	    \label{fig:KepRay_pkarea-a}
	    \includegraphics[width=.5\linewidth]{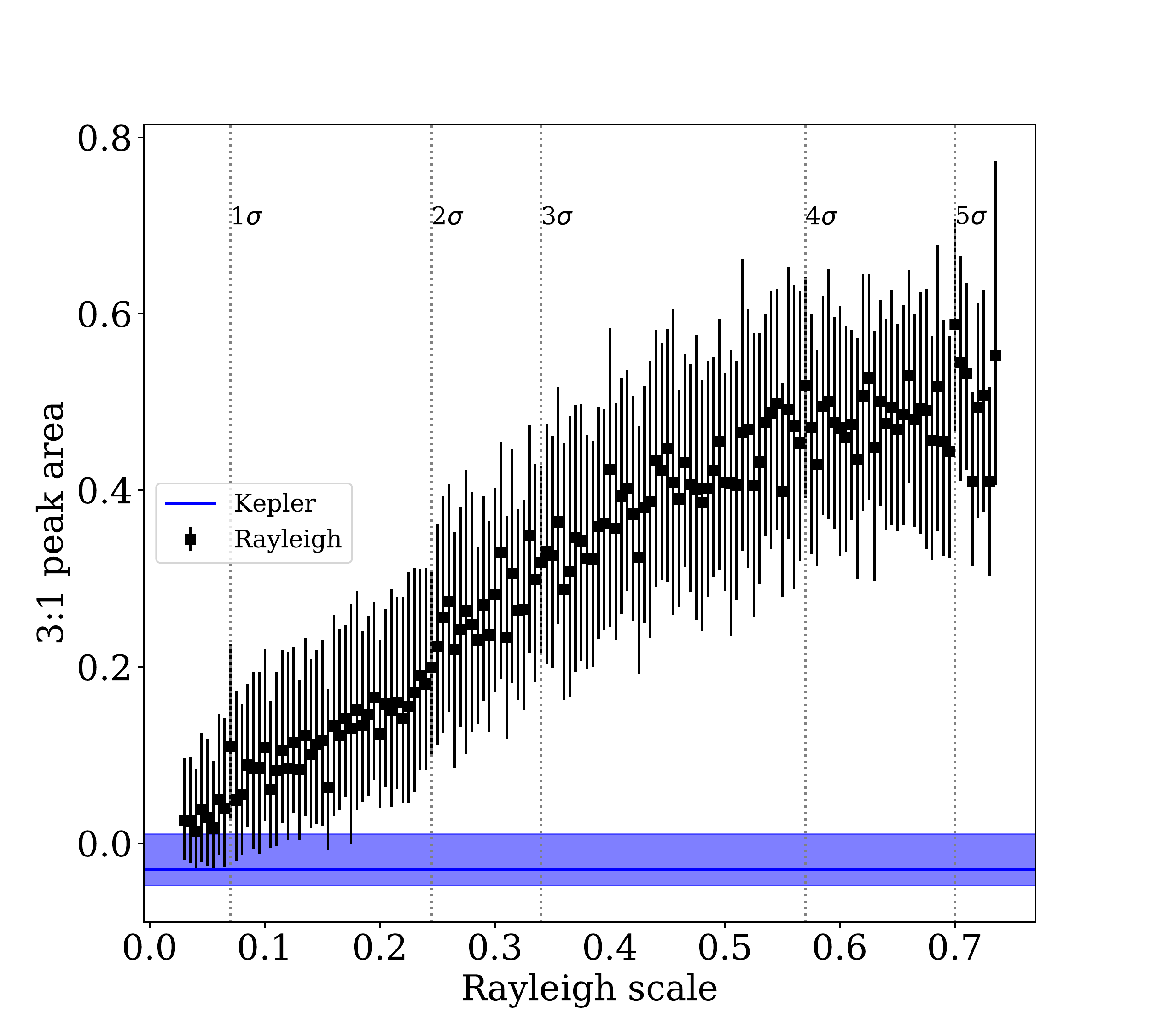}
	}
	\subfloat{
	    \label{fig:KepRay_pkarea-b}
	    \includegraphics[width=.5\linewidth]{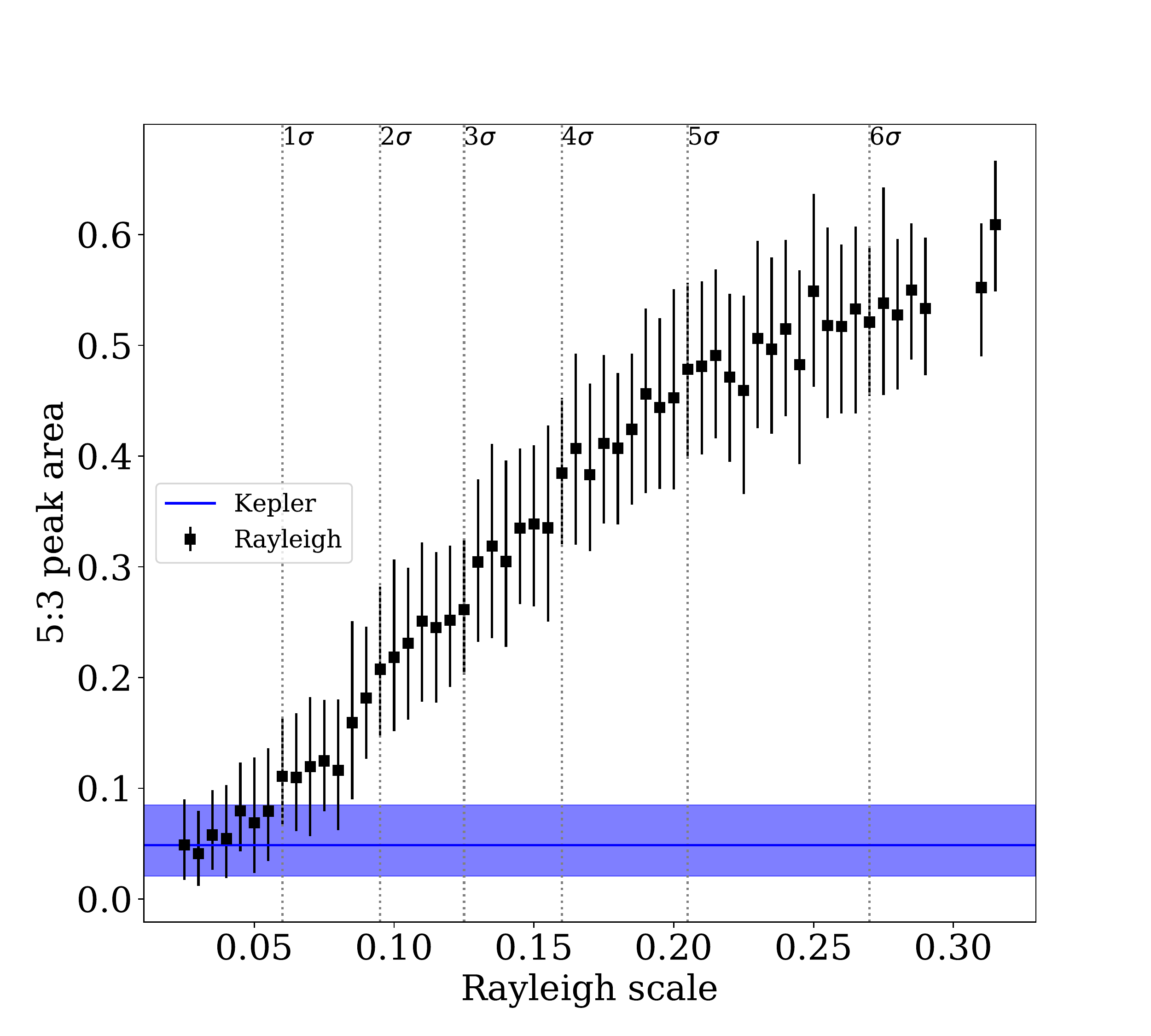}
	}
	\caption{Comparison of the estimated fractional peak area of the \subref{fig:KepRay_pkarea-a} 3:1 and \subref{fig:KepRay_pkarea-b} 5:3 mean-motion resonances for a variety of eccentricity scales. The Kepler sample is in blue and the simulated combined Rayleigh distribution samples in black. The median values are plotted with the 1-sigma values as error bars. The Kepler sample is not consistent with the higher-eccentricity distributions, but consistent with all those below a certain level, allowing for an upper limit constraint on the Kepler samples' eccentricity distributions.}
	\label{fig:KepRay_pkarea}
\end{figure*}

The combined results for all the fitted scales are summarized in Figure~\ref{fig:KepRay_pkarea}, using a representation of the posteriors of the peak area (error bars are the 68.2\% interval) at various scales and the Kepler data for comparison. 

We find that, at 95\% confidence, the upper limit for the Rayleigh scale consistent with the Kepler data is 0.245 for the 3:1 and 0.095 for the 5:3 resonance. The fitted parameters for a select number of the Rayleigh distribution sub-samples are shown in Table~\ref{tab:parameter_results}, with errors listed from the 68.2\% interval, along with the derived parameter of peak area. These selected scales are 1-, 2-, 3-, 4-, and 5-$\sigma$ discrepant with Kepler, and are also shown in Figure~\ref{fig:KepRay_pkarea} as dotted vertical lines.

\section{Analytics}\label{sec:analytics}

To better understand the physical underpinnings of our results from numerical simulations, we investigate an analytical approximation for the effects of a second-order mean-motion resonance on the observed period ratio distribution. We use the framework of the integrable model developed by \cite{2019Hadden}, hereafter H19. This model is applicable for MMRs of any order interior to 2:1, and so we focus our application here on the 5:3 resonance. While this analytical model is not suitable for the 3:1 resonance, we expect the actual dynamics to be similar, as seen as well in our simulations.

There are two relevant pieces of information that we obtain from the model: the resonance width (in period ratio space) and the libration timescale. We use the integrable one-degree-of-freedom Hamiltonian $\mathcal{H}(J,\theta;J^*)$ as given by Equation~19 of H19; see that paper for full details on the derivation and application. 

In brief, we first calculate the conserved quantity $J^*$ for a given system (from H19 Equations~9, 20, and the relation $J^*=J-kP$) and then find the unstable fixed point at $\theta=0$ by finding the value of the conjugate momentum $J$ that maximizes $\mathcal{H}$ and therefore determines the energy of the separatrix ($E_{sep}$). The maximum width is calculated by solving
$\mathcal{H}(J,\pi/2;J^*)=E_{sep}$ for $J$. The two $J$ values can be
converted back to orbital elements using provided relations (H19 Equation~21), giving a minimum and maximum period ratio for the resonance. To find the period of libration, we use an approximation that linearizes the equations of motion in the vicinity of the equilibrium, using the Andoyer module of \texttt{celmech}. This approximation is only accurate for small-amplitude librations, and a more detailed analytical model would develop this further.

To investigate the period ratio distribution, we use a simple model with the resonance width and the period of libration. First, we calculate the resonance width for a given system configuration of masses, periods, and eccentricities.  

If the period ratio of a planet pair is within the resonance width, the libration over time is represented by a sinusoid with the calculated libration period, initial phase as given by $k \theta_0$ (see H19 for details), and centered at $\frac{5}{3}$. The amplitude of the sinusoid is drawn randomly from a linear probability distribution between minimum at zero amplitude and maximum at an amplitude of half the resonance width. For period ratios outside the resonance width, the period ratio is assumed to be constant. This is a simplification, as the period ratios would likely change over time, but they would not follow the resonant model and so these changes are negligible to our result.

Finally, we take a time period to average over and ``observe'' the period ratio based on our libration model over that period $n$ times where $n=\textup{int}\left ( \frac{t_{avg}}{P_2} \right )$.

We use this analytical model in two complementary ways. First, we apply the model to a Kepler-like population of planet pairs for comparison with our N-body results. Second, we apply the model to a single fiducial planet pair and investigate the effect of changing individual parameters of the system.

The systems for the analytical population model are generated according to the same process as described in Section~\ref{sec:datagen}. These pairs are uniform in their initial period ratio between $\frac{5}{3} \pm \frac{7}{150}$, and each planet's eccentricity is randomly assigned using a Rayleigh distribution. We use three scales, a low-eccentricity $\sigma=0.05$, medium-eccentricity $\sigma=0.15$, and high-eccentricity $\sigma=0.50$.

\begin{figure*}[h]
	\centering
	\subfloat{
	    \label{fig:an_pop_pdf}
	    \includegraphics[width=.5\linewidth]{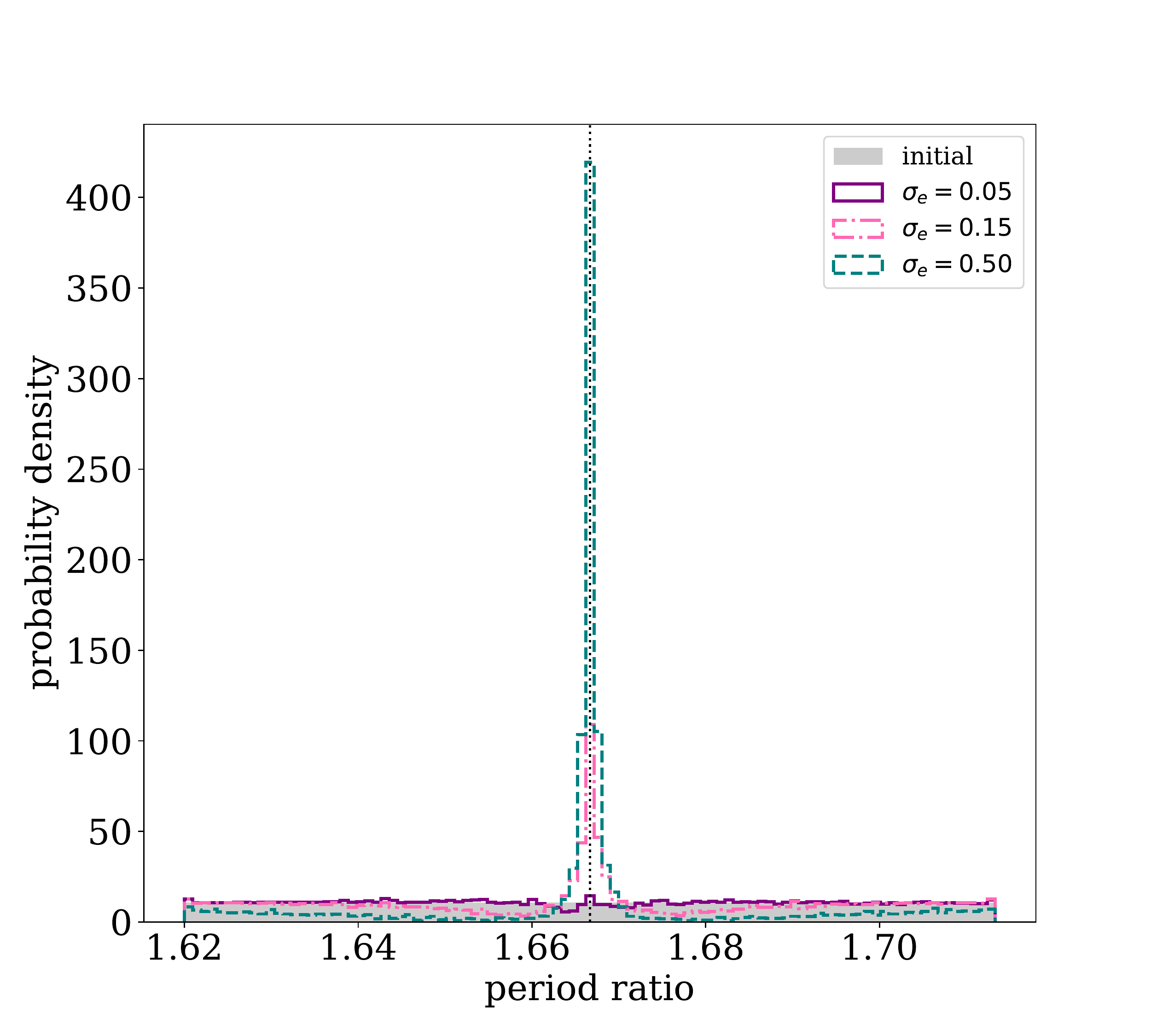}
	}
	\subfloat{
	    \label{fig:an_pop_cdf}
	    \includegraphics[width=.5\linewidth]{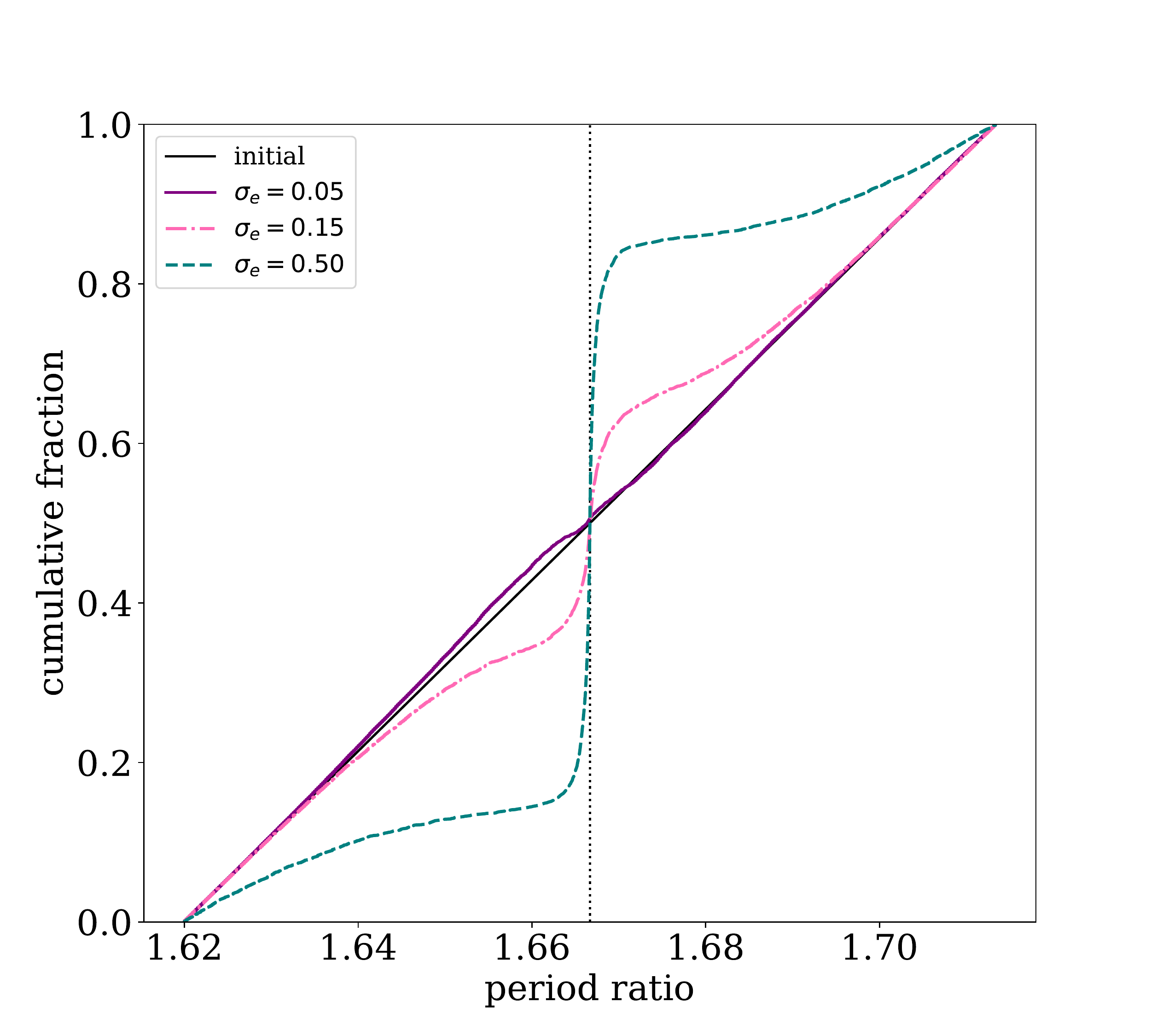}
	}
	\caption{The analytically-predicted \subref{fig:an_pop_pdf} probability density and \subref{fig:an_pop_cdf} cumulative distribution of time-averaged period ratios near the 5:3 mean-motion resonance for three populations of Kepler-like planet pairs. Each population has a different Rayleigh scale for the planets' individual eccentricities. The nominal resonance value is marked with a dotted line. As seen in the simulated data, there is a strong, very narrow peak associated with increased eccentricity.}
	\label{fig:analytical_population}
\end{figure*}

Figure~\ref{fig:analytical_population} shows the resulting probability distributions and cumulative distribution functions for the three populations. As expected, a very narrow peak forms at the exact resonance location, the height of which increases with increased eccentricity. Because the initial period ratio is exactly uniform, the addition of planet pairs to this peak must result in a lower density of planet pairs outside the peak. These troughs are seen in the distributions. However, because the resonant and eccentric pairs end up at a very narrow location but can start from a wide range of initial period ratios that varies based on the properties of each system, the troughs are very spread out compared to the peak. 

\begin{table}[]
\centering
\begin{tabular}{|c|c|c|}
\hline
\textbf{Parameter} & \textbf{Fiducial Value} & \textbf{Variable Range} \\ \hline
$M_\star$ ($M_\odot$) & 1 & 0.2 - 2 \\ \hline
$m_1$ ($M_\oplus$) & 10 & 0.5 - 20 \\ \hline
$m_2$ ($M_\oplus$) & 10 & 0.5 - 20 \\ \hline
$e_1$ & 0.1 & 0 - 0.9 \\ \hline
$e_2$ & 0.2 & 0 - 0.9 \\ \hline
$\varpi_1$ & 0 & $-\pi$ - $\pi$ \\ \hline
$\varpi_2$ & 0.5 & $-\pi$ - $\pi$ \\ \hline
$P_1$ (d) & 10 & 1 - 100 \\ \hline
$t_{avg}$ (yr) & 3.5 & 0.5 - 10 \\ \hline
\end{tabular}
\caption{Variable parameters used for analytical investigation. When varied, each parameter is varied between the minimum and maximum values in 20 linear steps. $P_2$ is determined for each system based on the period ratio.}
\label{tab:analytic_fiducial}
\end{table}

\begin{figure}[h]
	\centering
    \includegraphics[width=\linewidth]{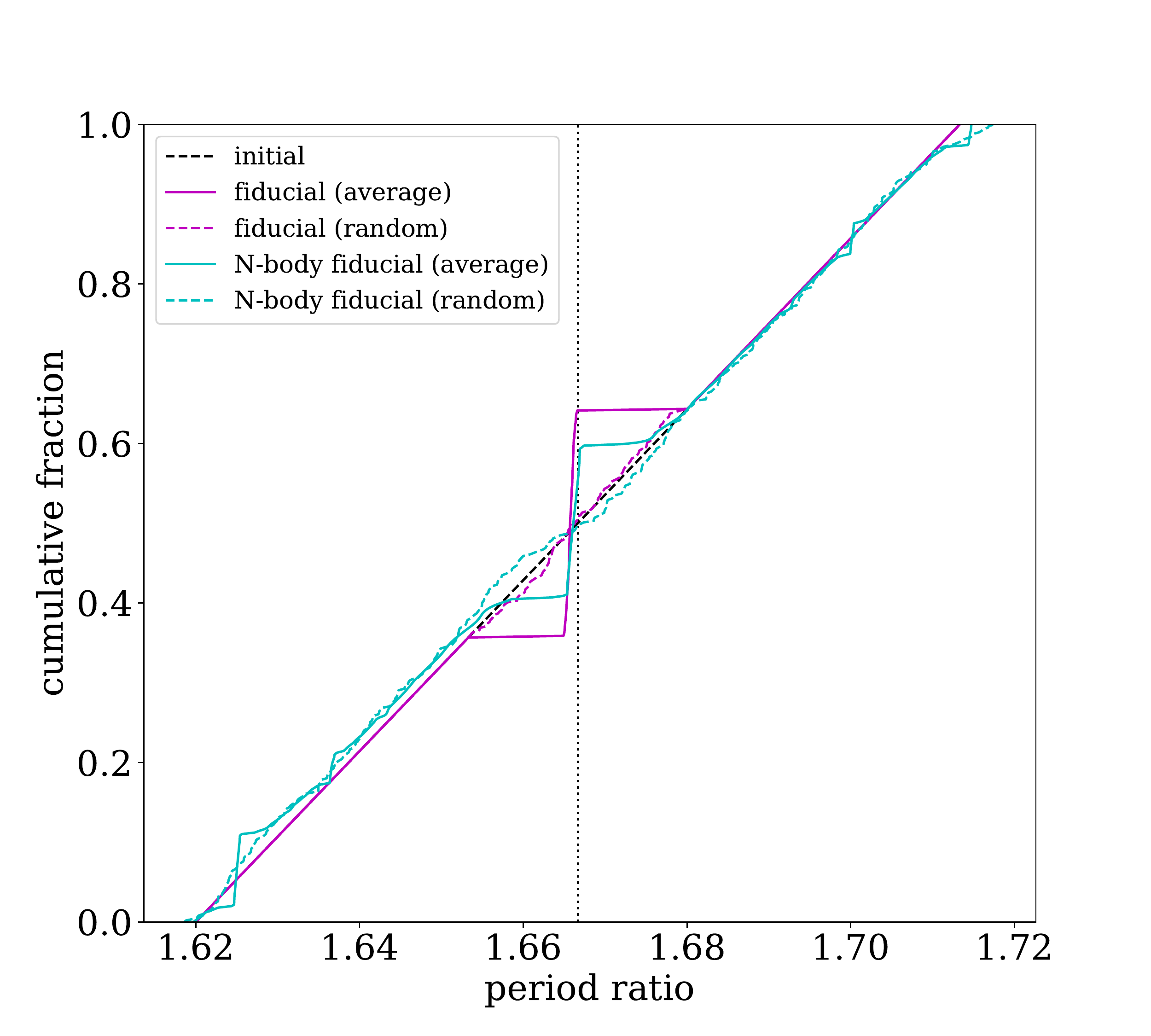}
	\caption{The analytically-predicted cumulative distribution function in period ratio for the fiducial system. The time-averaged period ratio is shown in solid magenta and a random period ratio from within the time period is shown in dashed magenta. The same distributions are shown for a comparable N-body simulation in cyan (solid is averaged and dashed is random). The black dashed line shows the initial flat period ratio distribution. The resonance location at $\frac{5}{3}$ is indicated with a vertical dotted line. There is reasonable agreement between the numerical and analytical distributions, and in both cases, the peak at resonance only appears when the period ratios are averaged.}
	\label{fig:analytic_fiducial_PRcdf}
\end{figure}

To examine the effect of various parameters on the period ratio distribution, we establish a fiducial system and then vary parameters one at a time. The fiducial system properties are given in Table~\ref{tab:analytic_fiducial}. The properties were chosen to be consistent with Kepler-like systems but without any special consideration. Then we investigate the period ratio libration for a grid of 500 period ratios (by changing the period of the outer planet) between our limits of interest, $\frac{5}{3} \pm \frac{7}{150}$. The resulting cumulative distribution function in period ratio is shown in Figure~\ref{fig:analytic_fiducial_PRcdf}, both for the average period ratio and a period ratio chosen randomly within the time period. As expected, a very clear and narrow peak forms when the period ratios are averaged over time. We also ran an N-body simulation with the same parameters for comparison. While the distributions are not exact, the agreement is good for this simplified model. It is also interesting to note in the N-body distribution what seem to be similar (though much weaker) signals at the 13:8, 18:11, and 17:10 resonances.

In this analytical model, we do not include a continuum such as is present in our model in Equation~\ref{eqn:model_pdf}. Instead, this analytical model examines only how the peak super-imposed on the continuum is affected by the changing system parameters. This, and the fact that all of the systems in this analytical model have the same properties aside from the period ratio, is the difference in the shape of the cumulative distribution functions in Figures~\ref{fig:alldata_cdf} and \ref{fig:analytic_fiducial_PRcdf}. The height and width of the peak is our interest, and thus what we investigate via this analytical model.

\begin{figure*}[h]
	\centering
    \includegraphics[width=\linewidth]{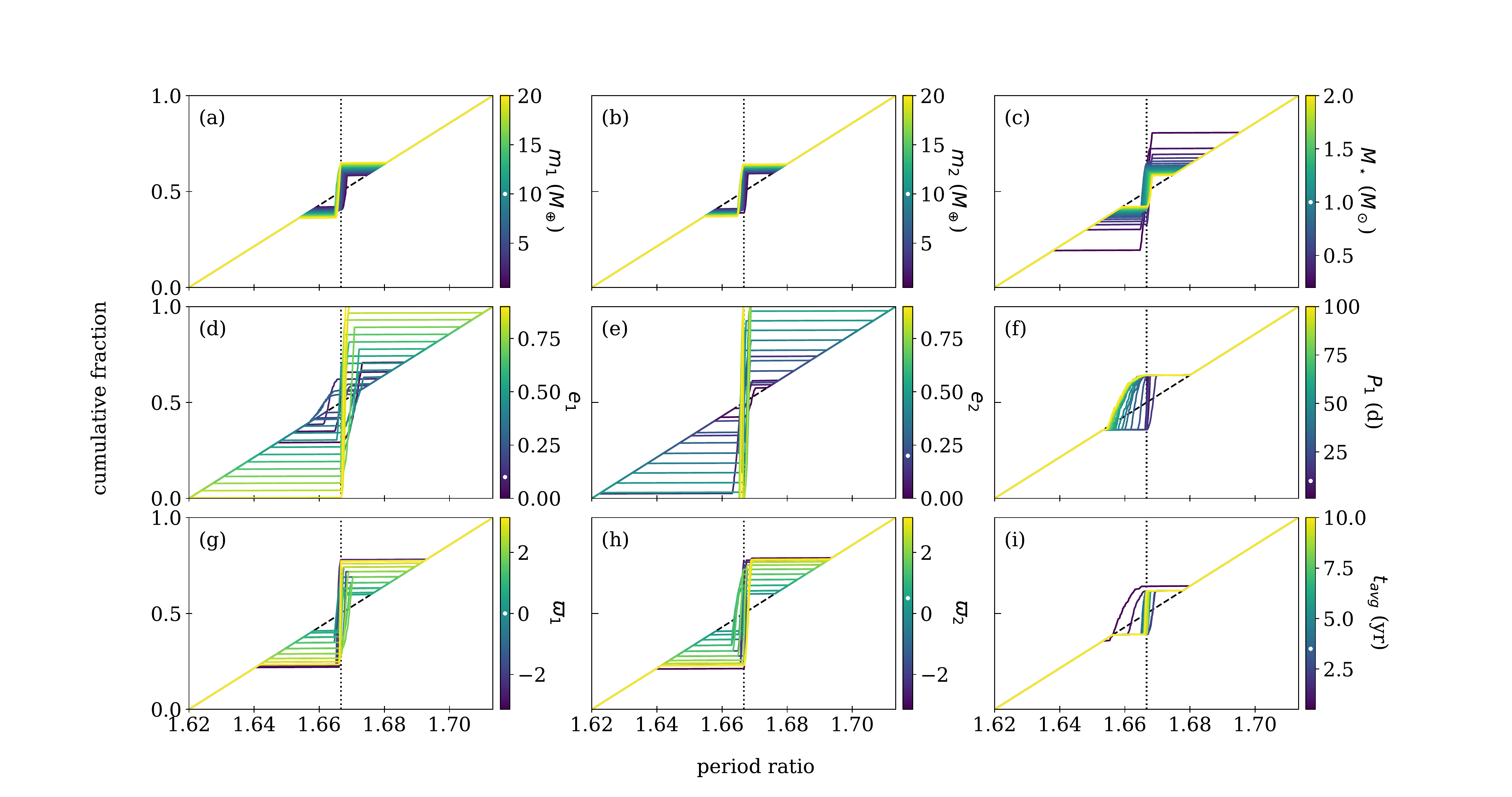}
	\caption{The analytically-predicted, time-averaged cumulative distribution functions in period ratio for various systems. The parameter being varied is indicated by the color bar to the right of each plot, where the fiducial value is plotted in white for reference. The resonance location at $\frac{5}{3}$ is indicated with a vertical dotted line. All the parameters affect the height and width of the peak, the most notable being the eccentricity of either planet, shown in panels (d) and (e).}
	\label{fig:analytic_variable_PRcdf}
\end{figure*}

The results of varying the nine parameters are shown in Figure~\ref{fig:analytic_variable_PRcdf}. Only the averaged period ratios are shown here for clarity, with the color indicating the parameter's value as indicated by the color bar to the right of each plot. For each parameter, the fiducial value is plotted as a white dot on the color bar for reference.

First we see that varying the masses of the planets (\ref{fig:analytic_variable_PRcdf}a and \ref{fig:analytic_variable_PRcdf}b) has a noticeable but small effect. The effect does not depend on which planet's mass changes, as the resonance is sensitive to the planets' combined mass. The amplitude of the peak does not change much, but the width does change slightly and the center of the peak shifts. This is due to a change in the libration period\textemdash if the libration period increases past $t_{avg}$, the averaged value will no longer lie at the center as a cycle has not completed. Changing the mass of the star (\ref{fig:analytic_variable_PRcdf}c) has a generally similar effect, though there is a stronger effect when reaching much larger mass ratios (i.e., a smaller stellar mass).

As expected, we see a very strong dependence of the peak on eccentricity (\ref{fig:analytic_variable_PRcdf}d and \ref{fig:analytic_variable_PRcdf}e). As eccentricity is increased, the amplitude of the peak at resonance becomes much larger. In this case, the effect is notably stronger when $e_2$ is increased. The orientation of the eccentricities (\ref{fig:analytic_variable_PRcdf}g and \ref{fig:analytic_variable_PRcdf}h) also has an effect, although less than from the magnitude. In our fiducial system, the eccentricities are 0.5 rad ($\sim 30^\circ$) out of alignment. As either of the $\varpi$ changes towards $\pm \pi$ we see an increase in the amplitude of the peak, suggesting a stronger signal for anti-aligned eccentric planet pairs.

Changing $P_1$ (\ref{fig:analytic_variable_PRcdf}f) has no apparent effect on the amplitude or width of the peak, as is expected. However, the location of the center of the peak changes significantly. This is because the libration period depends on the planets' periods, and so for longer period planet pairs, the libration cycle has not yet completed during this $t_{avg}$. We can similarly see this effect from changing $t_{avg}$ (\ref{fig:analytic_variable_PRcdf}i). As $t_{avg}$ decreases, fewer of the systems have completed their libration and so the center shifts away from the resonant location. In this case, the center of the peak shifts leftward for incomplete libration cycles. This depends on the initial phase of the libration. For different initial phases, the peak can also shift rightward, and so the direction of the peak shift should not be seen as significant.

We note a few remarks about these analytical results. First, we use a very simplified model for the libration. Several N-body comparisons were done to ensure this model was reasonable to use as an approximation, but it does not give exact behavior. Second, we use the simplified Hamiltonian from H19 that is based on a leading-order approximation in eccentricity. This introduces additional inaccuracies, especially for our high-eccentricity systems. Third, there are no stability cuts or considerations, and it is probable that not all of these systems are viable. Fourth, this analytical model assumes a planar configuration and no inclination effects.

Given these points, we do not provide this analytical investigation for a quantitative comparison. However, we expect the qualitative trends to be accurate and find the overall behavior to be in line with our expectations, confirming the resonant behavior as seen in our numerically simulated results.

\section{Discussion}\label{sec:disc}

It is particularly striking how narrow the resonance peak is. This arises from the time-averaging of the orbital parameters and is a clear feature of the second-order resonance. The narrowness of the peak is, for example, what allows us to ignore the effects of any associated troughs in our analysis.

Our sets of generated systems are intended to create sets of planet pairs with a large array of eccentricities limited only by stability. We are agnostic to formation mechanisms, i.e., whether such systems are physically realistic. This allows us to marginalize over other properties to isolate the effects of eccentricity on the period ratio distribution. Other potential effects are discussed below.

\subsection{Comparison with Previous Results}\label{sec:prevres}

The first results pertaining to the eccentricity of exoplanet populations were obtained via radial velocity measurements. These systems were typically composed of giant planets and generally excited in eccentricity. \cite{2009Wright} found a mean eccentricity of 0.22 for multi-planet systems, their data containing many more giant systems than in the Kepler observations. Similarly, \cite{2013Kipping} quantifies the eccentricity distribution of radial velocity planets with a beta distribution that gives a mean eccentricity of $\sim$0.22.

As more transiting planet data became available via Kepler, and different techniques were developed for constraining the eccentricity of these planets, a different picture began to emerge. \cite{2014Fabrycky} measured $\sigma\lesssim0.1$ for a Rayleigh distribution of the eccentricity of super-Earths. Studies done on subsets of the Kepler sample with well-characterized stellar properties found $\sigma=0.049 \pm 0.013$ \citep{2015VanEylen}, $\sigma = 0.03^{+0.02}_{-0.03}$  \citep{2016Xie}, $\sigma=0.061_{-0.012}^{+0.010}$ \citep{2019VanEylen}, and $\sigma \sim 0.04$ \citep{2019Mills}. \cite{2017Hadden} measured individual systems (their Figure~5) and quoted a typical scale of $\sigma=0.018^{+0.005}_{-0.004}$ from their prior work \citep{2014Hadden}. 

These results indicated much lower eccentricities are typical for small planets. \cite{2012Kane} found that for giant planets the Kepler observations were in agreement with previous radial velocity results, but that planets smaller than Neptune became “rapidly and significantly more circular.”

Our constraints on the planets near the 3:1 and 5:3 are consistent with the results obtained previously for small, multi-planet systems observed by Kepler. Our 5:3 constraint, which would predict a mean eccentricity of no more than 0.12 at 95\% confidence (for a Rayleigh distribution, $ \langle E \rangle = \sigma \sqrt{\frac{\pi}{2}}$), is not consistent with the eccentricities of giant planets such as those analyzed by \cite{2009Wright}. We find a 1.8$\sigma$ (3:1) and 3.2$\sigma$ (5:3) disagreement with the beta distribution of \cite{2013Kipping} (see Figure~\ref{fig:Kepbeta_pkarea}).

\cite{2020He} find a strong multiplicity dependence for the eccentricity distribution, with the median eccentricity decreasing as $n$ increases. This multiplicity dependence has also been found by \cite{2021BachMoller}. Additional evidence for a difference between single-planet and multi-planet eccentricities was found in \cite{2016Xie}, \cite{2019Mills}, and \cite{2019VanEylen}. This dichotomy may be due to unobserved outer companions \citep{2020Poon}. As our method depends on measuring a period ratio, which must have two planets, we probe the observed $n\geq2$ Kepler sample.

\cite{2019He} and \cite{2020He} use forward modeling to constrain properties of the intrinsic population of Kepler planets. While in this work we are constraining the eccentricities of the observed population, rather than the intrinsic population, we find the comparison of interest. \cite{2019He} find low eccentricities, with characteristic Rayleigh scales $\sigma \simeq 0.01 - 0.02$. \cite{2020He} uses an angular momentum deficit approach for the modeling and finds a median eccentricity of 0.138 for two-planet systems, decreasing down 0.029 for five-planet systems and to 0.009 for ten-planet systems (see their Table~5). These results are consistent with ours, despite the differing populations being described.

Interestingly, \cite{2020He} find that a log-normal distribution is a better fit for the eccentricity distribution, rather than a Rayleigh distribution which is more typically assumed (e.g. \citealt{2019He}, \citealt{2014Fabrycky}). \cite{2015Tremaine} and \cite{2016Shabram} also recommended a non-Rayleigh distribution, while \cite{2019VanEylen} consider several models and do not advocate one over another. We leave the investigation of additional eccentricity distributions to future work.

Our results, then, are consistent with other studies and further evidence that the eccentricity distribution of small (i.e. not-giant) planets in multi-planet systems have a different underlying eccentricity distribution than single-planet or giant-planet systems.

\subsection{Tides}\label{sec:tides}

Tides might act to remove some of the planet pairs from resonance or decrease their eccentricity over time \citep{2003Novak,2014Delisle}. Here we consider the impact of tides on these relatively close-in planets.

\begin{figure}[h]
	\centering
    \includegraphics[width=\linewidth]{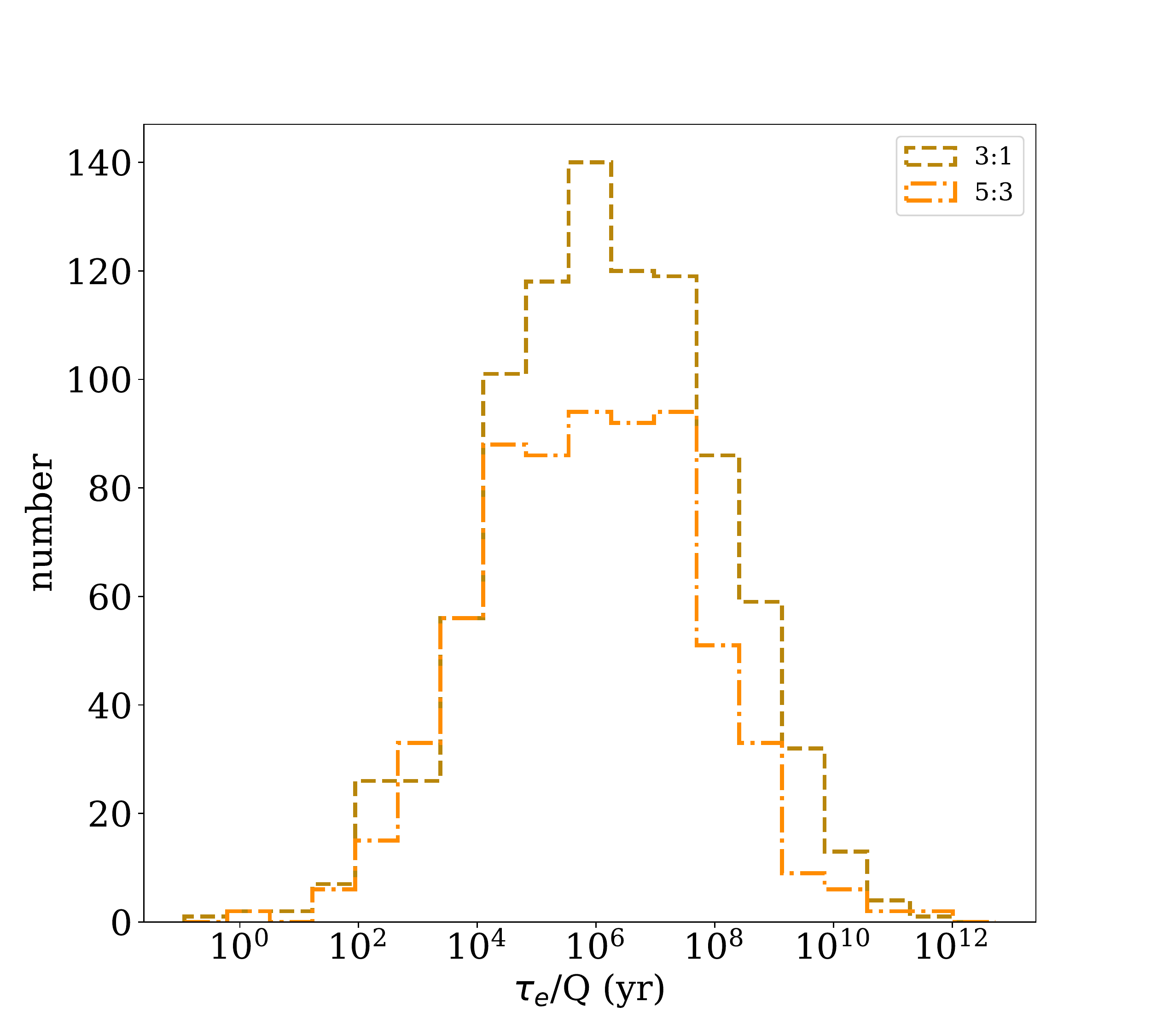}
	\caption{The circularization timescale of the inner planet of our generated populations, scaled by Q. Based on a typical Q value of $\sim$100 for terrestrial planets, we expect that perhaps half of our sample may have undergone one tidal circularization timescale in 100 Myr, indicating that maintaining these planets at such high eccentricities may be difficult.}
	\label{fig:tides}
\end{figure}

We calculate the circularization timescale for our planet pairs using Equation~9 of \cite{1996Rasio}. The resulting timescales, scaled by Q, are shown in Figure~\ref{fig:tides}. For terrestrial planets, typical Q values range from 10 to 500 \citep{1966Goldreich}. We would thus expect approximately half of our sample would have undergone one tidal circularization timescale in 100 Myr.

However, this does not affect our results, which are focused solely on the effect of eccentricity in shaping the period ratio distribution and not in the feasibility of such an eccentricity distribution. This tidal timescale is a consideration for other limits on the stable eccentricity of a similar population.

Furthermore, we can split our sample sets into two equal parts based on the scaled circularization timescale to see if systems with faster circularization timescales have different properties in this context that might be affecting our results. Instead, we find that the resulting averaged period ratio and eccentricity distributions are not statistically different. The p-values from a 2-sample Kolmogorov–Smirnov test for the split samples of the 3:1 set are 0.70 for the averaged period ratio distribution and 0.17 for the averaged combined eccentricity distribution. For the 5:3 set, the corresponding p-values are 0.13 and 0.34, respectively. Thus, we expect tides to be an interesting but negligible consideration for our results.

\subsection{Eccentricity at Formation}\label{sec:form}

In our approach to creating our sets of planet pairs, we did not use a physical motivation for the eccentricities of the planets. Indeed, we purposely attempted to cover a large eccentricity space to allow for sub-sampling by eccentricity. We did, however, use a stability analysis to cut unstable planet pairs from our sample (see Section~\ref{sec:datagen}). This stability cut produced some shaping of the eccentricity distribution (i.e. Figure~\ref{fig:eif}), in particular preferentially removing high-eccentricity pairs.

The stability-shaped eccentricity distribution is still much too eccentric to match the Kepler data, however. Because stability alone cannot explain the low observed eccentricities, this implies some other process must prevent high-eccentricity pairs, likely related to formation mechanisms.

The eccentricity distribution predicted by formation models varies depending on many parameters, particularly related to the properties of the protoplanetary disk \citep{2013Fortier}. Effects that serve to excite eccentricity include reduced solid surface density \citep{2019VanEylen}, shallower solid surface density profiles \citep{2016Moriarty}, the presence of additional companions \citep{2017Huang,2017Anderson}, and others. Conversely, there are effects that can dampen eccentricity, such as the presence of planetesimals to provide dynamical friction during formation or gas in the disk \citep{2019Mulders}. Recent findings by \cite{2021Yee} indicate that the observed eccentricities (of those measured by \cite{2017Hadden}) are lower than predicted stability limits or from giant impact formation theory. Fitting the exact formation conditions consistent with the Kepler planets is the beyond the scope of this paper.

\subsection{Mass Dependence}\label{sec:masses}

Mass and eccentricity both have strong effects on the behavior of planets in mean-motion resonance. This can be seen, for example, in Figure~\ref{fig:analytic_variable_PRcdf}. By using a distribution of masses and mass ratios in our set of simulated planet pairs that approximates that seen in the Kepler data, we intend to marginalize over the mass effect to isolate the effect from eccentricity. To illustrate how the mass and eccentricity effects interact, we split our set of planet pairs into 2 groups in mass ratio and 2 groups in eccentricity. The split was chosen at the median values in each case; $e_1+e_2 = 0.58$ and $\textup{log}\frac{m_1+m_2}{M_\star} = -4.42$ for the 3:1, $e_1+e_2 = 0.28$ and $\textup{log}\frac{m_1+m_2}{M_\star} = -4.44$ for the 5:3.

\begin{figure*}[h]
	\centering
    \includegraphics[width=\linewidth]{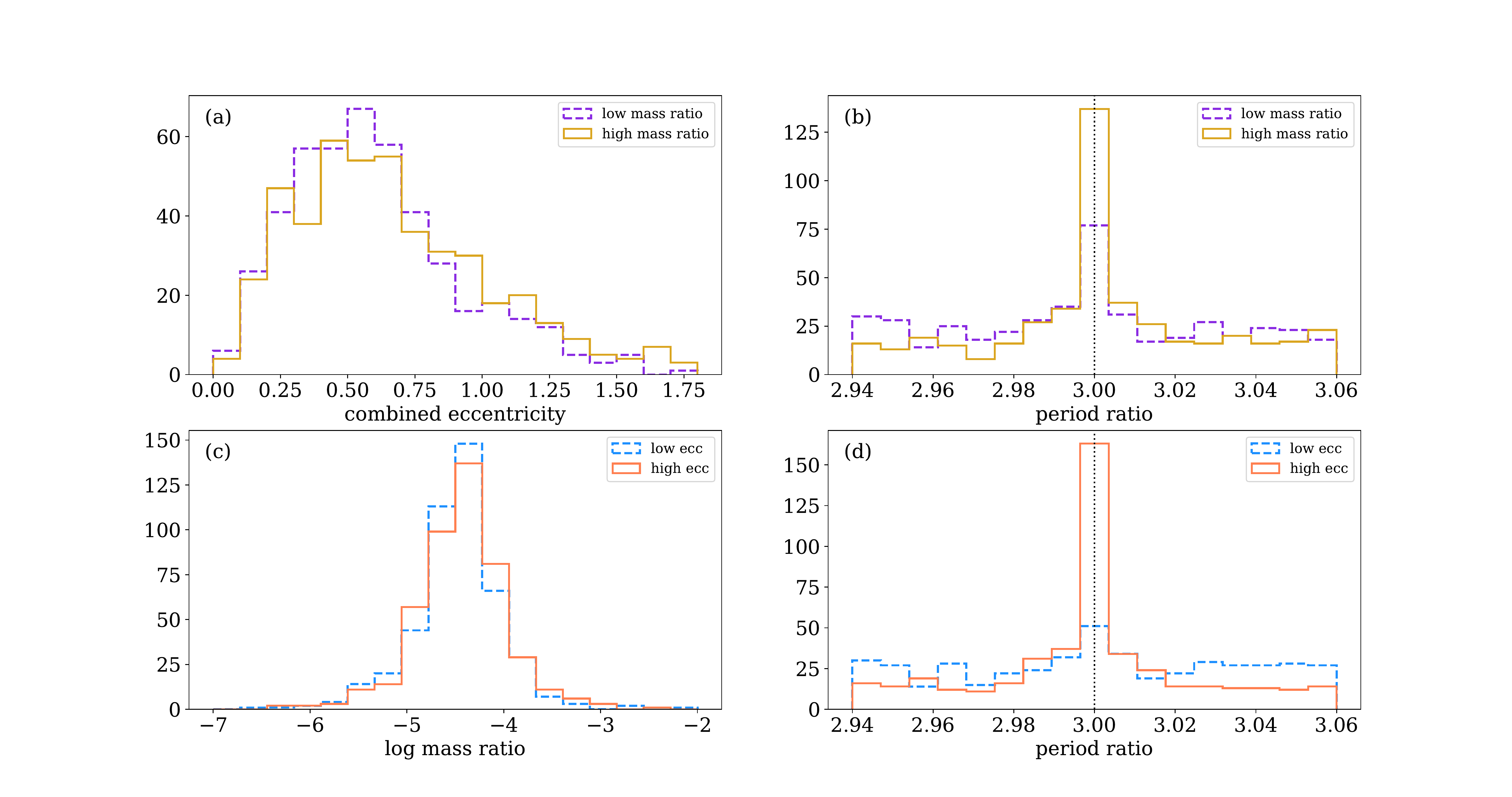}
	\caption{For our generated planet pairs near the 3:1 resonance, we split the population into two halves. First by the log of the mass ratio (log $\frac{m_1+m_2}{M_\star}$), shown in the top panels as the combined eccentricity (a) and period ratio (b) distributions for these two halves of the population. In the bottom panels, we split the population into two halves by the combined eccentricity and show the log mass ratio (c) and period ratio (d) distributions for these two halves. While both mass and eccentricity affect the period ratio distribution, the time-averaged peak at resonance is still present even in the low-mass sample, but it disappears in the low-eccentricity sample, indicating the eccentricity is more strongly contributing to the effect.}
	\label{fig:massecc31}
\end{figure*}

\begin{figure*}[h]
	\centering
    \includegraphics[width=\linewidth]{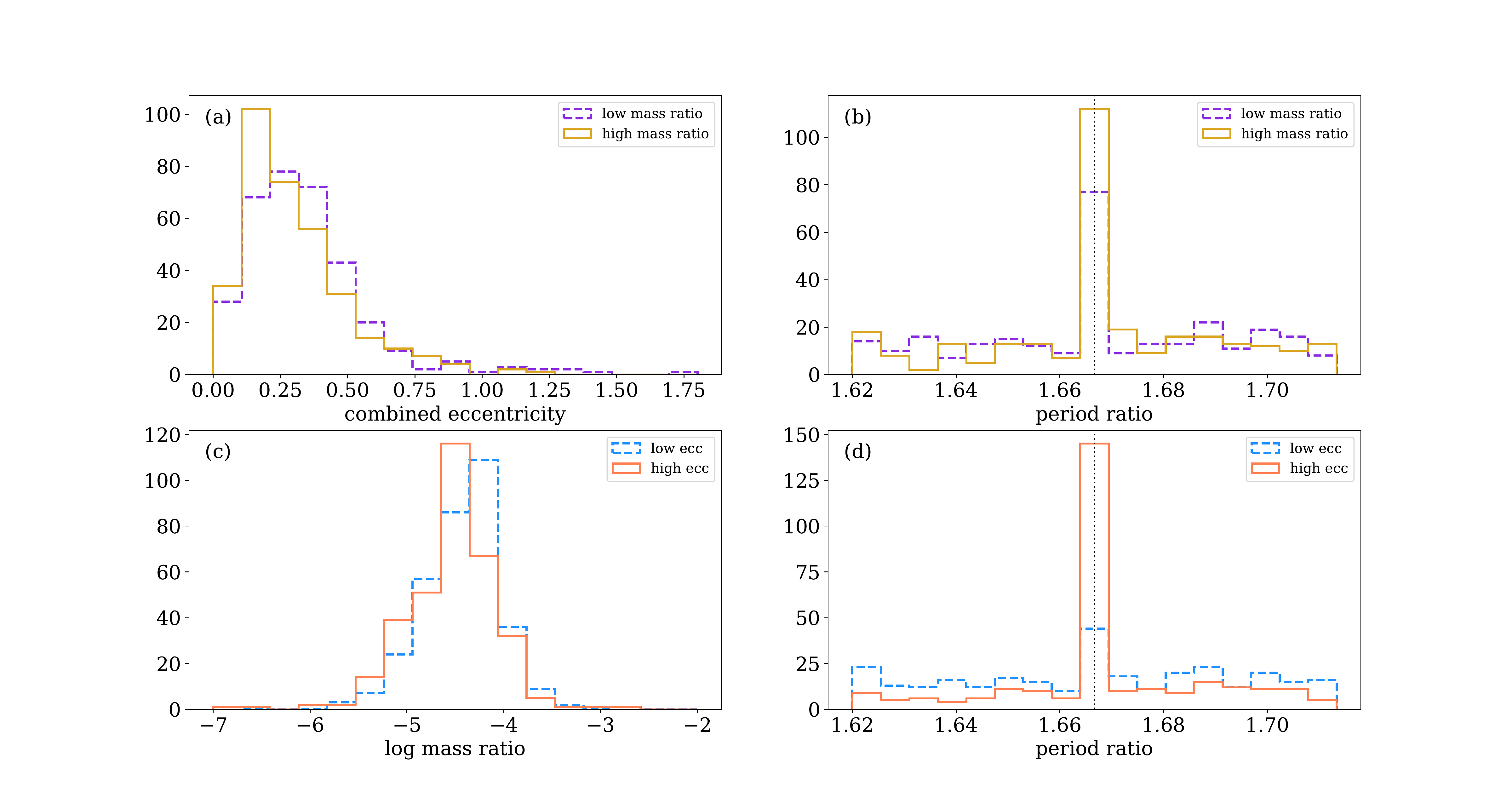}
	\caption{The same as Figure~\ref{fig:massecc31} but for the 5:3 resonance.}
	\label{fig:massecc53}
\end{figure*}

Figures~\ref{fig:massecc31} and \ref{fig:massecc53} show the distributions for these split sets of planet pairs. The eccentricity distribution is largely unaffected by the mass ratio cut (\ref{fig:massecc31}a/\ref{fig:massecc53}a), and the same is true for the mass ratio distribution and the eccentricity cut (\ref{fig:massecc31}c/\ref{fig:massecc53}c). However, strong differences are seen in the distributions of the period ratios. When looking across all eccentricities, the higher mass ratio half of the simulated systems has a more notable peak at the exact commensurability location, but the peak is still clear for the lower mass ratio planet pairs (\ref{fig:massecc31}b/\ref{fig:massecc53}b). Conversely, when looking across all mass ratios, the higher eccentricity half of the data has almost the entirety of the effect at the commensurate period ratio, and the peak almost vanishes for the lower eccentricity half of the data (\ref{fig:massecc31}d/\ref{fig:massecc53}d).

Thus, while mass and eccentricity both have strong effects on the resonant behavior, the eccentricity effect is much stronger. Given this difference in strength and the marginalization over the apparent mass ratio distribution of Kepler planets, our results robustly isolate the eccentricity effect near the 3:1 and 5:3 resonances.

\subsection{Resonance Probability}\label{sec:resprob}

Using our fitted model for the Kepler period ratio distribution, we can calculate a resonance probability for a given period ratio by comparing the height of the full model at that period ratio to the height the model predicts for the continuum at that period ratio. 

When applied to the Kepler 3:1 sample, the resonance probability for all the planet pairs is zero. This is due to the fact that our fitted model predicts no peak at the resonance.  

\begin{figure}[h]
	\centering
    \includegraphics[width=\linewidth]{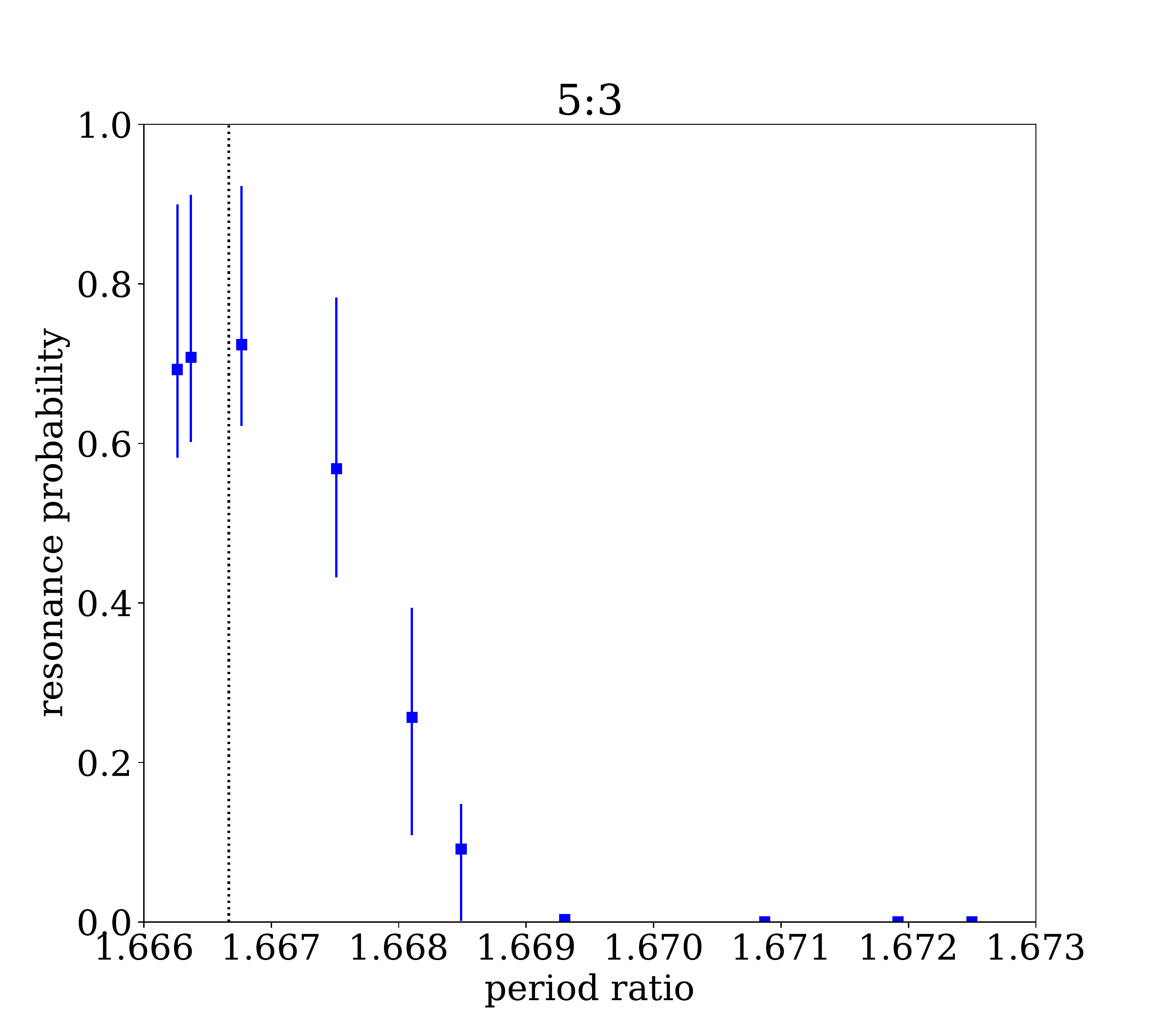}
	\caption{The probability of the Kepler planet pairs being part of an excess peak associated with 5:3 mean-motion resonance and not part of the background continuum of period ratios. All pairs with a non-zero median probability are plotted; see Table~\ref{tab:5to3resprob} for planet names and values.}
	\label{fig:5to3resprob}
\end{figure}

For the Kepler 5:3 sample, there are 10 KOI pairs with non-zero resonance probability. These are listed in Table~\ref{tab:5to3resprob} with their calculated resonant probability. The resonance probabilities are plotted in Figure~\ref{fig:5to3resprob}. In Section~\ref{sec:K02261} we investigate one of these KOI pairs, K02261, in more detail.

\begin{table*}[]
\centering
\resizebox{\textwidth}{!}{%
\begin{tabular}{|c|c|c|c|c|}
\hline
\multicolumn{2}{|c|}{\textbf{KOIs}} & \multicolumn{2}{c|}{\textbf{Kepler Names}} & \textbf{Resonance Probability} \\ \hline
K01574.01 & K01574.02 & Kepler-87 b & Kepler-87 c & $0.7^{+0.1}_{-0.2}$ \\ \hline
K02261.01 & K02261.02 & Kepler-1164 b & - & $0.7^{+0.1}_{-0.2}$ \\ \hline
K04034.01 & K04034.02 & Kepler-1543 b & - & $0.7^{+0.1}_{-0.2}$ \\ \hline
K02163.01 & K02163.02 & Kepler-365 b & Kepler-365 c & $0.6^{+0.1}_{-0.2}$ \\ \hline
K02086.01 & K02086.03 & Kepler-60 b & Kepler-60 d & $0.3 \pm 0.1$ \\ \hline
K04136.01 & K04136.02 & - & - & $0.1 \pm 0.1$ \\ \hline
K02169.01 & K02169.02 & Kepler-1130 A b & - & $0.003^{+0.007}_{-0.002}$ \\ \hline
K02693.02 & K02693.03 & Kepler-398 c & Kepler-398 d & $10^{+121}_{-9} \times 10^{-8}$ \\ \hline
K00582.01 & K00582.03 & Kepler-191 d & Kepler-191 b & $7^{+381}_{-7} \times 10^{-12}$ \\ \hline
K01151.01 & K01151.04 & Kepler-271 b & - & $2^{+198}_{-2} \times 10^{-14}$ \\ \hline
\end{tabular}%
}
\caption{KOIs near the 5:3 with non-zero resonance probability. The associated Kepler name is listed where applicable.}
\label{tab:5to3resprob}
\end{table*}

\subsection{Investigating the Case of K02261 (Kepler-1164) via Timing Analysis}\label{sec:K02261}

There are two KOIs in this system: K02261.01 and K02261.02. The former is confirmed (Kepler-1164 b, \citealt{2016Morton}); the latter is a candidate. The K02261 system is also known as Kepler-1164.

To investigate this system, we use the transit numbers, times, and variations from \cite{2015Rowe}. There are 315 transits of K02261.01 and 157 transits of K02261.02 covering a time period of approximately 1455 days. The pair's average period ratio over this time is 1.666(4).

Transit data cannot give an instantaneous value of the osculating period ratio as measured exactly from an N-body simulation. Instead, we approximate the period ratio at any given point in time using the last two observed transits of each planet prior to that point. The average period for each planet is approximated as the difference in transit times divided by the difference in transit numbers, and these periods are used to calculate an approximate period ratio for that point in time.

\begin{figure*}[h]
	\centering
    \includegraphics[width=\linewidth]{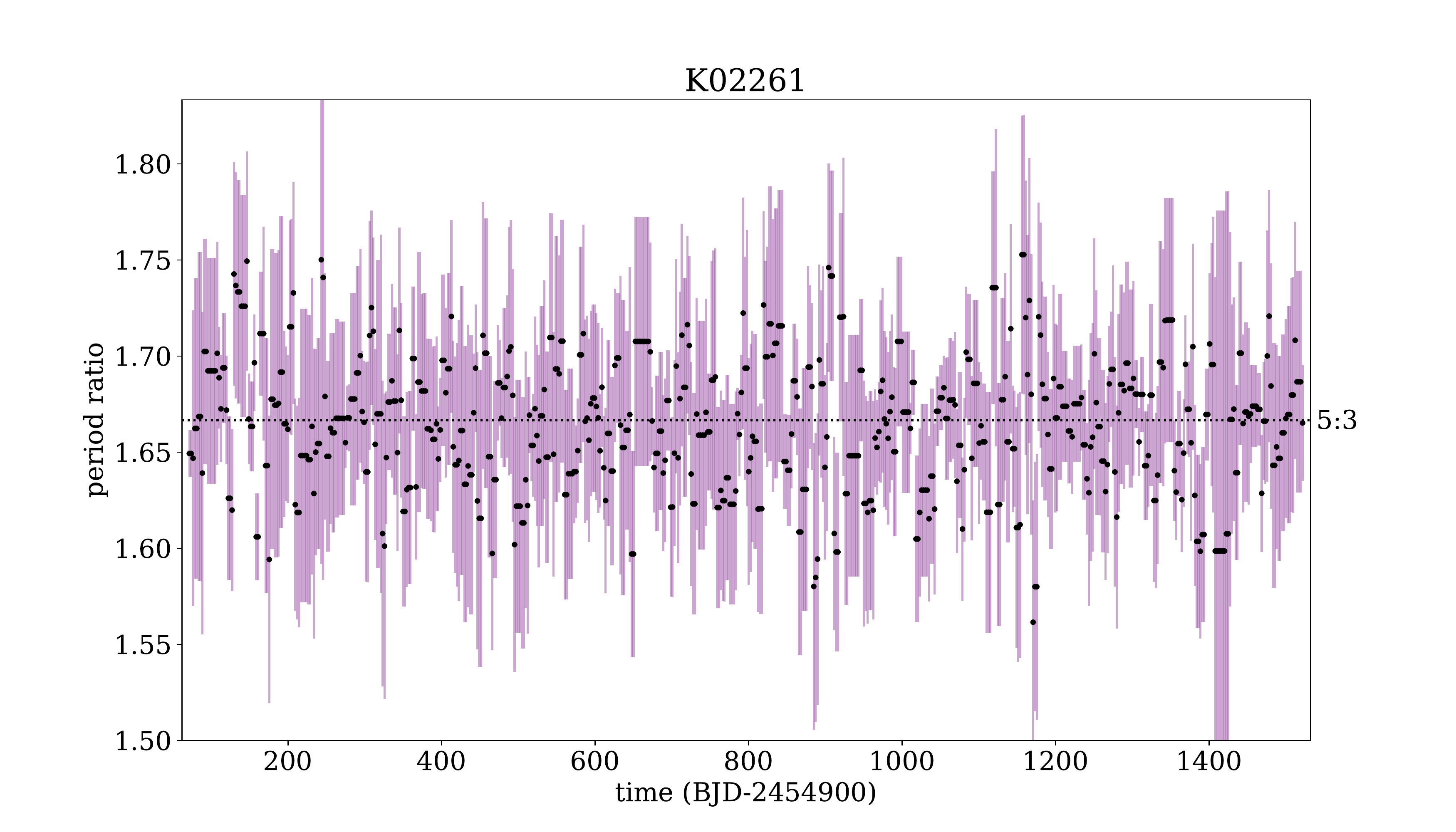}
	\caption{The period ratio over time for K02261 estimated from transit time data from \cite{2015Rowe}. Black points are calculated from the given transit times, and errors are the standard deviation of 1000 transit time realizations drawn using the uncertainty on the TTVs. The uncertainty from observations is too large to reveal any structure in the behavior of the period ratio over time.}
	\label{fig:2261PRtime}
\end{figure*}

The variation of period ratio over time for K02261 according to this method is shown in Figure~\ref{fig:2261PRtime}. Errors are incorporated numerically by repeating the calculation 1000 times, drawing transit times randomly from a Gaussian described by the transit time and TTV uncertainty for each transit from \cite{2015Rowe}. We use 600 evenly spaced times from the observing period to calculate the period ratio.

Due to the scatter and uncertainty of the observational data, it is possible but not evident that the period ratio of the planet pair may be oscillating around the exact commensurability over time. This is the case even when binning the data or attempting to phase fold on various periods. We use N-body models of the planet pair to investigate further.

First, we need to have a defined set of system parameters in order to produce an N-body model. Given the uncertainty of the data, we do not aim to constrain a fit but merely find sample systems that are consistent with the observational TTV data.

From the least squares optimization function in \texttt{SciPy} \citep{SciPyref}, we obtain a fit and covariance matrix using the observed transit times. We use these to generate $10^5$ samples from a multivariate normal distribution, enforcing a requirement that the samples must be physical: that is, not allowing negative masses or eccentricities greater than 1. We use the analytical stability criterion from \cite{2018Hadden} to ensure these samples are stable, with $Z < Z_{crit}$, where $Z$ is the complex relative eccentricity (see their Equations~15 and 19).

From these stable, physical samples that are consistent with the observational data, we can select examples to illustrate the possible behavior of K02261. We select a subset of 10 random samples that show libration in the resonant arguments of the 5:3 resonance and a subset of 10 random samples that do not. From these two example models, we repeat the period ratio over time calculation that was done for the observed transit times in \cite{2015Rowe}. This model uses the same transit numbers as \cite{2015Rowe}, and so any observational gaps in the data will be re-created and associated systematic uncertainty in this method of approximating the period ratio from transits is maintained. There is no uncertainty associated with the transit times themselves, however, being calculated exactly from the N-body simulation. For comparison, we also directly take the osculating period ratios from the N-body simulation at the same points in time.

\begin{figure}[h]
	\centering
    \includegraphics[width=\linewidth]{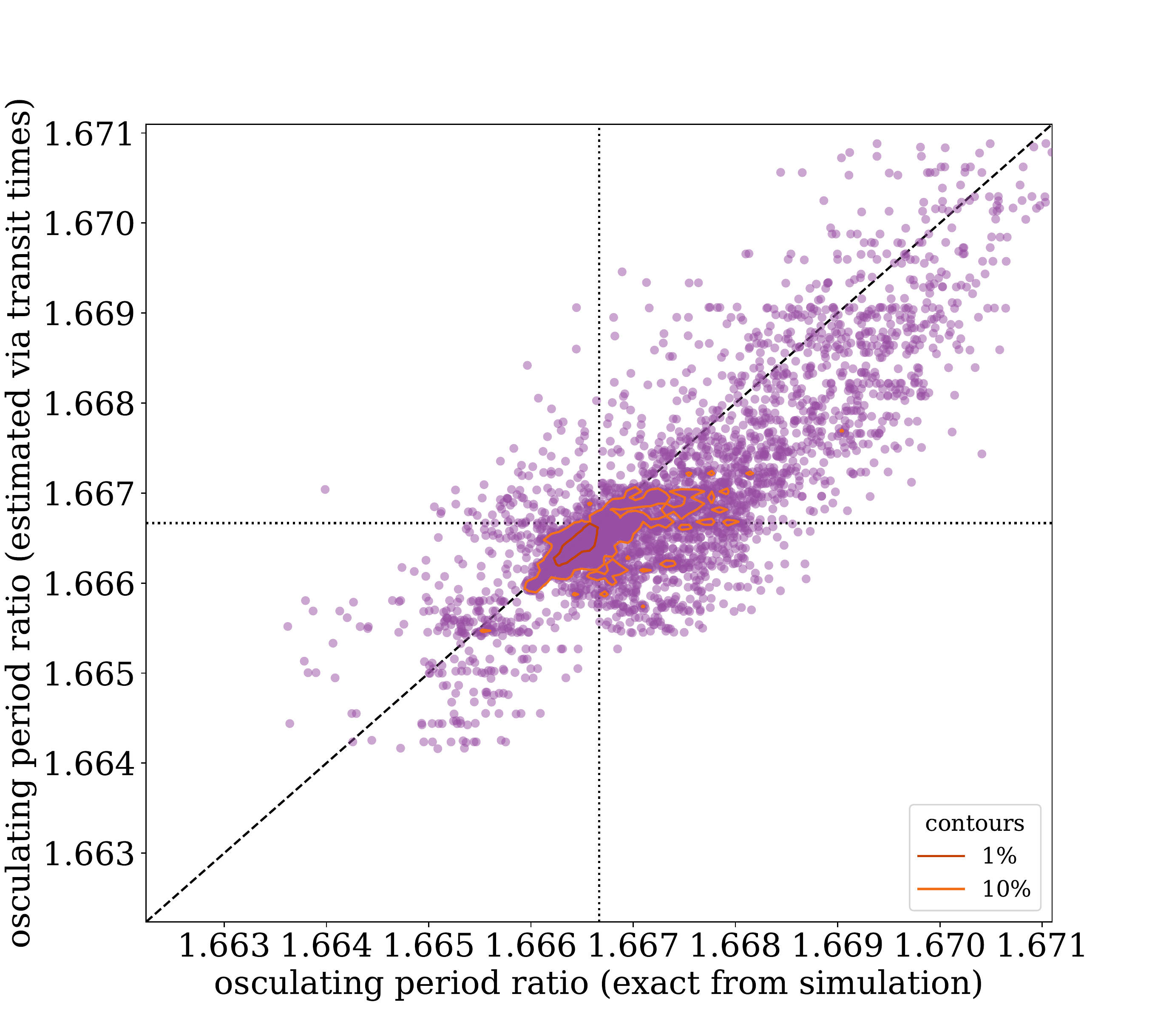}
	\caption{Comparison of the measured period ratio over time using exact osculating periods from an N-body simulation and the estimated osculating period ratio using the previous two transits at a given point in time to estimate the planet periods. The Pearson correlation coefficient is 0.834, indicating a highly linear correlation. Exact correlation is shown by the black line for reference. The comparison is done for twenty realizations of the K02261 that are consistent with the observed data, ten which show signs of libration in the resonant argument and ten that do not. Contours of the top 1\% and 10\% of the point density are shown to aid in interpreting the densest regions. The high correlation indicates that this estimation approach is a useful measurement of the period ratio changing over time directly from the transit times. }
	\label{fig:2261PRcomp}
\end{figure}

From these twenty sample systems, we can compare the osculating period ratios estimated from the transit times and taken exactly from the simulation. We compare these two period ratio measurements in Figure~\ref{fig:2261PRcomp} and find that the estimation from the transit times is highly correlated with the exact period ratios over time, with a Pearson correlation coefficient of 0.834 for the entire set of twenty. Looking at the same relation in the sample systems categorized as resonant, the Pearson correlation coefficient is 0.925, and for those categorized as non-resonant, the Pearson correlation coefficient is 0.777. These high correlations indicate that this estimation approach is a useful measurement of the period ratio changing over time, without requiring an assumption about the resonant state.

\begin{figure*}[h]
	\centering
    \includegraphics[width=\linewidth]{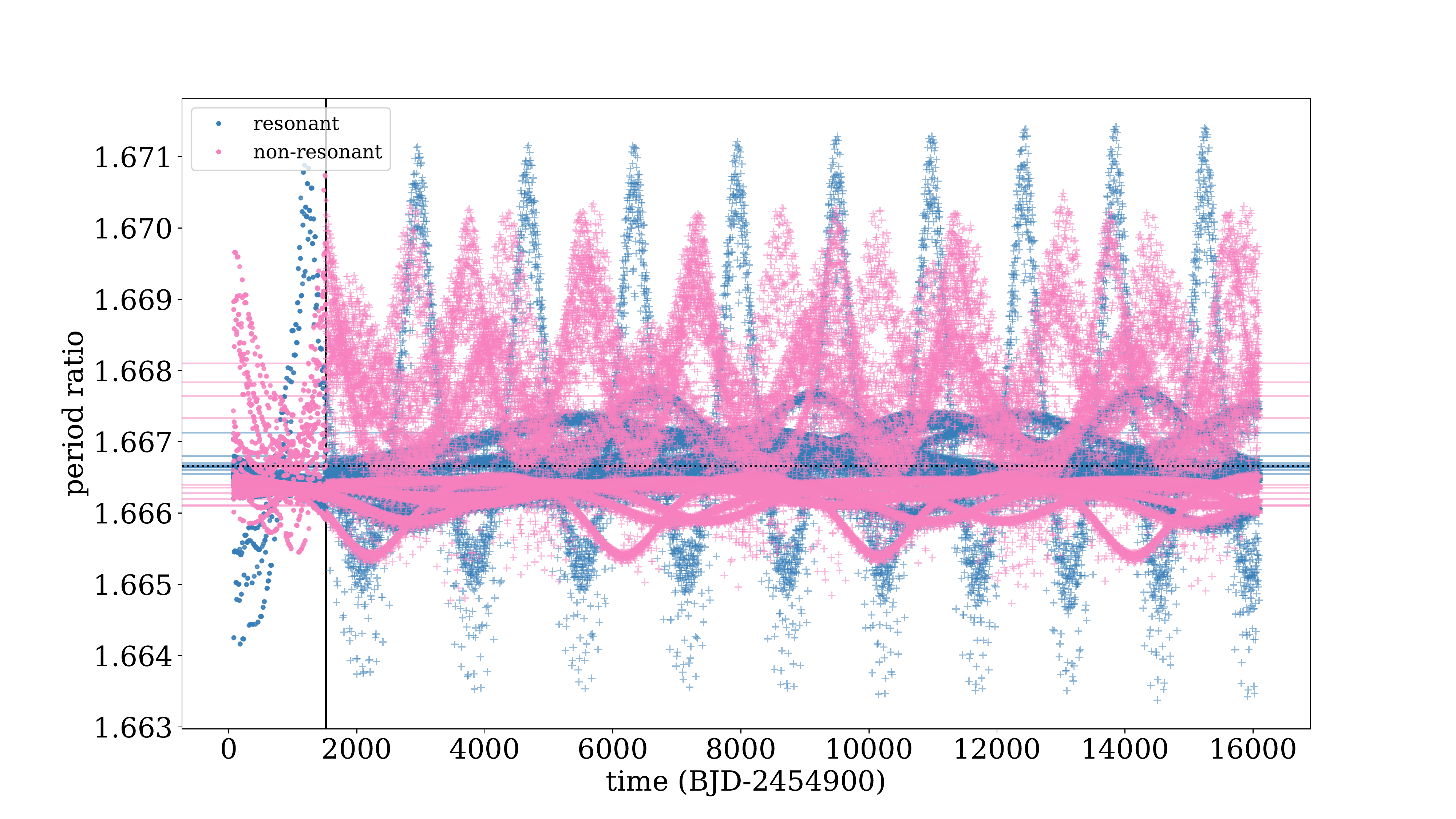}
	\caption{The period ratio over time for 20 sample systems of K02261 that are consistent with the observed TTVs. We use the libration or non-libration in the resonant arguments of the 5:3 resonance to categorize the samples as resonant or non-resonant and extrapolate beyond the Kepler observational period (the time before the solid vertical line) to visualize the periodic behavior in this system. The horizontal lines show the average period ratio over the entire time for each system. The behavior of each individual system varies, but the resonant systems cluster much closer to the exact resonance over time, as expected. }
	\label{fig:2261PRtimemulti}
\end{figure*}

Next, we use these samples to examine how the period ratio changes over time. During the Kepler observation period, there is little structure to the period ratio change. This is likely due to the small expected masses of the planets, both of which are consistent with zero from our stable, physical samples resulting from the least squares fit. However, if we extend the time period another ten Kepler observation periods, structure begins to emerge, as shown in Figure~\ref{fig:2261PRtimemulti}. The period ratio varies periodically, although not always sinusoidally, and while there are many types of variation within these 20 samples, some difference between the resonant and non-resonant samples is visible. In particular, the mean of the period ratio over the entire time is shown with a horizontal line for each sample, and the resonant samples clearly cluster closer to the 5:3 commensurability while the non-resonant samples tend to vary around different period ratios.

However, when we compare these models to the observational data, it is clear the errors from measuring the transit times dominate over any of these effects\textemdash contrast the range in the y-axis in Figures~\ref{fig:2261PRtime} and \ref{fig:2261PRtimemulti}. From this, we infer that the observational data from Kepler for specific systems near the 5:3 resonance do not contradict our conclusions in this paper, but that more precise transit times observations are needed to place any substantive constraints on these individual systems.

\section{Conclusion}\label{sec:conc}

In this work, we demonstrated that highly-eccentric planets produce a very narrow peak near the second-order resonance locations of the 5:3 and 3:1 MMRs when averaged over sufficient time (i.e., longer than the typical libration period), primarily via numerical simulations but also supported by an analytical framework from \cite{2019Hadden}. While we focused on second-order resonances in this work, as they are the widest and strongest resonances of this shape and therefore the likeliest to have a detectable signal, we note that this type of narrow, time-averaged peak shaping is expected to be present for higher-order resonances as well.

We fit a model to the 5:3 and 3:1 peaks in the Kepler period ratio distribution, finding a small peak at the 5:3 and no evidence of a peak at the 3:1.

The lack of these peaks in the planet pairs observed by Kepler therefore places an upper limit on the eccentricity distribution of these planets. We place that upper limit at $\sigma=0.245$ (3:1) and $\sigma=0.095$ (5:3) at 95\% confidence for a Rayleigh distribution with scale $\sigma$ for each planet.

Using our model of the 5:3 Kepler data, we can calculate a resonance probability for the planet pairs located near the resonance. For the K02261, a potentially resonant planet pair identified by this calculation, we investigate the observed transit times from \cite{2015Rowe} and conclude that it is possible to estimate the change in period ratio over time based on the transit times. However, in this case the observational errors in the TTVs are too large to place any constraints on the individual system.

Given that not all of the highly eccentric planets in our numerical simulations are ruled out by stability, we infer that some other mechanism, likely from formation processes, drives the observed low eccentricities of the planet pairs observed by Kepler.

Our method is an independent approach to constraining the eccentricities of the population observed by Kepler, and we find results consistent with other methods: the small planets typical of Kepler observations tend to have low eccentricities, lower than those typically observed of giant planets.

\acknowledgments

This research has made use of the NASA Exoplanet Archive, which is operated by the California Institute of Technology, under contract with the National Aeronautics and Space Administration under the Exoplanet Exploration Program. We thank the anonymous referee for comments that greatly improved this work. We thank Darin Ragozzine for a helpful discussion on quantifying our model via equation~\ref{eqn:model_pdf} and Sam Hadden for assistance in using the \texttt{celmech} package and his analytical model.

\software{
\texttt{REBOUND}\footnote{http://github.com/hannorein/rebound},
\texttt{Forecaster}\footnote{https://github.com/chenjj2/forecaster},
\texttt{emcee}\footnote{https://github.com/dfm/emcee},
\texttt{celmech}\footnote{https://github.com/shadden/celmech},
\texttt{SciPy}\footnote{https://www.scipy.org/}}

\appendix

\section{KOIs}\label{app:KOIs}

The KOIs used in the analysis of the Kepler population in this paper are listed in Table~\ref{tab:KOIs}. 

\begin{table*}[]
\centering
\caption{KOIs in the vicinity of the 3:1 and 5:3 resonances. KOI data were taken from the NASA Exoplanet Archive for all candidate and confirmed KOIs on October 28, 2020. Observational errors are not listed for the periods (and therefore period ratios) as they are generally of an order of one part per million and may not be representative of the accuracy of the listed periods (i.e. \citealt{2020Lissauer}).}
\label{tab:KOIs}
\resizebox{\textwidth}{!}{%
\begin{tabular}{cccccc|c|c|c|c|c|cccccc}
\cline{1-5} \cline{7-17}
\multicolumn{5}{|c|}{\textbf{3:1}} &  & \multicolumn{11}{c|}{\textbf{5:3}} \\ \cline{1-5} \cline{7-17} 
\multicolumn{2}{|c|}{\textbf{KOIs}} & \multicolumn{1}{c|}{\textbf{Period Ratio}} & \multicolumn{2}{c|}{\textbf{Periods (d)}} &  & \multicolumn{2}{c|}{\textbf{KOIs}} & \textbf{Period Ratio} & \multicolumn{2}{c|}{\textbf{Periods (d)}} & \multicolumn{1}{c|}{\textbf{}} & \multicolumn{2}{c|}{\textbf{KOIs}} & \multicolumn{1}{c|}{\textbf{Period Ratio}} & \multicolumn{2}{c|}{\textbf{Periods (d)}} \\ \cline{1-5} \cline{7-11} \cline{13-17} 
\multicolumn{1}{|c|}{K01929.01} & \multicolumn{1}{c|}{K01929.02} & \multicolumn{1}{c|}{2.944} & \multicolumn{1}{c|}{9.693} & \multicolumn{1}{c|}{3.293} &  & K00413.01 & K00413.02 & 1.6202 & 15.2290 & 24.6746 & \multicolumn{1}{c|}{} & \multicolumn{1}{c|}{K01151.01} & \multicolumn{1}{c|}{K01151.04} & \multicolumn{1}{c|}{1.672} & \multicolumn{1}{c|}{10.435} & \multicolumn{1}{c|}{17.453} \\ \cline{1-5} \cline{7-11} \cline{13-17} 
\multicolumn{1}{|c|}{K00490.03} & \multicolumn{1}{c|}{K00490.04} & \multicolumn{1}{c|}{2.944} & \multicolumn{1}{c|}{7.406} & \multicolumn{1}{c|}{21.804} &  & K02352.01 & K02352.03 & 1.6220 & 13.3916 & 8.2563 & \multicolumn{1}{c|}{} & \multicolumn{1}{c|}{K00993.01} & \multicolumn{1}{c|}{K00993.02} & \multicolumn{1}{c|}{1.673} & \multicolumn{1}{c|}{21.854} & \multicolumn{1}{c|}{13.061} \\ \cline{1-5} \cline{7-11} \cline{13-17} 
\multicolumn{1}{|c|}{K00834.03} & \multicolumn{1}{c|}{K00834.04} & \multicolumn{1}{c|}{2.944} & \multicolumn{1}{c|}{6.156} & \multicolumn{1}{c|}{2.091} &  & K00117.02 & K00117.04 & 1.6237 & 4.9014 & 7.9583 & \multicolumn{1}{c|}{} & \multicolumn{1}{c|}{K02704.01} & \multicolumn{1}{c|}{K02704.03} & \multicolumn{1}{c|}{1.674} & \multicolumn{1}{c|}{4.871} & \multicolumn{1}{c|}{8.153} \\ \cline{1-5} \cline{7-11} \cline{13-17} 
\multicolumn{1}{|c|}{K01364.01} & \multicolumn{1}{c|}{K01364.02} & \multicolumn{1}{c|}{2.953} & \multicolumn{1}{c|}{20.834} & \multicolumn{1}{c|}{7.056} &  & K01589.03 & K01589.05 & 1.6239 & 27.4346 & 44.5520 & \multicolumn{1}{c|}{} & \multicolumn{1}{c|}{K02693.01} & \multicolumn{1}{c|}{K02693.03} & \multicolumn{1}{c|}{1.675} & \multicolumn{1}{c|}{4.081} & \multicolumn{1}{c|}{6.834} \\ \cline{1-5} \cline{7-11} \cline{13-17} 
\multicolumn{1}{|c|}{K01301.01} & \multicolumn{1}{c|}{K01301.02} & \multicolumn{1}{c|}{2.954} & \multicolumn{1}{c|}{12.699} & \multicolumn{1}{c|}{37.514} &  & K02029.01 & K02029.02 & 1.6243 & 16.3327 & 10.0555 & \multicolumn{1}{c|}{} & \multicolumn{1}{c|}{K00314.01} & \multicolumn{1}{c|}{K00314.02} & \multicolumn{1}{c|}{1.675} & \multicolumn{1}{c|}{13.781} & \multicolumn{1}{c|}{23.089} \\ \cline{1-5} \cline{7-11} \cline{13-17} 
\multicolumn{1}{|c|}{K00152.02} & \multicolumn{1}{c|}{K00152.04} & \multicolumn{1}{c|}{2.958} & \multicolumn{1}{c|}{27.402} & \multicolumn{1}{c|}{81.064} &  & K00912.01 & K00912.02 & 1.6269 & 10.8485 & 6.6683 & \multicolumn{1}{c|}{} & \multicolumn{1}{c|}{K00972.01} & \multicolumn{1}{c|}{K00972.02} & \multicolumn{1}{c|}{1.677} & \multicolumn{1}{c|}{13.119} & \multicolumn{1}{c|}{7.822} \\ \cline{1-5} \cline{7-11} \cline{13-17} 
\multicolumn{1}{|c|}{K02199.01} & \multicolumn{1}{c|}{K02199.02} & \multicolumn{1}{c|}{2.959} & \multicolumn{1}{c|}{9.033} & \multicolumn{1}{c|}{3.053} &  & K00566.01 & K00566.03 & 1.6271 & 25.8550 & 42.0686 & \multicolumn{1}{c|}{} & \multicolumn{1}{c|}{K00571.02} & \multicolumn{1}{c|}{K00571.04} & \multicolumn{1}{c|}{1.679} & \multicolumn{1}{c|}{13.343} & \multicolumn{1}{c|}{22.408} \\ \cline{1-5} \cline{7-11} \cline{13-17} 
\multicolumn{1}{|c|}{K00566.03} & \multicolumn{1}{c|}{K00566.02} & \multicolumn{1}{c|}{2.971} & \multicolumn{1}{c|}{42.069} & \multicolumn{1}{c|}{14.157} &  & K02243.01 & K02243.02 & 1.6311 & 5.1856 & 8.4581 & \multicolumn{1}{c|}{} & \multicolumn{1}{c|}{K04383.01} & \multicolumn{1}{c|}{K04383.02} & \multicolumn{1}{c|}{1.680} & \multicolumn{1}{c|}{4.342} & \multicolumn{1}{c|}{2.584} \\ \cline{1-5} \cline{7-11} \cline{13-17} 
\multicolumn{1}{|c|}{K00245.01} & \multicolumn{1}{c|}{K00245.03} & \multicolumn{1}{c|}{2.977} & \multicolumn{1}{c|}{39.792} & \multicolumn{1}{c|}{13.367} &  & K01940.01 & K01940.02 & 1.6320 & 10.9947 & 6.7368 & \multicolumn{1}{c|}{} & \multicolumn{1}{c|}{K00490.01} & \multicolumn{1}{c|}{K00490.03} & \multicolumn{1}{c|}{1.686} & \multicolumn{1}{c|}{4.393} & \multicolumn{1}{c|}{7.406} \\ \cline{1-5} \cline{7-11} \cline{13-17} 
\multicolumn{1}{|c|}{K00497.01} & \multicolumn{1}{c|}{K00497.02} & \multicolumn{1}{c|}{2.981} & \multicolumn{1}{c|}{13.193} & \multicolumn{1}{c|}{4.425} &  & K00719.02 & K00719.03 & 1.6323 & 28.1224 & 45.9030 & \multicolumn{1}{c|}{} & \multicolumn{1}{c|}{K00351.02} & \multicolumn{1}{c|}{K00351.07} & \multicolumn{1}{c|}{1.686} & \multicolumn{1}{c|}{210.601} & \multicolumn{1}{c|}{124.923} \\ \cline{1-5} \cline{7-11} \cline{13-17} 
\multicolumn{1}{|c|}{K00582.01} & \multicolumn{1}{c|}{K00582.02} & \multicolumn{1}{c|}{2.984} & \multicolumn{1}{c|}{5.945} & \multicolumn{1}{c|}{17.738} &  & K02704.01 & K02704.02 & 1.6324 & 4.8712 & 2.9842 & \multicolumn{1}{c|}{} & \multicolumn{1}{c|}{K00547.03} & \multicolumn{1}{c|}{K00547.02} & \multicolumn{1}{c|}{1.686} & \multicolumn{1}{c|}{12.386} & \multicolumn{1}{c|}{7.347} \\ \cline{1-5} \cline{7-11} \cline{13-17} 
\multicolumn{1}{|c|}{K03029.01} & \multicolumn{1}{c|}{K03029.02} & \multicolumn{1}{c|}{2.990} & \multicolumn{1}{c|}{18.976} & \multicolumn{1}{c|}{6.347} &  & K03083.01 & K03083.02 & 1.6340 & 10.1832 & 6.2321 & \multicolumn{1}{c|}{} & \multicolumn{1}{c|}{K00664.01} & \multicolumn{1}{c|}{K00664.02} & \multicolumn{1}{c|}{1.688} & \multicolumn{1}{c|}{13.137} & \multicolumn{1}{c|}{7.782} \\ \cline{1-5} \cline{7-11} \cline{13-17} 
\multicolumn{1}{|c|}{K00117.02} & \multicolumn{1}{c|}{K00117.01} & \multicolumn{1}{c|}{3.009} & \multicolumn{1}{c|}{4.901} & \multicolumn{1}{c|}{14.749} &  & K01920.01 & K01920.02 & 1.6345 & 16.5710 & 10.1385 & \multicolumn{1}{c|}{} & \multicolumn{1}{c|}{K03470.01} & \multicolumn{1}{c|}{K03470.02} & \multicolumn{1}{c|}{1.689} & \multicolumn{1}{c|}{15.062} & \multicolumn{1}{c|}{25.434} \\ \cline{1-5} \cline{7-11} \cline{13-17} 
\multicolumn{1}{|c|}{K01835.02} & \multicolumn{1}{c|}{K01835.03} & \multicolumn{1}{c|}{3.010} & \multicolumn{1}{c|}{2.248} & \multicolumn{1}{c|}{6.767} &  & K00945.02 & K00945.03 & 1.6404 & 40.7156 & 66.7893 & \multicolumn{1}{c|}{} & \multicolumn{1}{c|}{K04846.02} & \multicolumn{1}{c|}{K04846.01} & \multicolumn{1}{c|}{1.689} & \multicolumn{1}{c|}{32.527} & \multicolumn{1}{c|}{19.255} \\ \cline{1-5} \cline{7-11} \cline{13-17} 
\multicolumn{1}{|c|}{K00664.02} & \multicolumn{1}{c|}{K00664.03} & \multicolumn{1}{c|}{3.013} & \multicolumn{1}{c|}{7.782} & \multicolumn{1}{c|}{23.443} &  & K02612.01 & K02612.02 & 1.6419 & 4.6123 & 7.5730 & \multicolumn{1}{c|}{} & \multicolumn{1}{c|}{K03398.01} & \multicolumn{1}{c|}{K03398.03} & \multicolumn{1}{c|}{1.692} & \multicolumn{1}{c|}{7.319} & \multicolumn{1}{c|}{4.327} \\ \cline{1-5} \cline{7-11} \cline{13-17} 
\multicolumn{1}{|c|}{K02092.02} & \multicolumn{1}{c|}{K02092.03} & \multicolumn{1}{c|}{3.015} & \multicolumn{1}{c|}{25.564} & \multicolumn{1}{c|}{77.086} &  & K00733.02 & K00733.04 & 1.6427 & 11.3493 & 18.6435 & \multicolumn{1}{c|}{} & \multicolumn{1}{c|}{K00352.01} & \multicolumn{1}{c|}{K00352.02} & \multicolumn{1}{c|}{1.692} & \multicolumn{1}{c|}{27.082} & \multicolumn{1}{c|}{16.007} \\ \cline{1-5} \cline{7-11} \cline{13-17} 
\multicolumn{1}{|c|}{K02218.01} & \multicolumn{1}{c|}{K02218.02} & \multicolumn{1}{c|}{3.022} & \multicolumn{1}{c|}{5.535} & \multicolumn{1}{c|}{16.726} &  & K01165.01 & K01165.02 & 1.6433 & 7.0539 & 4.2926 & \multicolumn{1}{c|}{} & \multicolumn{1}{c|}{K01060.01} & \multicolumn{1}{c|}{K01060.03} & \multicolumn{1}{c|}{1.693} & \multicolumn{1}{c|}{12.110} & \multicolumn{1}{c|}{20.497} \\ \cline{1-5} \cline{7-11} \cline{13-17} 
\multicolumn{1}{|c|}{K00952.02} & \multicolumn{1}{c|}{K00952.04} & \multicolumn{1}{c|}{3.022} & \multicolumn{1}{c|}{8.752} & \multicolumn{1}{c|}{2.896} &  & K01598.01 & K01598.02 & 1.6445 & 56.4758 & 92.8747 & \multicolumn{1}{c|}{} & \multicolumn{1}{c|}{K01221.01} & \multicolumn{1}{c|}{K01221.02} & \multicolumn{1}{c|}{1.694} & \multicolumn{1}{c|}{30.160} & \multicolumn{1}{c|}{51.077} \\ \cline{1-5} \cline{7-11} \cline{13-17} 
\multicolumn{1}{|c|}{K00542.01} & \multicolumn{1}{c|}{K00542.02} & \multicolumn{1}{c|}{3.031} & \multicolumn{1}{c|}{41.885} & \multicolumn{1}{c|}{13.817} &  & K06209.01 & K06209.02 & 1.6449 & 12.7144 & 7.7297 & \multicolumn{1}{c|}{} & \multicolumn{1}{c|}{K04500.01} & \multicolumn{1}{c|}{K04500.03} & \multicolumn{1}{c|}{1.695} & \multicolumn{1}{c|}{8.702} & \multicolumn{1}{c|}{14.751} \\ \cline{1-5} \cline{7-11} \cline{13-17} 
\multicolumn{1}{|c|}{K00435.01} & \multicolumn{1}{c|}{K00435.05} & \multicolumn{1}{c|}{3.032} & \multicolumn{1}{c|}{20.550} & \multicolumn{1}{c|}{62.303} &  & K04435.01 & K04435.02 & 1.6452 & 10.8614 & 17.8694 & \multicolumn{1}{c|}{} & \multicolumn{1}{c|}{K02842.02} & \multicolumn{1}{c|}{K02842.03} & \multicolumn{1}{c|}{1.696} & \multicolumn{1}{c|}{5.149} & \multicolumn{1}{c|}{3.036} \\ \cline{1-5} \cline{7-11} \cline{13-17} 
\multicolumn{1}{|c|}{K01832.02} & \multicolumn{1}{c|}{K01832.03} & \multicolumn{1}{c|}{3.034} & \multicolumn{1}{c|}{12.762} & \multicolumn{1}{c|}{38.715} &  & K02263.03 & K02263.02 & 1.6452 & 15.5944 & 9.4785 & \multicolumn{1}{c|}{} & \multicolumn{1}{c|}{K00283.01} & \multicolumn{1}{c|}{K00283.03} & \multicolumn{1}{c|}{1.696} & \multicolumn{1}{c|}{16.092} & \multicolumn{1}{c|}{9.488} \\ \cline{1-5} \cline{7-11} \cline{13-17} 
\multicolumn{1}{|c|}{K06097.01} & \multicolumn{1}{c|}{K06097.03} & \multicolumn{1}{c|}{3.034} & \multicolumn{1}{c|}{93.057} & \multicolumn{1}{c|}{282.343} &  & K02236.01 & K02236.02 & 1.6483 & 19.9866 & 12.1258 & \multicolumn{1}{c|}{} & \multicolumn{1}{c|}{K01364.02} & \multicolumn{1}{c|}{K01364.03} & \multicolumn{1}{c|}{1.698} & \multicolumn{1}{c|}{7.056} & \multicolumn{1}{c|}{11.979} \\ \cline{1-5} \cline{7-11} \cline{13-17} 
\multicolumn{1}{|c|}{K02038.01} & \multicolumn{1}{c|}{K02038.04} & \multicolumn{1}{c|}{3.036} & \multicolumn{1}{c|}{8.305} & \multicolumn{1}{c|}{25.215} &  & K00505.01 & K00505.04 & 1.6491 & 13.7671 & 8.3482 & \multicolumn{1}{c|}{} & \multicolumn{1}{c|}{K00658.01} & \multicolumn{1}{c|}{K00658.02} & \multicolumn{1}{c|}{1.698} & \multicolumn{1}{c|}{3.163} & \multicolumn{1}{c|}{5.371} \\ \cline{1-5} \cline{7-11} \cline{13-17} 
\multicolumn{1}{|c|}{K01713.02} & \multicolumn{1}{c|}{K01713.01} & \multicolumn{1}{c|}{3.047} & \multicolumn{1}{c|}{2.240} & \multicolumn{1}{c|}{6.828} &  & K00070.04 & K00070.02 & 1.6500 & 6.0986 & 3.6961 & \multicolumn{1}{c|}{} & \multicolumn{1}{c|}{K03068.01} & \multicolumn{1}{c|}{K03068.02} & \multicolumn{1}{c|}{1.698} & \multicolumn{1}{c|}{3.917} & \multicolumn{1}{c|}{6.652} \\ \cline{1-5} \cline{7-11} \cline{13-17} 
\multicolumn{1}{|c|}{K00232.01} & \multicolumn{1}{c|}{K00232.04} & \multicolumn{1}{c|}{3.048} & \multicolumn{1}{c|}{12.466} & \multicolumn{1}{c|}{37.996} &  & K00707.01 & K00707.04 & 1.6527 & 21.7757 & 13.1756 & \multicolumn{1}{c|}{} & \multicolumn{1}{c|}{K00082.01} & \multicolumn{1}{c|}{K00082.03} & \multicolumn{1}{c|}{1.700} & \multicolumn{1}{c|}{16.146} & \multicolumn{1}{c|}{27.454} \\ \cline{1-5} \cline{7-11} \cline{13-17} 
\multicolumn{1}{|c|}{K01589.02} & \multicolumn{1}{c|}{K01589.04} & \multicolumn{1}{c|}{3.050} & \multicolumn{1}{c|}{12.883} & \multicolumn{1}{c|}{4.225} &  & K00271.01 & K00271.02 & 1.6545 & 48.6304 & 29.3934 & \multicolumn{1}{c|}{} & \multicolumn{1}{c|}{K00408.01} & \multicolumn{1}{c|}{K00408.02} & \multicolumn{1}{c|}{1.702} & \multicolumn{1}{c|}{7.382} & \multicolumn{1}{c|}{12.561} \\ \cline{1-5} \cline{7-11} \cline{13-17} 
\multicolumn{1}{|c|}{K04500.02} & \multicolumn{1}{c|}{K04500.03} & \multicolumn{1}{c|}{3.050} & \multicolumn{1}{c|}{44.985} & \multicolumn{1}{c|}{14.751} &  & K02433.02 & K02433.06 & 1.6565 & 10.0438 & 6.0633 & \multicolumn{1}{c|}{} & \multicolumn{1}{c|}{K01413.01} & \multicolumn{1}{c|}{K01413.02} & \multicolumn{1}{c|}{1.702} & \multicolumn{1}{c|}{12.645} & \multicolumn{1}{c|}{21.526} \\ \cline{1-5} \cline{7-11} \cline{13-17} 
\multicolumn{1}{|c|}{K01563.01} & \multicolumn{1}{c|}{K01563.04} & \multicolumn{1}{c|}{3.051} & \multicolumn{1}{c|}{5.487} & \multicolumn{1}{c|}{16.739} &  & K06145.01 & K06145.02 & 1.6600 & 19.6723 & 32.6570 & \multicolumn{1}{c|}{} & \multicolumn{1}{c|}{K00220.01} & \multicolumn{1}{c|}{K00220.02} & \multicolumn{1}{c|}{1.703} & \multicolumn{1}{c|}{2.422} & \multicolumn{1}{c|}{4.125} \\ \cline{1-5} \cline{7-11} \cline{13-17} 
\multicolumn{1}{|c|}{K00597.03} & \multicolumn{1}{c|}{K00597.01} & \multicolumn{1}{c|}{3.051} & \multicolumn{1}{c|}{52.815} & \multicolumn{1}{c|}{17.308} &  & K04034.01 & K04034.02 & 1.6663 & 6.9670 & 11.6089 & \multicolumn{1}{c|}{} & \multicolumn{1}{c|}{K03158.02} & \multicolumn{1}{c|}{K03158.04} & \multicolumn{1}{c|}{1.703} & \multicolumn{1}{c|}{4.546} & \multicolumn{1}{c|}{7.743} \\ \cline{1-5} \cline{7-11} \cline{13-17} 
\multicolumn{1}{|c|}{K00082.01} & \multicolumn{1}{c|}{K00082.05} & \multicolumn{1}{c|}{3.054} & \multicolumn{1}{c|}{16.146} & \multicolumn{1}{c|}{5.287} &  & K02261.01 & K02261.02 & 1.6664 & 3.9760 & 6.6255 & \multicolumn{1}{c|}{} & \multicolumn{1}{c|}{K00248.02} & \multicolumn{1}{c|}{K00248.04} & \multicolumn{1}{c|}{1.704} & \multicolumn{1}{c|}{10.913} & \multicolumn{1}{c|}{18.596} \\ \cline{1-5} \cline{7-11} \cline{13-17} 
 &  &  &  &  &  & K01574.01 & K01574.02 & 1.6668 & 114.7366 & 191.2391 & \multicolumn{1}{c|}{} & \multicolumn{1}{c|}{K01306.02} & \multicolumn{1}{c|}{K01306.03} & \multicolumn{1}{c|}{1.705} & \multicolumn{1}{c|}{3.468} & \multicolumn{1}{c|}{5.914} \\ \cline{7-11} \cline{13-17} 
 &  &  &  &  &  & K02163.01 & K02163.02 & 1.6675 & 10.6649 & 17.7838 & \multicolumn{1}{c|}{} & \multicolumn{1}{c|}{K02926.01} & \multicolumn{1}{c|}{K02926.03} & \multicolumn{1}{c|}{1.706} & \multicolumn{1}{c|}{12.285} & \multicolumn{1}{c|}{20.957} \\ \cline{7-11} \cline{13-17} 
 &  &  &  &  &  & K02086.01 & K02086.03 & 1.6681 & 7.1329 & 11.8985 & \multicolumn{1}{c|}{} & \multicolumn{1}{c|}{K01194.01} & \multicolumn{1}{c|}{K01194.02} & \multicolumn{1}{c|}{1.707} & \multicolumn{1}{c|}{8.708} & \multicolumn{1}{c|}{14.862} \\ \cline{7-11} \cline{13-17} 
 &  &  &  &  &  & K04136.01 & K04136.02 & 1.6685 & 2.4175 & 4.0336 & \multicolumn{1}{c|}{} & \multicolumn{1}{c|}{K02018.01} & \multicolumn{1}{c|}{K02018.02} & \multicolumn{1}{c|}{1.708} & \multicolumn{1}{c|}{27.496} & \multicolumn{1}{c|}{16.099} \\ \cline{7-11} \cline{13-17} 
 &  &  &  &  &  & K02169.01 & K02169.02 & 1.6693 & 5.4530 & 3.2666 & \multicolumn{1}{c|}{} & \multicolumn{1}{c|}{K00446.01} & \multicolumn{1}{c|}{K00446.02} & \multicolumn{1}{c|}{1.709} & \multicolumn{1}{c|}{16.709} & \multicolumn{1}{c|}{28.552} \\ \cline{7-11} \cline{13-17} 
 &  &  &  &  &  & K02693.02 & K02693.03 & 1.6709 & 11.4194 & 6.8344 & \multicolumn{1}{c|}{} & \multicolumn{1}{c|}{K01563.01} & \multicolumn{1}{c|}{K01563.03} & \multicolumn{1}{c|}{1.712} & \multicolumn{1}{c|}{5.487} & \multicolumn{1}{c|}{3.205} \\ \cline{7-11} \cline{13-17} 
 &  &  &  &  &  & K00582.01 & K00582.03 & 1.6719 & 5.9450 & 9.9396 &  &  &  &  &  &  \\ \cline{7-11}
\end{tabular}%
}
\end{table*}

\section{Population Comparisons}\label{app:pop}

In this section, we illustrate various properties of the synthetic population of planet pairs that we generated as described in Section~\ref{sec:datagen}. When applicable, the corresponding properties of the Kepler sample are shown as well to highlight the general similarity in parameter coverage. Data for the Kepler sample, including errors, were taken from the NASA Exoplanet Archive for all candidate and confirmed KOIs on October 28, 2020. This similarity is important to ensure that we are marginalizing over other contributing factors to isolate the eccentricity effect on the period ratio distribution near second-order MMRs.

\begin{figure}[h]
	\centering
	\subfloat{
	    \label{fig:logmratio-cdf-a}
	    \includegraphics[width=.5\linewidth]{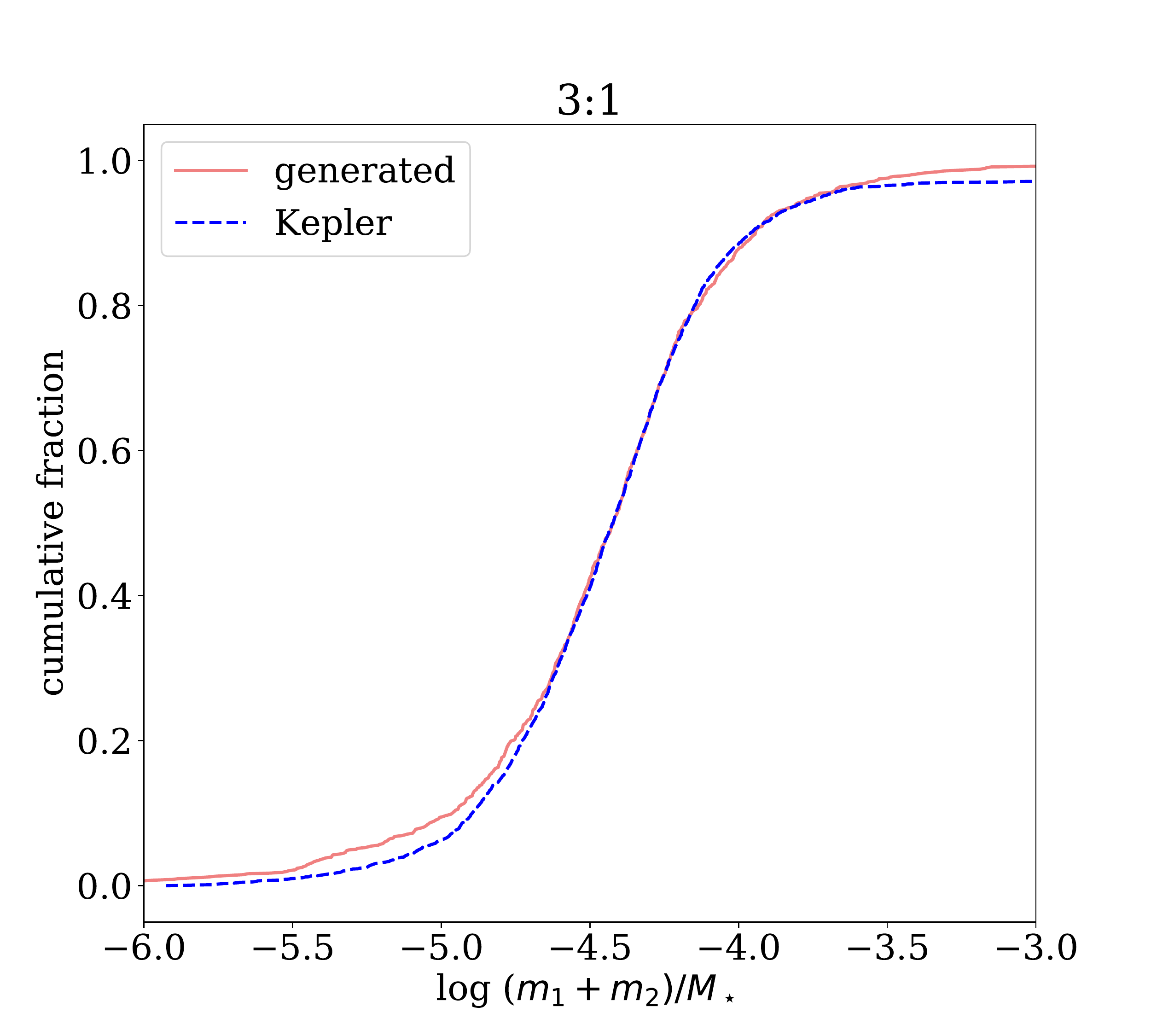}
	}
	\subfloat{
	    \label{fig:logmratio-cdf-b}
	    \includegraphics[width=.5\linewidth]{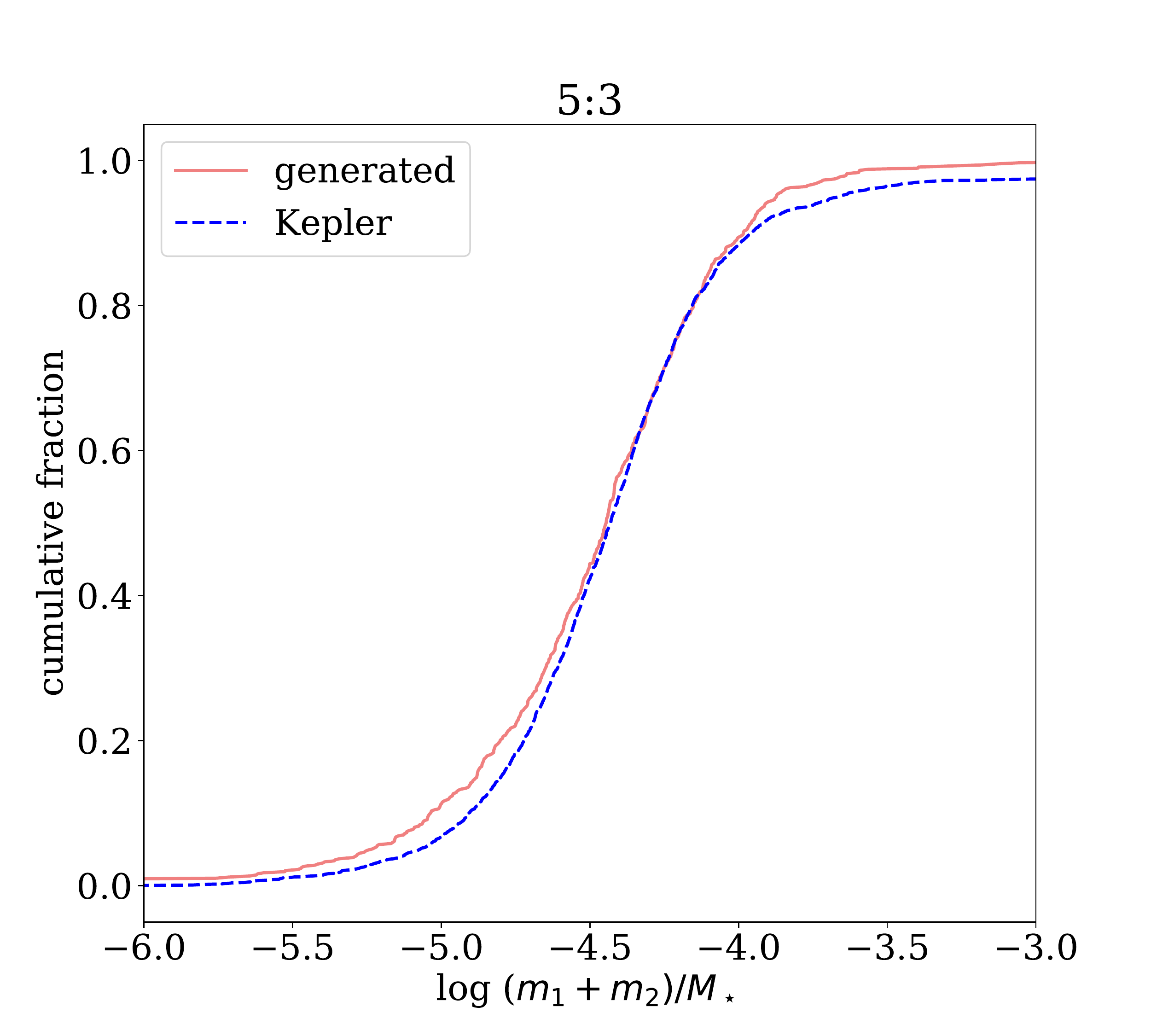}
	}
	\caption{Cumulative fraction of the logarithm of the mass ratio of the planets to the star for the generated population of planet pairs (solid coral) and for planet pairs in the Kepler sample (dashed blue) for the \subref{fig:logmratio-cdf-a} 3:1 and \subref{fig:logmratio-cdf-b} 5:3 resonances.}
	\label{fig:logmratio-cdf}
\end{figure}

Figure~\ref{fig:logmratio-cdf} shows the cumulative distribution of the logarithm of the planet-to-star mass ratios for the generated planet pairs and Kepler planet pairs. The planet masses for the Kepler planets were forecast from their radii using the \texttt{Forecaster} package. Observational uncertainty and the probabilistic nature of the \texttt{Forecaster} mass-radius relation were incorporated numerically by drawing 10 radius and stellar mass samples for each planet.

\begin{figure}[h]
	\centering
	\subfloat{
	    \label{fig:P1-a}
	    \includegraphics[width=.5\linewidth]{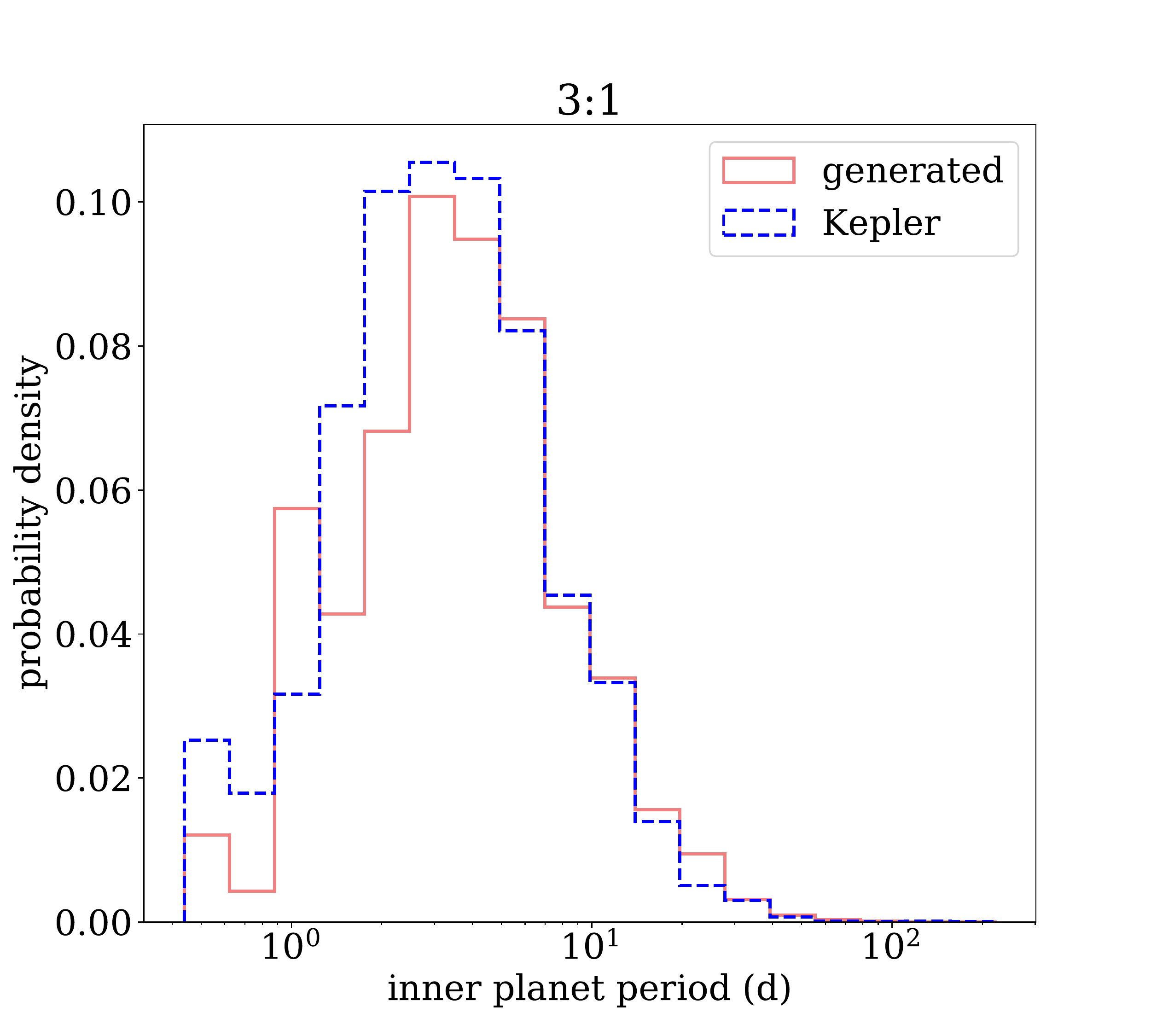}
	}
	\subfloat{
	    \label{fig:P1-b}
	    \includegraphics[width=.5\linewidth]{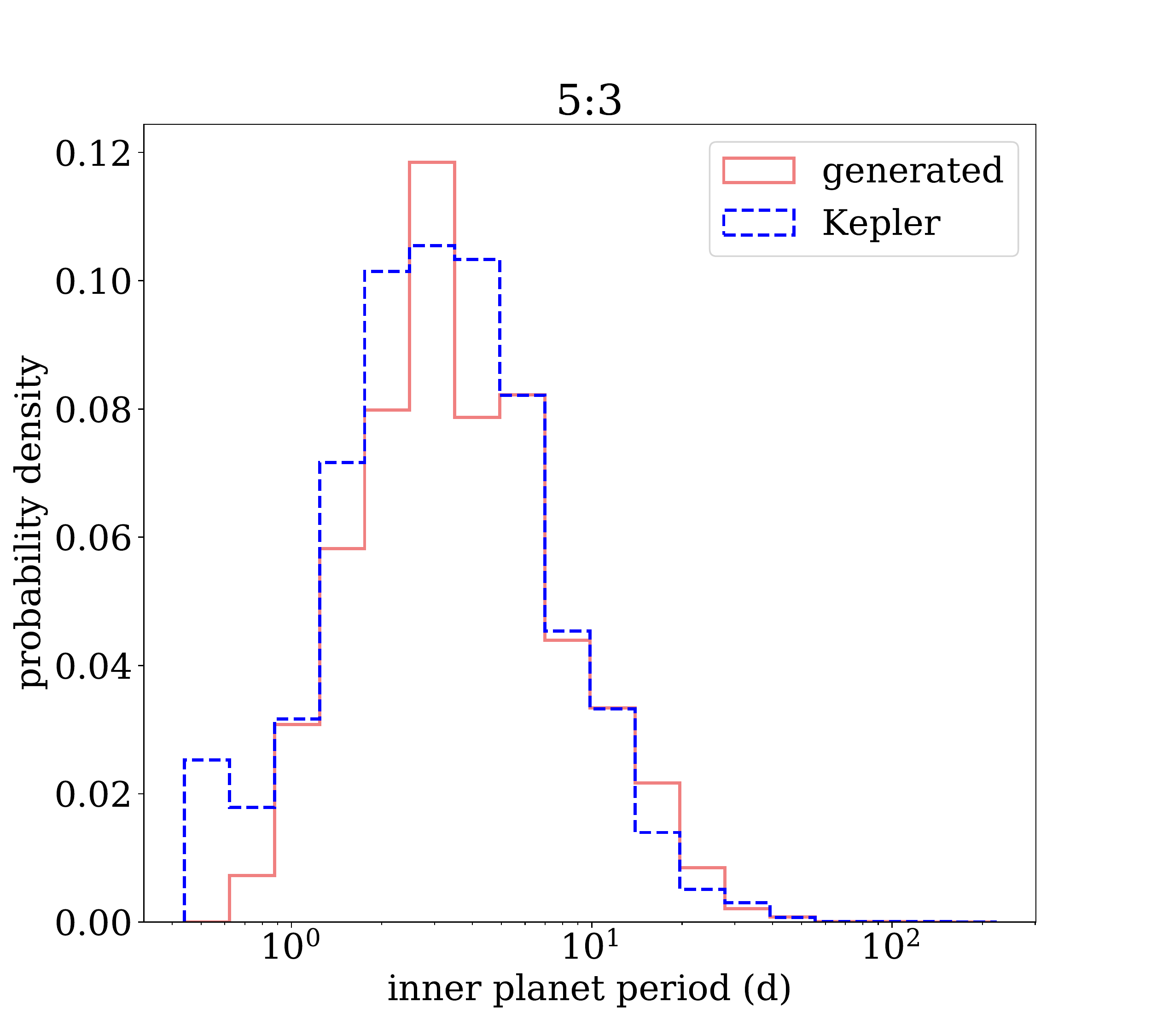}
	}
	\caption{Distribution of the inner planet's period for the generated population of planet pairs (solid coral) and for planet pairs in the Kepler sample (dashed blue) for the \subref{fig:P1-a} 3:1 and \subref{fig:P1-b} 5:3 resonances.}
	\label{fig:P1}
\end{figure}

Figure~\ref{fig:P1} shows the distribution of the inner planet's period for the generated planet pairs and Kepler multi-planet systems. This period generally sets the dynamical timescale of the systems, including tidal circularization and the period ratio libration periods.

\begin{figure}[h]
	\centering
	\subfloat{
	    \label{fig:PRif-a}
	    \includegraphics[width=.5\linewidth]{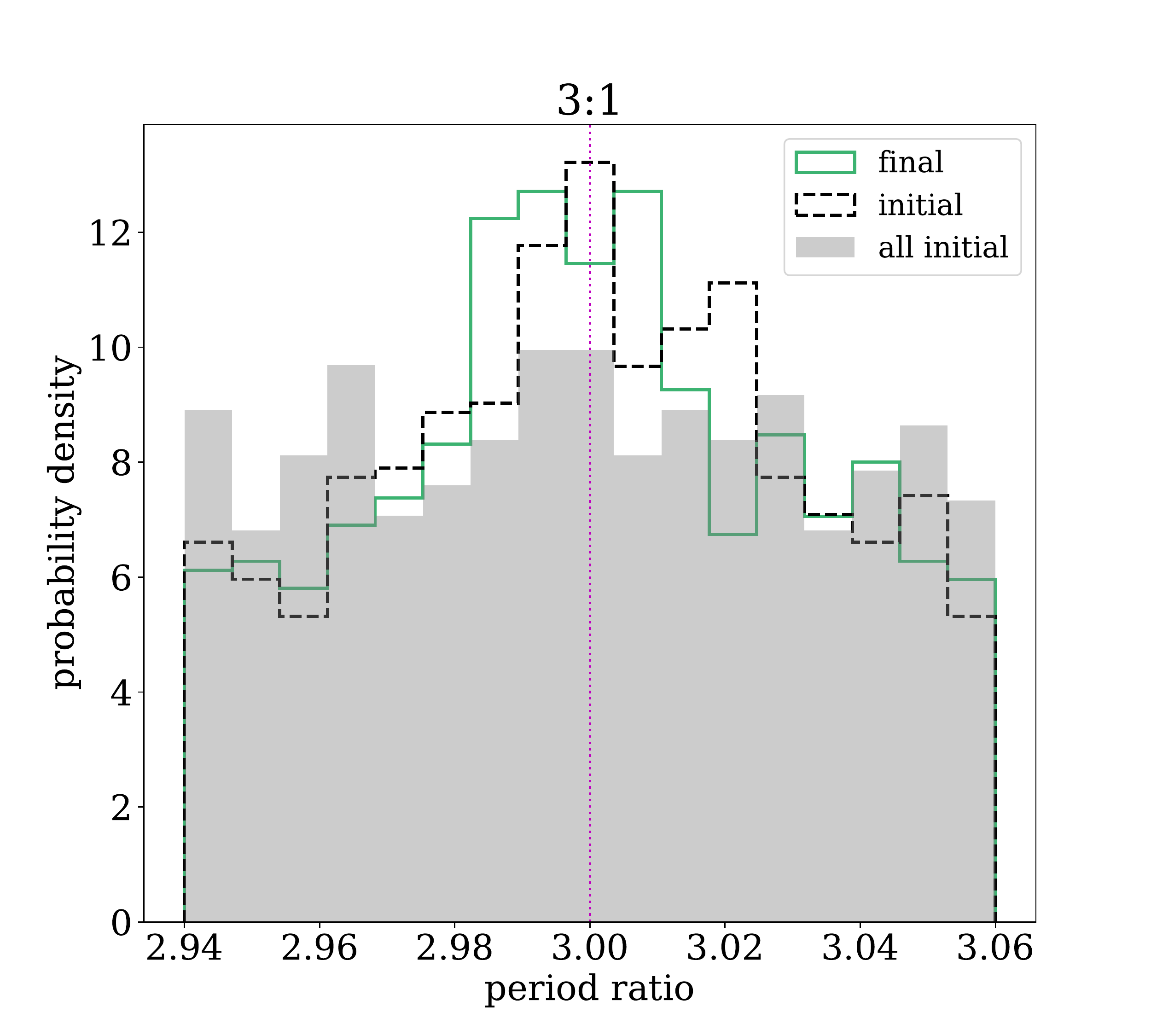}
	}
	\subfloat{
	    \label{fig:PRif-b}
	    \includegraphics[width=.5\linewidth]{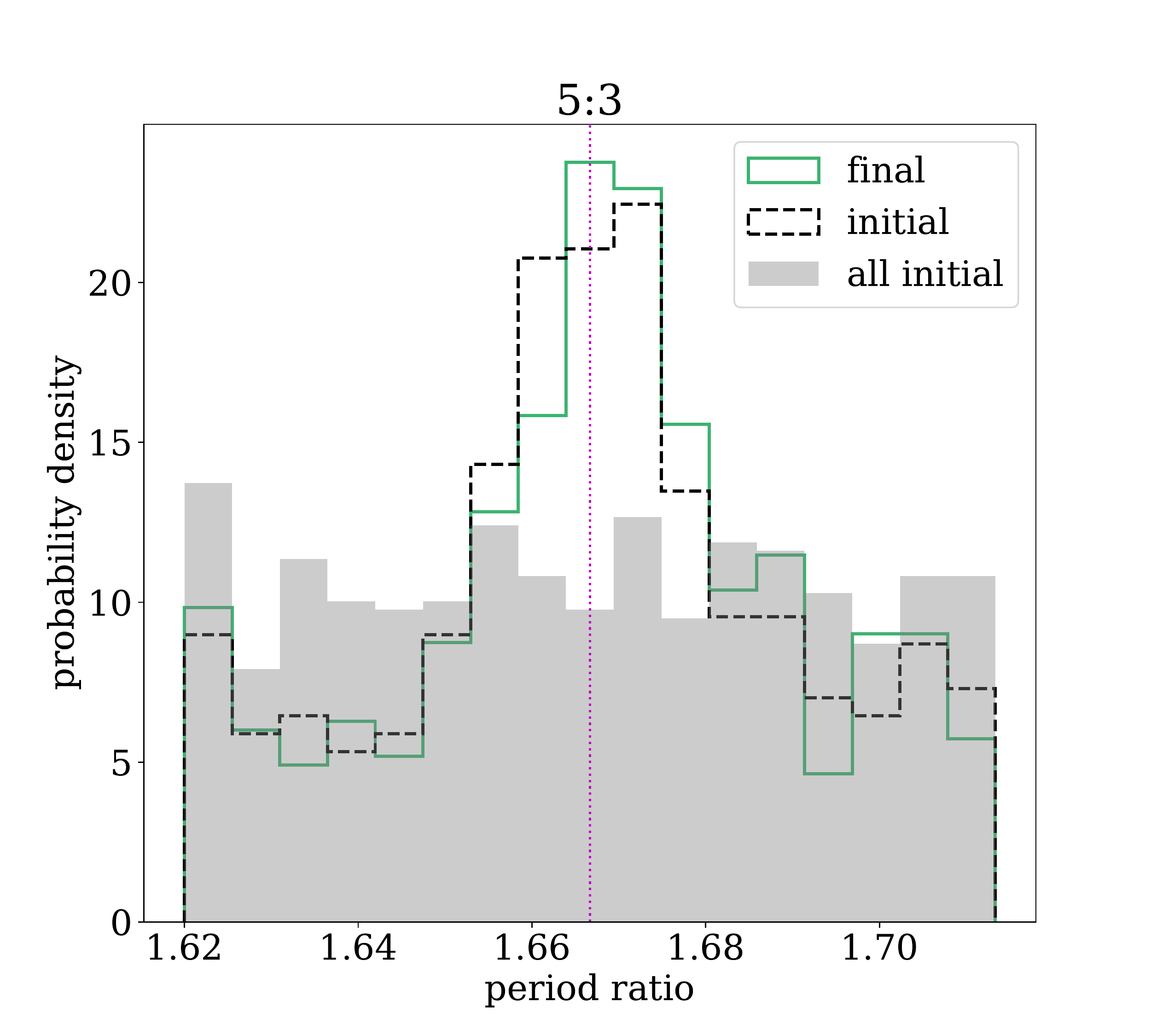}
	}
	\caption{Distribution of the instantaneous period ratio of the final selected planet pairs at the beginning (dashed black) and end (solid green) of the stability run. The gray background shows the initial period ratios in this range for all planet pairs, including those that went unstable and were excluded from the final selection. Distributions are shown for the \subref{fig:PRif-a} 3:1 and \subref{fig:PRif-b} 5:3 resonances, and the exact commensurabilities are shown with a dotted pink line.}
	\label{fig:PRif}
\end{figure}

Figure~\ref{fig:PRif} shows the distribution of the period ratio for the generated planet pairs before and after the stability run. More systems went unstable near to but not right at the exact commensurabilities, leading to some peaking even in the instantaneous (osculating) period ratios. This effect is intensified when time-averaged (see Figure~\ref{fig:PR_distro_pdf}).

\begin{figure}[h]
	\centering
	\subfloat{
	    \label{fig:mutincif-a}
	    \includegraphics[width=.5\linewidth]{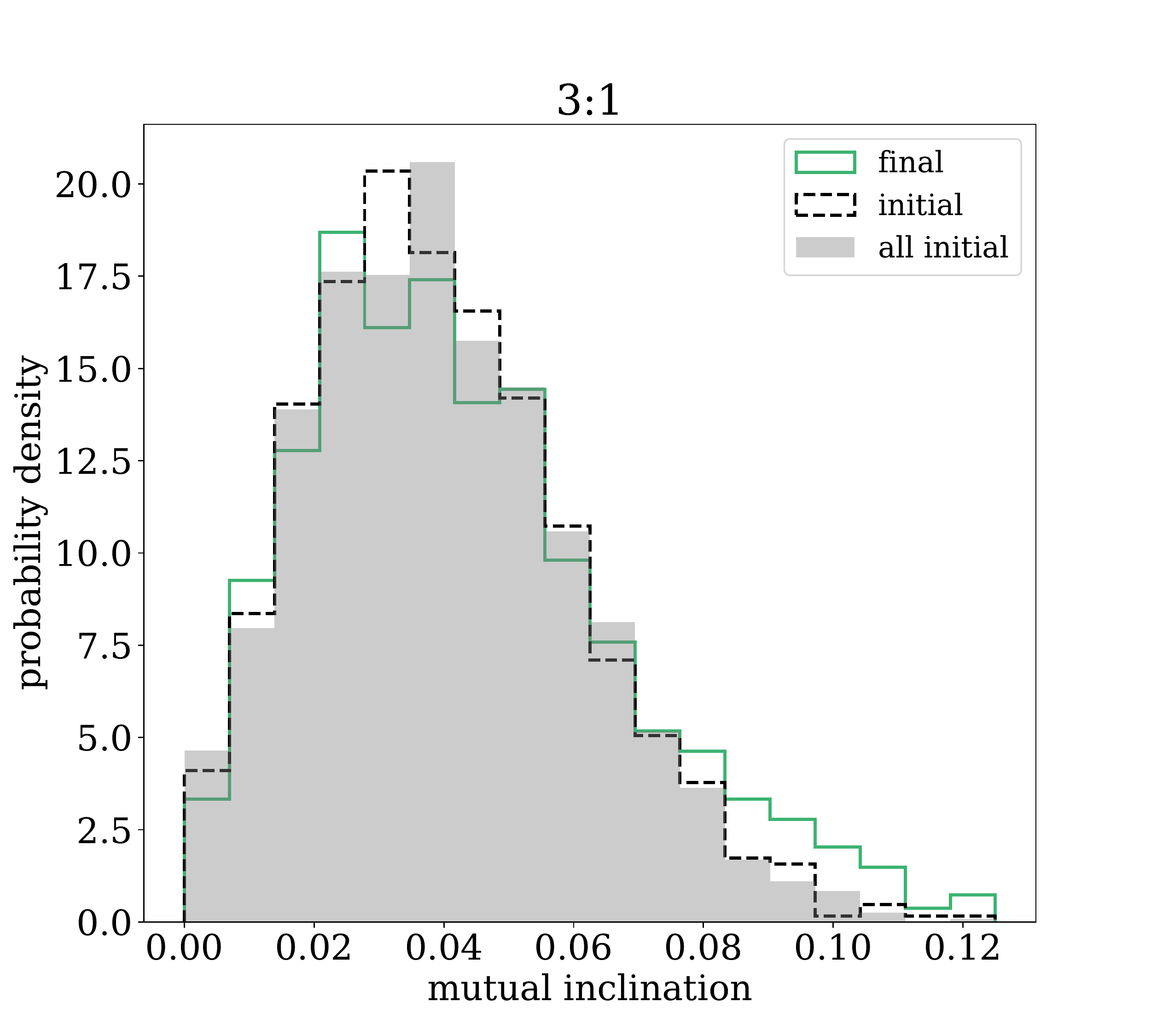}
	}
	\subfloat{
	    \label{fig:mutincif-b}
	    \includegraphics[width=.5\linewidth]{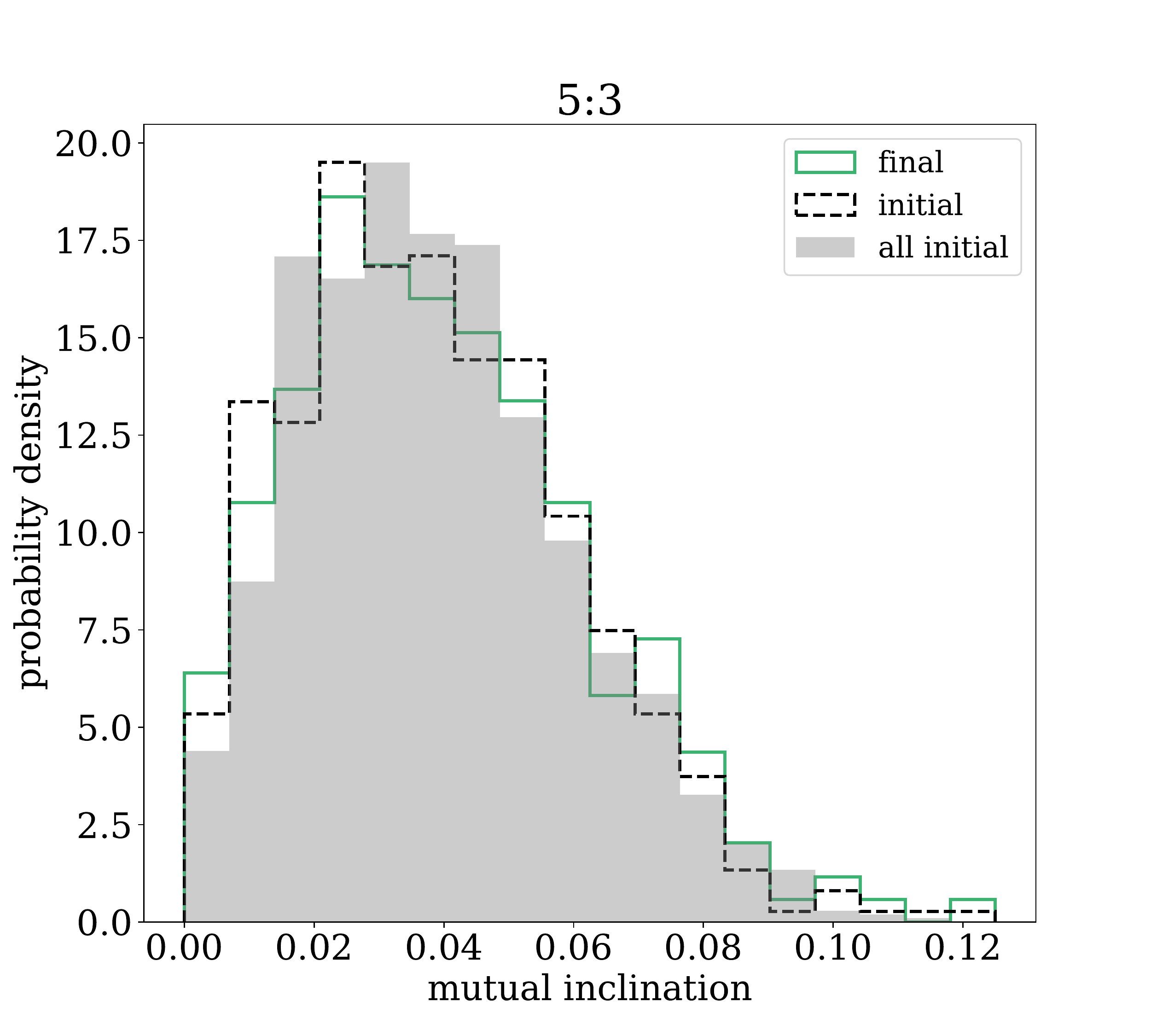}
	}
	\caption{Distribution of the instantaneous mutual inclination of the final selected planet pairs at the beginning (dashed black) and end (solid green) of the stability run. The gray background shows the initial mutual inclinations for all planet pairs, including those that went unstable and were excluded from the final selection. Distributions are shown for the \subref{fig:PRif-a} 3:1 and \subref{fig:PRif-b} 5:3 resonances.}
	\label{fig:mutincif}
\end{figure}

Figure~\ref{fig:mutincif} shows the distribution of the mutual inclination for the generated planet pairs before and after the stability run. There is little difference from stability sculpting, and we do not anticipate mutual inclination to play a significant role in this resonance behavior. The mutual inclination does affect the ability to detect both planets via transit; however, as we are modeling the observed population and not the intrinsic population, this effect is not examined in this work.

\begin{figure}[h]
	\centering
	\subfloat{
	    \label{fig:pomegaif-a}
	    \includegraphics[width=.5\linewidth]{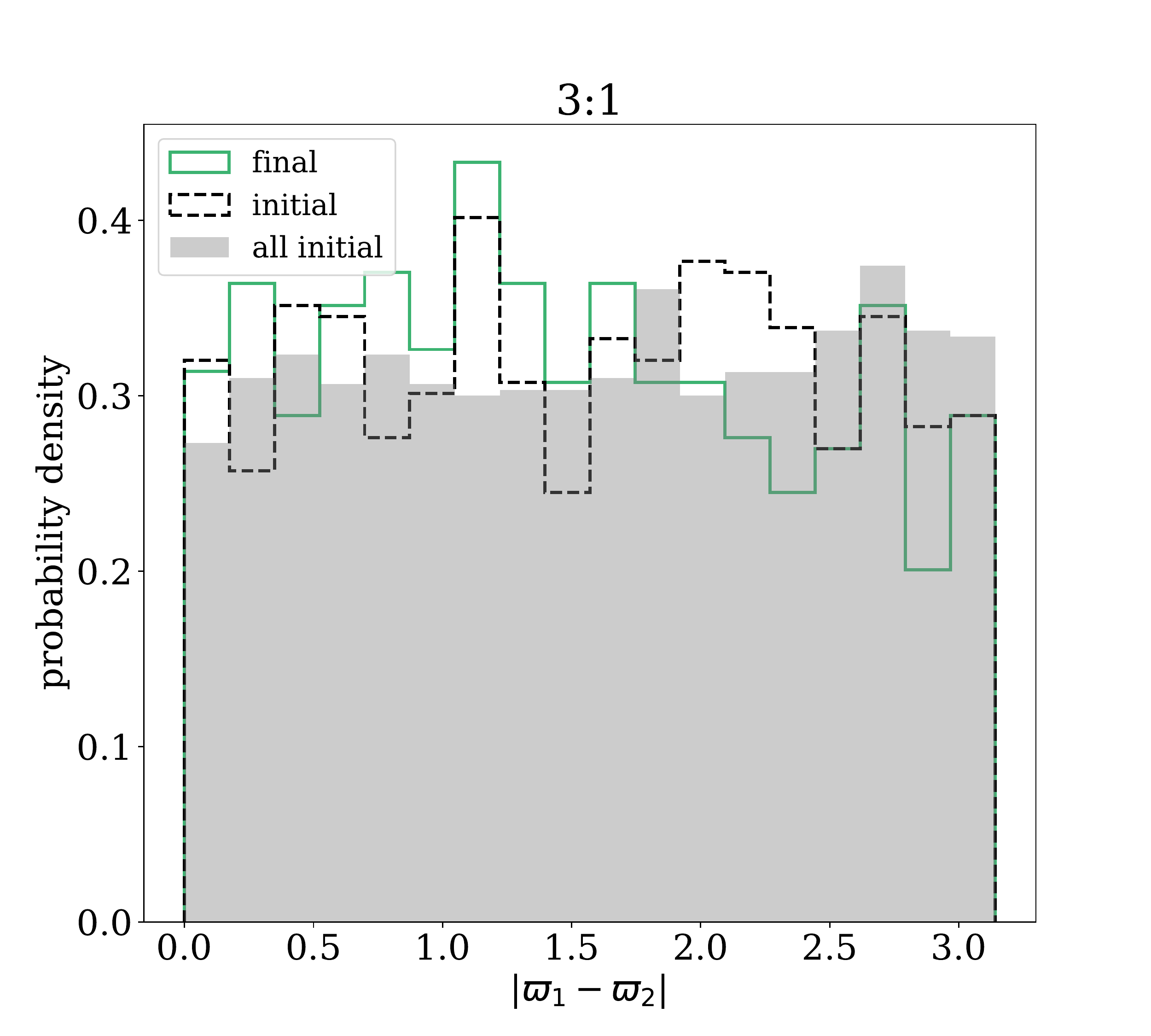}
	}
	\subfloat{
	    \label{fig:pomegaif-b}
	    \includegraphics[width=.5\linewidth]{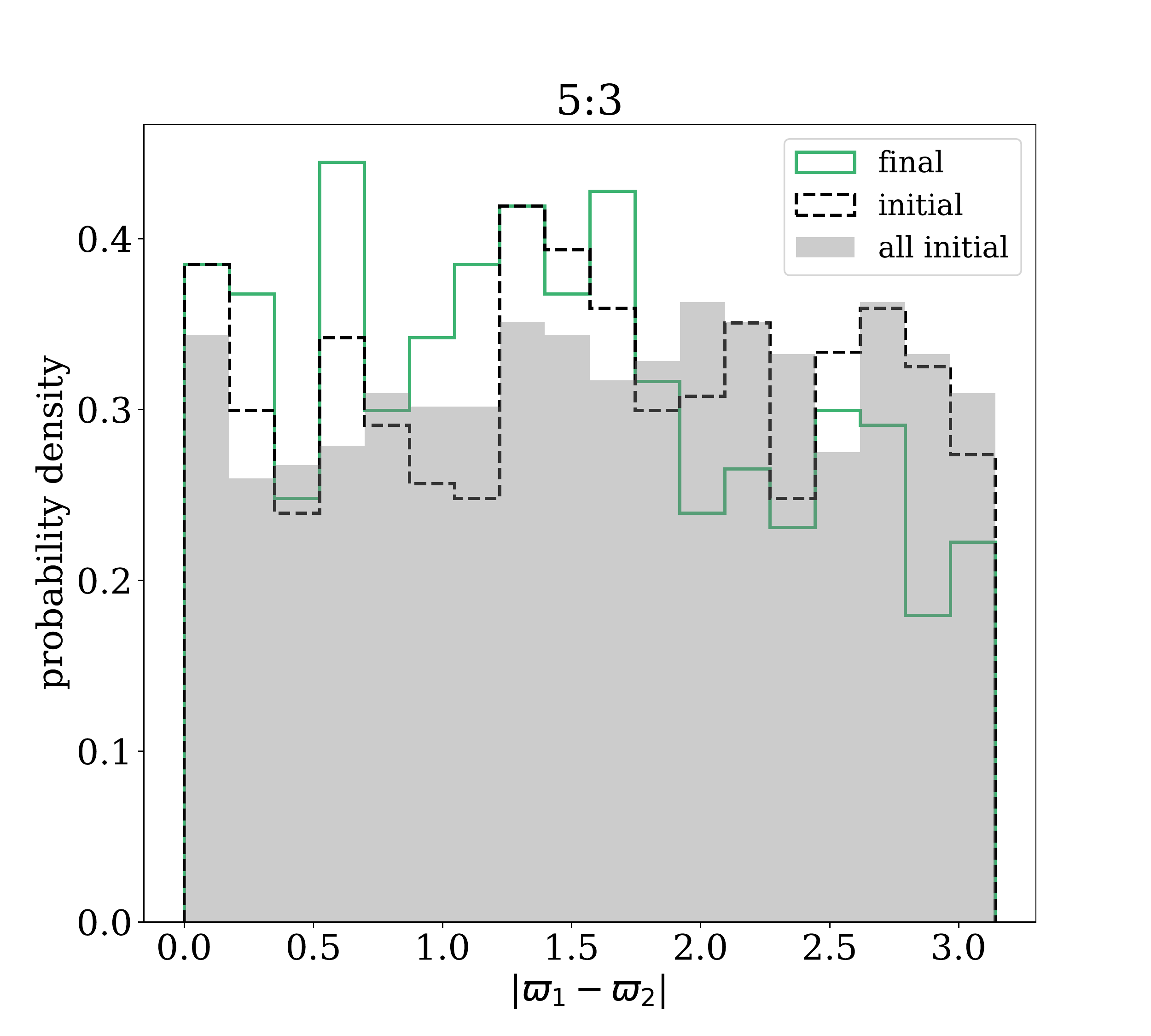}
	}
	\caption{Distribution of the instantaneous magnitude of the difference in the longitudes of the pericenter (such that 0 is aligned and $\pi$ is anti-aligned) of the final selected planet pairs at the beginning (dashed black) and end (solid green) of the stability run. The gray background shows the initial magnitude of the difference in the longitudes of the pericenter for all planet pairs, including those that went unstable and were excluded from the final selection. Distributions are shown for the \subref{fig:pomegaif-a} 3:1 and \subref{fig:pomegaif-b} 5:3 resonances. Their agreement shows the phasing of the orbits in the input systems and the randomized phasing share the same distributions.}
	\label{fig:pomegaif}
\end{figure}

\begin{figure}[h]
	\centering
	\subfloat{
	    \label{fig:omegaif-a}
	    \includegraphics[width=.5\linewidth]{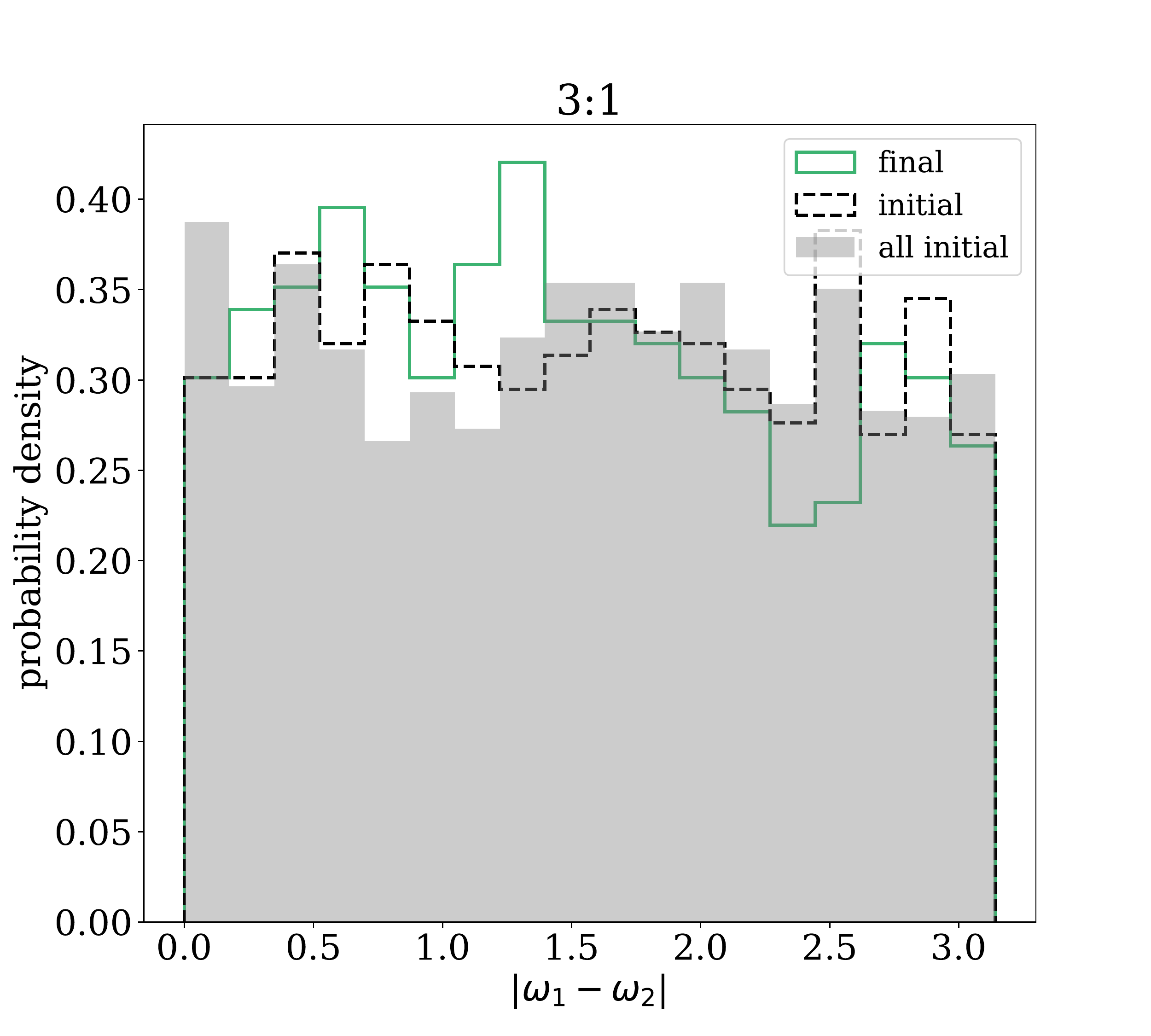}
	}
	\subfloat{
	    \label{fig:omegaif-b}
	    \includegraphics[width=.5\linewidth]{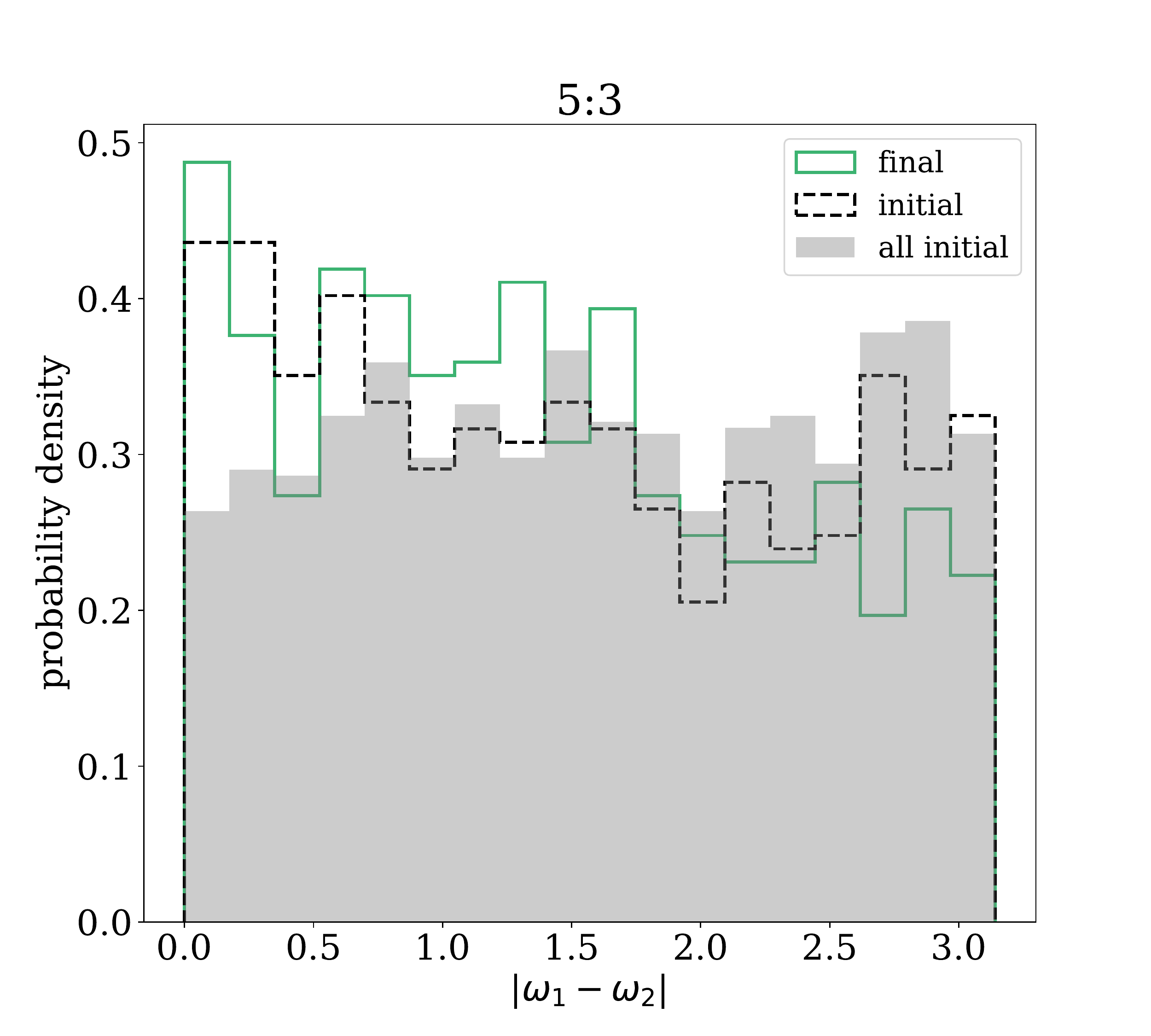}
	}
	\caption{Distribution of the instantaneous magnitude of the difference in the arguments of the pericenter (such that 0 is aligned and $\pi$ is anti-aligned) of the final selected planet pairs at the beginning (dashed black) and end (solid green) of the stability run. The gray background shows the initial magnitude of the difference in the arguments of the pericenter for all planet pairs, including those that went unstable and were excluded from the final selection. Distributions are shown for the \subref{fig:omegaif-a} 3:1 and \subref{fig:omegaif-b} 5:3 resonances.}
	\label{fig:omegaif}
\end{figure}

Figure~\ref{fig:pomegaif} shows the distribution of the magnitude of the difference in the longitudes of the pericenter ($|\varpi_1-\varpi_2|$, such that 0 is aligned and $\pi$ is anti-aligned) for the generated planet pairs, and Figure~\ref{fig:omegaif} the same for the arguments of the pericenter ($|\omega_1-\omega_2|$). There is little indication of stability sculpting in these distributions, except perhaps a slight preference for aligned arguments of the pericenter in stable pairs near the 5:3 resonance. The effect of the alignment on the dynamical behavior on non-secular timescales is expected to be small.

The distribution of the combined eccentricities for the generated planet pairs can be found in text in Figure~\ref{fig:eif}.

\bibliography{PR2MMR}{}
\bibliographystyle{aasjournal}

\end{document}